\documentclass[AMA,STIX1COL,doublespace]{WileyNJD-v2}

\usepackage{textcomp}
\usepackage{bbm}
\usepackage{comment}
\usepackage{graphicx, graphics}
\usepackage{algorithm}
\usepackage{url}
\usepackage{enumitem}
\usepackage{psfrag,pst-node,subfigure, amssymb, setspace, picture, epsfig, amsfonts, upgreek,amsmath} 
\usepackage{mathtools}
\usepackage{helvet}
\usepackage{booktabs}
\usepackage{siunitx}
\usepackage{tikz} 
\usetikzlibrary{shapes,backgrounds,shapes.misc, positioning,shapes.geometric,arrows,matrix,fit,calc,arrows.meta, trees, hobby, 
decorations.pathreplacing,calligraphy,decorations.markings}
\usepackage[figuresright]{rotating}
\usepackage{amsmath,amssymb,amsfonts,amsthm}

\usepackage{diagbox}

\usepackage{calc}

\usepackage{color}

\usepackage{multirow, makecell}

\usepackage{arydshln}

\usepackage{mathtools}

\usepackage{xr}

\newcommand{\argmin}{\operatornamewithlimits{argmin}}

\newcommand\bovermat[2]{%
	\makebox[0pt][l]{$\smash{\overbrace{\phantom{%
					\begin{matrix}#2\end{matrix}}}^{\text{#1}}}$}#2}

\DeclareSIUnit\periods{periods}
\DeclareSIUnit\weeks{week}
\DeclareSIUnit\days{days}
\DeclareSIUnit\hours{hours}
\DeclareSIUnit\minutes{minutes}

\def\bbeta{\boldsymbol \beta}

\def\wtbbeta{\widetilde{\boldsymbol \beta}}

\def\bgamma{\boldsymbol \gamma}
\def\bepsilon{\boldsymbol \epsilon}
\def\wtbepsilon{\widetilde{\boldsymbol \epsilon}}

\def\bSigma{\boldsymbol \Sigma}
\def\bnu{\boldsymbol \nu}

\def\betahat{\widehat{\boldsymbol \beta}}
\def\betatrue{\boldsymbol \beta^0}

\def\bfy{\mathbf{y}}
\def\bfY{\boldsymbol Y}
\def\bfF{\boldsymbol F}
\def\bfA{\boldsymbol A}
\def\bfB{\boldsymbol B}
\def\bfC{\boldsymbol C}
\def\bfc{\boldsymbol c}
\def\bfD{\boldsymbol D}
\def\bfI{\boldsymbol I}
\def\bfX{\boldsymbol X}

\def\bfH{\boldsymbol H}

\def\bfE{\boldsymbol E}
\def\calF{\mathcal{F}}
\def\calC{\mathcal{C}}
\def\calS{\mathcal{S}}
\def\calH{\mathcal{H}}
\def\calJ{\mathcal{J}}
\def\calI{\mathcal{I}}
\def\calE{\mathcal{E}}

\def\wtbfX{\widetilde{\boldsymbol X}}
\def\wtbfY{\widetilde{\boldsymbol Y}}

\def\bfZ{\boldsymbol Z}

\def\bfV{\boldsymbol V}
\def\bfx{\mathbf{x}}
\def\bfu{\mathbf{u}}
\def\bfzero{\boldsymbol 0}
\def\bzero{\boldsymbol 0}

\def\bQ{\boldsymbol Q}
\def\bfQ{\boldsymbol Q}

\def\bfI{\boldsymbol I}

\def\P{\mathrm{P}}

\def\tr{\text{tr}}

\def\calG{\mathcal{G}}

\def\calK{\mathcal{K}}

\def\vec{\text{vec}}

\tikzset{
	mybackground/.style={execute at end picture={
			\begin{scope}[on background layer]
				\draw[draw=blue!10,fill=blue!10,rounded corners=1.5ex] (current bounding box.south west)
				rectangle (current bounding box.north east);
				\node[draw,anchor=west,inner sep=1pt,minimum width=4ex] at (current bounding box.north
				west){#1};
			\end{scope}
	}},
}

\tikzset{
	buffer/.style={
		isosceles triangle,
		isosceles triangle apex angle=66,
		shape border rotate=90,
		draw,
		thick,
		fill=blue!20,
		node distance=5cm,
		rounded corners=60pt,
		opacity=0.6,
		minimum height=6.25cm
	}
}

\pgfdeclarelayer{foreground}
\pgfdeclarelayer{background}
\pgfsetlayers{background,%
	main,%
	foreground%
}

\newcommand{\cb}[1]{\textcolor{blue}{#1}}

\articletype{Article Type}%

\received{26 April 20xx}
\revised{6 June 20xx}
\accepted{6 June 20xx}

\begin{document}



\title{Doubly structured sparsity for grouped multivariate responses with application to functional outcome score modeling}

\author[1]{Jared D. Huling*}

\author[2,3]{Jennifer P. Lundine}

\author[4,5]{Julie C. Leonard}

\authormark{Huling, J.D. \textsc{et al}}

\address[1]{\orgdiv{Division of Biostatistics}, \orgname{University of Minnesota}, \orgaddress{\state{Minnesota}, \country{U.S.A.}}}

\address[2]{\orgdiv{Department of Speech and Hearing Science}, \orgname{The Ohio State University}, \orgaddress{\state{Ohio}, \country{U.S.A.}}}

\address[3]{\orgdiv{Division of Clinical Therapies and Inpatient Rehabilitation Program}, \orgname{Nationwide Children’s Hospital}, \orgaddress{\state{Ohio}, \country{U.S.A.}}}

\address[4]{\orgdiv{Division of Emergency Medicine, Department of Pediatrics}, \orgname{The Ohio State University College of Medicine}, \orgaddress{\state{Ohio}, \country{U.S.A.}}}

\address[5]{\orgdiv{Abigail Wexner Research Institute}, \orgname{Nationwide Children’s Hospital}, \orgaddress{\state{Ohio}, \country{U.S.A.}}}

\corres{*Jared D. Huling \email{huling@umn.edu}}

\abstract[Summary]{This work is motivated by the need to accurately model a vector of responses related to pediatric functional status using administrative health data from inpatient rehabilitation visits. The components of the responses have known and structured interrelationships. To make use of these relationships in modeling, we develop a two-pronged regularization approach to borrow information across the responses. The first component of our approach encourages joint selection of the effects of each variable across possibly overlapping groups related responses and the second component encourages shrinkage of effects towards each other for related responses. As the responses in our motivating study are not normally-distributed, our approach does not rely on an assumption of multivariate normality of the responses. We show that with an adaptive version of our penalty, our approach results in the same asymptotic distribution of estimates as if we had known in advance which variables were non-zero and which variables have the same effects across some outcomes. We demonstrate the performance of our method in extensive numerical studies and in an application in the prediction of functional status of pediatric patients using administrative health data in a population of children with neurological injury or illness at a large children's hospital.}


\maketitle


\section{Introduction}\label{sec:introduction}

Each year in the United States more than 20,000 children are diagnosed with an acute neurologic injury or illness that result in debilitating physical and cognitive complications, and reduced quality-of-life \citep{lo2009pediatric, patel2014pediatric, taylor2017traumatic, dhillon2017us}. These events are costly.  As an example, it is estimated that one billion dollars are spent annually on management of pediatric traumatic brain injury (TBI)-associated hospitalizations \citep{schneier2006incidence}. Though neurologic illnesses and injuries as a category are a leading cause of morbidity in children, each individual diagnosis is uncommon. These low disease-specific incidences make the study of rehabilitation interventions for children with neurologic injuries or illnesses challenging. Further, since many of the material effects manifest in the long term and can change with later child development, it is important to be able to track outcomes over time.

The WeeFIM\textregistered{} is a validated scoring system for appraising functional ability in children and provides valuable information about a multitude of components of health in children with neurologic injury or illness. WeeFIM\textregistered{} is administered and scored by trained assessors and has been demonstrated to have high interrater reliability. WeeFIM\textregistered{} has been shown to be predictive of longitudinal functional recovery of children with neurological disorders and can be used for discharge planning, prediction of functional outcomes, and documentation of functional performance of children over time to assess recovery or decline. 
Yet, because the WeeFIM\textregistered{} system requires trained assessors, scores for all children are not always available. 
WeeFIM\textregistered{} training and subscription is expensive and time-intensive and thus there is significant interest in being able to assess broadly the functional ability of children across a health system without the need for explicit scoring by trained assessors.  
One avenue for doing so is to relate administrative data sources to WeeFIM\textregistered{} scores to facilitate care management and identify individuals who may require additional care or rehabilitation. Our interest is thus in building predictive models to understand functional ability of children using administrative health data in scenarios where WeeFIM\textregistered{} is not used. WeeFIM\textregistered{} scores have particular characteristics and organization. In this work we seek to leverage these characteristics to facilitate the building of accurate and interpretable WeeFIM\textregistered{} risk models. We now describe these characteristics and how we aim to utilize them in this paper. 

\begin{figure}[!htpb]
	\centering
	\includegraphics[width=0.75\textwidth]{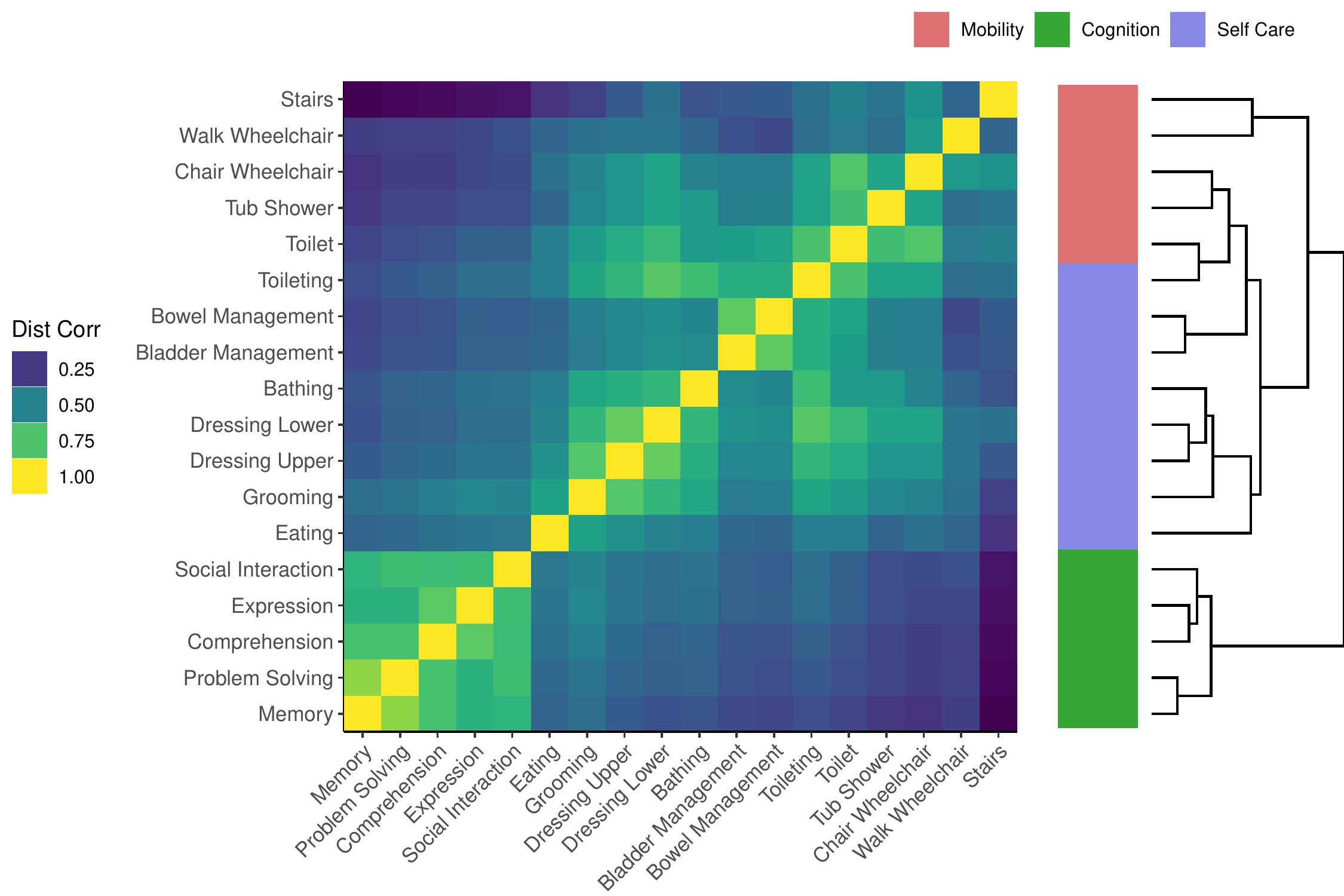}
	\caption{Heatmap of the distance correlations of the WeeFIM\textregistered{} scores. The dendrogram from hierarchical clustering of the scores shows that they group naturally by the three major WeeFIM\textregistered{} domains: self care, mobility, and cognition; thus the pre-defined domains align well with the natural variation in the components in our motivating study.}	
	\label{fig:weefim_corrs}
\end{figure}

WeeFIM\textregistered{} is comprised of 18 component scores, each of which is measured on a 7 point Likert scale with 7 indicating complete independence and 1 indicating complete dependence on others to perform various tasks. The WeeFIM\textregistered{} component scores are categorized into three main domains: mobility, cognition, and self-care, and represent three distinct clinical concepts related to functional outcomes. We describe the components of each domain in Section \ref{sec:application}.
 In our dataset, the correlations of the WeeFIM\textregistered{} component scores align closely with these domains. We computed the distance correlations \citep{szekely2007measuring} of all pairs of the 18 WeeFIM\textregistered{} components to form an $18\times 18$ distance matrix and ran a hierarchical clustering algorithm on the resulting distance matrix. As can be seen in Figure \ref{fig:weefim_corrs}, the scores in the cognition domain are immediately completely separated from the scores in mobility and self-care;  self-care and mobility separate from each other later in the hierarchical clustering.

 In our motivating study, we aim to use information commonly present in administrative health data including International Classification of Diseases (ICD) -9 and -10 codes, Current Procedural Terminology (CPT) codes, pharmaceutical codes, data indicating durable medical equipment use, and basic demographic information to predict and evaluate functional outcomes in children admitted to an inpatient rehabilitation unit with a diagnosis of neurologic injury or illness at a large children's hospital. Our code information is high-dimensional, with well over 500 billing codes as potential predictors of a child's functional status.

To deal with high-dimensionality, one could fit lasso-penalized regression models \citep{tibshirani96} for each of the 18 scores separately, however this would ignore the inherent relationships and similarities between the components. First, as some WeeFIM\textregistered{} components are highly related (e.g. dressing lower and dressing upper, bowel management and bladder management, bathing and tub shower), it is plausible that some variables (e.g. billing codes) may have the same or similar effects across some of the WeeFIM\textregistered{} components. Similarly, some variables may not have the same effect size across some WeeFIM\textregistered{} components, but are likely to be jointly predictive for multiple components, especially for related components, or even for all components. A complication is that while linearity of effects of covariates on the components may be reasonable, an assumption of normality of the responses or error terms is not, as the WeeFIM\textregistered{} component scores are not normally-distributed.

There exists some work for building models for multivariate responses using high-dimensional predictors.
Rothman, et al. \cite{rothman2010sparse} introduced a lasso-based estimation procedure that incorporates variable selection and covariance estimation for multivariate normal responses and developed an iterative estimation procedure that makes use of the covariance structure among the responses by penalizing the inverse covariance matrix using the methods of Yuan and Lin \cite{yuan2007model}.  However, this approach does not make use of similar predictor-response relationships across related responses and only borrows information across responses through their covariance structure. Sofer, et al. \cite{sofer2014variable} extended these ideas to incorporate a wider variety of penalties such as the SCAD penalty \citep{fan2001variable} and the MC+ penalty\cite{zhang2010nearly} and developed new algorithms for estimation. Li, et al. \cite{li2015multivariate} incorporated group structure both in responses and among predictors using an overlapping group lasso penalty \citep{jenatton2011}, which allows for borrowing of strength across responses through joint selection of variables across related responses. 

In our work, we borrow strength across responses in two ways through two different structured sparsity inducing penalties. In particular, we first aim to incorporate joint selection and removal of variables across all responses \textit{and} joint selection and removal of variables across related responses (i.e. responses that are in the same functional domain). In our application, the group structure is the natural groups formed by the WeeFIM\textregistered{} components displayed in Figure \ref{fig:weefim_corrs}.  Thus, we require use of a group lasso \citep{yuan2006} with overlapping groups \citep{jenatton2011} similar to Li, et al. \cite{li2015multivariate}. Second, we use a fused lasso \citep{tibshirani2005sparsity} for shrinkage of the effects of a variable across related outcomes to borrow strength more explicitly in estimation by partially collapsing models across different responses into a single model for individual variables separately. The fused lasso penalty allows a variable's effects across related responses to be estimated to be exactly the same. 
We prove that with adaptively chosen weights \cite{zou2006, wang2008note} but for the overlapping group lasso as in Huling, et al. \cite{huling2018risk} and fused lasso as in Viallon, et al. \cite{viallon2013adaptive}, our doubly-structured sparsity inducing estimator has an oracle property. The theoretical results are general in that they allow for any arbitrary overlapping group structure for the group lasso penalty and any arbitrary fused lasso penalty. The oracle property we show demonstrates that estimation for the non-zero coefficients has the same asymptotic distribution as if we had known in advance both i) which coefficients are non-zero and ii) which coefficients are equal to each other and fit models with only the non-zero coefficients and forcing the truly equal coefficients to be equal to each other. We also show selection consistency for the non-zero coefficients and selection consistency for which coefficients are equal to each other. Neither our methodology nor our theory requires an assumption of normality or independence of the responses, as we work under a semiparametric multivariate linear model assumption, where we make no parametric distributional assumptions about the error terms. Further, our theory, methodology, and computational approach allows for general use of the overlapping group lasso in combination with the fused lasso beyond our application to multivariate response regression.

The remainder of our paper is organized as follows. In Section \ref{sec:methods_def} we introduce the key definition of our methodology. In Section \ref{sec:asymptotics} we develop theory for our methodology using adaptive regularization under semiparametric multivariate linear models and in Section \ref{sec:simulation} we investigate the operating characteristics of our method in small samples using simulation experiments. In Section \ref{sec:application} we demonstrate the use of our methodology on our motivating application involving pediatric functional outcomes score modeling. Finally, we conclude with discussion.


\section{Doubly Structured Sparsity for Multivariate Responses}\label{sec:methods}

\subsection{The Overlapping Group + Fused Method}\label{sec:methods_def}

We assume that the observed multivariate response $\bfy_i = (y_{i1}, \dots, y_{iK})^\top$ follows the semiparametric linear model
\begin{align}
	\bfy_i = {\betatrue}^\top\bfx_i + \boldsymbol\epsilon_i \text{ for } i=1,\dots,N, \label{lin_model}
\end{align}
where $\bfx_i = (x_{i1}, \dots, x_{ip})^\top$ is a length $p$ vector of predictors, $\betatrue$ is a $p\times K$ matrix of regression coefficients with $k$th column equal to $\betatrue_{\cdot,k} = (\beta^0_{1,k}, \dots, \beta^0_{p,k})^\top$, and $\bepsilon_i = (\epsilon_{i1}, \dots, \epsilon_{iK})^\top$ is a vector of errors with mean zero and finite, positive definite variance-covariance matrix $\bSigma$; no requirement that the response vector be continuous is required, as we do not assume a particular distribution for $\boldsymbol\epsilon_i$. Denote the $j$th variable's coefficients across the $K$ response as $\betatrue_{j,\cdot} = (\beta^0_{j,1}, \dots, \beta^0_{j,K})^\top$. Let $\bfX$ denote the $N\times p$ matrix of predictors with $i$th row as $\bfx_i^\top$ and $j$th column as $\bfx_{\cdot,j}$ and let $\bfY$ denote the $N\times K$ response matrix with $i$th row as $\bfy_i^\top$ and denote the $k$th column of $\bfY$ as $\bfY_{\cdot,k}$.
The typical least squares estimator is the solution of 
\begin{align*}
	\argmin_{\bbeta} (2N)^{-1}\tr\left[(\bfY - \bfX\bbeta)(\bfY - \bfX\bbeta)^\top\right],
\end{align*}
which is simply $(\bfX^\top\bfX)^{-1}\bfX^\top\bfY$, where for a matrix $\bfA$, $\tr(\bfA)$ indicates its trace. 

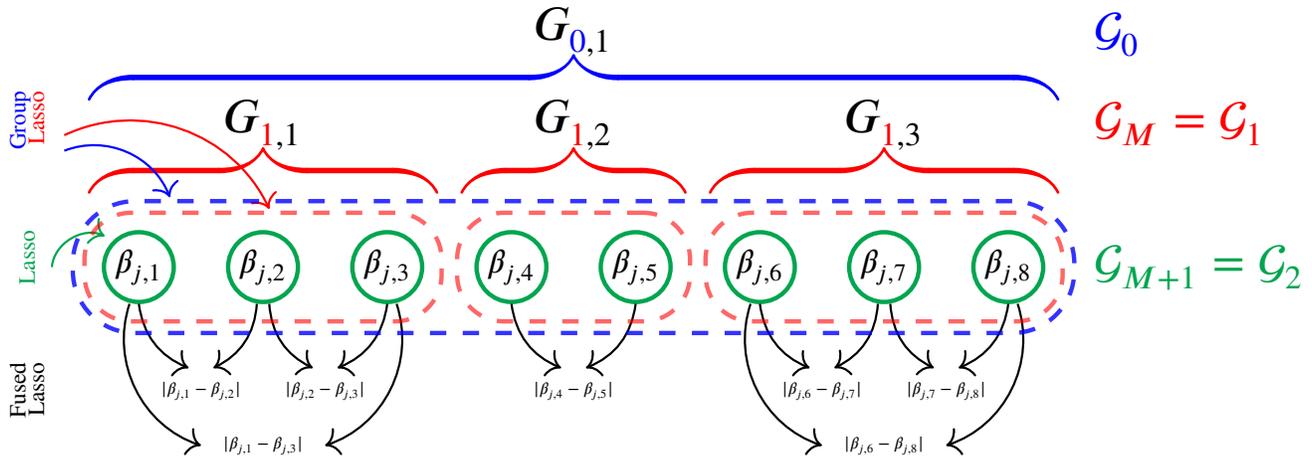
\begin{figure}[!htpb]
	\centering
	\resizebox{1\textwidth}{!}{
		\tikzstyle{background rectangle}=
		[draw=blue!8,fill=blue!8,rounded corners=1.5ex]
		\begin{tikzpicture}[y=1cm, x=0.8cm, thick, font=\footnotesize,
			vc/.style = {
				circle, draw, thick, fill=#1,
				minimum width=12mm, opacity=0.7},
			vg/.style args = {#1/#2}{
				minimum height=12mm,
				minimum width=#1+\pgfkeysvalueof{/pgf/minimum height},
				thick,
				draw, rounded corners=\pgfkeysvalueof{/pgf/minimum height}/2, 
				opacity=1,
				sloped},
			frames/.style args = {#1/#2}{minimum height=#1,
				minimum width=#2+\pgfkeysvalueof{/pgf/minimum height},
				draw, thick, rounded corners=3mm, opacity=0.6, color=red, dashed,
				sloped},
			framesbig/.style args = {#1/#2}{minimum height=#1,
				minimum width=#2+\pgfkeysvalueof{/pgf/minimum height},
				draw, thick, rounded corners=4mm, opacity=0.8, color=blue, dashed,
				sloped}]
			\usetikzlibrary{arrows,decorations.pathreplacing}
			
			\tikzset{number line/.style={}}
			
			\tikzset{
				brace_top/.style={
					pen colour=blue,
					line width=1.0pt,
					decoration={calligraphic brace,amplitude=8pt},
					decorate
				},
				brace_top_red/.style={
					pen colour=red,
					line width=1.0pt,
					decoration={calligraphic brace,amplitude=8pt},
					decorate
				},
				brace_bottom/.style={
					pen colour=red,
					decoration={calligraphic brace,amplitude=8pt, mirror},
					decorate
				}
			}
			
			\begin{scope}[xshift=0.9cm]
				\begin{scope}[xshift=-0.27cm]

					\foreach \i in {1,...,8}
					{
						\node[draw,thick,color=Green,circle, inner sep=0.1, minimum size=1.5em,scale=0.85] (coef\i) at (\i-1,0){$\textcolor{black}{\beta_{j,\i}}$};
					}
					
					\node (start_week) at (-0.4,0.95) {};
					\node (end_week) at (7.4,0.95) {};
					\draw [brace_top] (start_week.north) -- node [above=4pt, pos=0.5] {\scriptsize $G_{\textcolor{blue}{0},1}$} (end_week.north);

					\coordinate [label={[black, align=left,scale=1.15]right: $\textcolor{blue}{\calG_0}$}] (fl) at (7.5,1.5);
					\coordinate [label={[black, align=left,scale=1.15]right: $\textcolor{red}{\calG_M = \calG_1}$}] (fl) at (7.5,0.925);
					\coordinate [label={[black, align=left,scale=1.15]right: $\textcolor{Green}{\calG_{M+1} = \calG_2}$}] (fl) at (7.5,0);

					\node (start_week) at (-0.4,0.35) {};
					\node (end_week) at (2.4,0.35) {};
					\draw [brace_top_red] (start_week.north) -- node [above=4pt, pos=0.5] {\scriptsize $G_{\textcolor{red}{1},1}$} (end_week.north);
					
					\node (start_week) at (2.6,0.35) {};
					\node (end_week) at (4.4,0.35) {};
					\draw [brace_top_red] (start_week.north) -- node [above=4pt, pos=0.5] {\scriptsize $G_{\textcolor{red}{1},2}$} (end_week.north);
					
					\node (start_week) at (4.6,0.35) {};
					\node (end_week) at (7.4,0.35) {};
					\draw [brace_top_red] (start_week.north) -- node [above=4pt, pos=0.5] {\scriptsize $G_{\textcolor{red}{1},3}$} (end_week.north);

					\coordinate [label={[black, align=right,rotate=90,scale=0.85]left:\tiny Fused\\[-5pt]\tiny Lasso}] (fl) at (-0.9,-0.5);
					
					\coordinate [label={[black, align=right,rotate=90,scale=0.85]left:\tiny \textcolor{Green}{Lasso}}] (fl) at (-0.9,0.35);
					
					\coordinate [label={[black, align=right,rotate=90,scale=0.85]left:\tiny \textcolor{blue}{Group}\\[-5pt]\tiny \textcolor{red}{Lasso}}] (fl) at (-0.9,1.25);
					
					\draw [->,thin,Green] (-0.7,0)  to [bend right=-55] (-0.275,0.2);

					\draw [->,thin,red] (-0.6,0.85)  to [bend right=-45] (1.05,0.375);
					\draw [->,thin,blue] (-0.6,0.75)  to [bend right=-45] (0.25,0.45);


					\draw [->,thin] (coef1.south)++(0,0cm)  to [bend right] (0.4,-0.65) node[below right, draw=none] {};
					\draw [->,thin] (coef2.south)++(-0.05,0cm)  to [bend left] (0.6,-0.65) node[below left, draw=none] {};
					
					\node[scale=0.425] (fuse12) at (0.5,-0.8) {$|\beta_{j,1}-\beta_{j,2}|$};
					
					\draw [->,thin] (coef2.south)++(0.05,0cm)  to [bend right] (1.4,-0.65) node[below right, draw=none] {};
					\draw [->,thin] (coef3.south)++(0,0cm)  to [bend left] (1.6,-0.65) node[below left, draw=none] {};
					
					\node[scale=0.425] (fuse23) at (1.5,-0.8) {$|\beta_{j,2}-\beta_{j,3}|$};
					
					\draw [->,thin] (coef1.south)++(-0.075,0cm)  to [bend right=45] (0.5,-1.15) node[below right, draw=none] {};
					\draw [->,thin] (coef3.south)++(0.075,0cm)  to [bend left=45] (1.5,-1.15) node[below left, draw=none] {};
					
					\node[scale=0.425] (fuse13) at (1,-1.15) {$|\beta_{j,1}-\beta_{j,3}|$};
					
					\draw [->,thin] (coef4.south)++(0,0cm)  to [bend right] (3.4,-0.65) node[below right, draw=none] {};
					\draw [->,thin] (coef5.south)++(0,0cm)  to [bend left] (3.6,-0.65) node[below left, draw=none] {};
					
					\node[scale=0.425] (fuse45) at (3.5,-0.8) {$|\beta_{j,4}-\beta_{j,5}|$};
					
					\draw [->,thin] (coef6.south)++(0,0cm)  to [bend right] (5.4,-0.65) node[below right, draw=none] {};
					\draw [->,thin] (coef7.south)++(-0.05,0cm)  to [bend left] (5.6,-0.65) node[below left, draw=none] {};
					
					\node[scale=0.425] (fuse67) at (5.5,-0.8) {$|\beta_{j,6}-\beta_{j,7}|$};
					
					\draw [->,thin] (coef7.south)++(0.05,0cm)  to [bend right] (6.4,-0.65) node[below right, draw=none] {};
					\draw [->,thin] (coef8.south)++(0,0cm)  to [bend left] (6.6,-0.65) node[below left, draw=none] {};
					
					\node[scale=0.425] (fuse78) at (6.5,-0.8) {$|\beta_{j,7}-\beta_{j,8}|$};
					
					\draw [->,thin] (coef6.south)++(-0.075,0cm)  to [bend right=45] (5.5,-1.15) node[below right, draw=none] {};
					\draw [->,thin] (coef8.south)++(0.075,0cm)  to [bend left=45] (6.5,-1.15) node[below left, draw=none] {};
					
					\node[scale=0.425] (fuse68) at (6.0,-1.15) {$|\beta_{j,6}-\beta_{j,8}|$};
					
				\end{scope}

			\end{scope}

			\begin{pgfonlayer}{main}
				\path   let \p2 = ($(coef1.center)-(coef8.center)$),
				\n2 = {veclen(\y2,\x2)} in
				(coef1) -- node[framesbig=8.5mm/\n2] {} (coef8);
				\path   let \p2 = ($(coef6.center)-(coef8.center)$),
				\n2 = {veclen(\y2,\x2)} in
				(coef6) -- node[frames=7mm/\n2] {} (coef8);
				\path   let \p2 = ($(coef1.center)-(coef3.center)$),
				\n2 = {veclen(\y2,\x2)} in
				(coef1) -- node[frames=7mm/\n2] {} (coef3);
				\path   let \p2 = ($(coef4.center)-(coef5.center)$),
				\n2 = {veclen(\y2,\x2)} in
				(coef4) -- node[frames=7mm/\n2] {} (coef5);
			\end{pgfonlayer}
		\end{tikzpicture}
	}
	\caption{An illustration of the structure of the two components of the penalization methods used to induce group-wise selection of covariate effects and fuse them to be more similar. \textcolor{blue}{\textbf{Blue}} indicates joint selection across all responses via the group $\calG_0$, \textcolor{red}{\textbf{red}} indicates joint selection by response group $\calG_M = \calG_1$, \textcolor{Green}{\textbf{green}} indicates  individual coefficient selection achieved via the group $\calG_{M+1}$,  and \textbf{black} indicates shrinkage towards a common effect within response group $G_{1,k}$.}
	\label{fig:regularizer_structure}
\end{figure}

In our setting, the $K$ outcomes are grouped into $M$ different, possibly overlapping groupings that describe inherent relationships between the outcomes. In our motivating data, the WeeFIM outcomes are grouped into three pre-defined domains that represent different categories of functional status. In other settings, these domains may be further subcategorized. In this manner, the groupings  generally correspond to hierarchically-organized categorizations of the outcomes. In this work, we make use of the natural grouping of the outcomes by adding regularizers that induce the coefficients of each different variable to be selected or removed jointly across an entire group of the responses.  The $j$th grouping $\calG_j$ consists of a set of $g_j$ groups $G_{j,1}, \dots, G_{j,g_j}$ such that the union of the groups includes all of the $K$ outcomes.  
	Generally, since the groupings will be hierarchically defined,  $g_j$ will be larger than $g_{j+1}$ and groups in $\calG_{j}$ can be expressed as unions of groups in $\calG_{j+1}$.  
More formally, the groupings are written as $\calG_1 = \{G_{1,1}, \dots, G_{1,g_1}\}, \dots, \calG_M = \{G_{M,1}, \dots, G_{M,g_M}\}$, where $G_{m,g} \in \{1,\dots,K\} \equiv \calK$ and $\bigcup_{G\in\calG_m}G = \calK$ for $m=1,\dots, M$. 
To encourage a variable to be selected jointly across \textit{all} responses, we define the trivial group $\calG_0 = \{\calK\} $, thus $G_{0,1} = \calK  = \{1, \dots, K \}$. To allow variables to be selected or removed individually, we further include the group $\calG_{M+1} = \{\{1\}, \dots, \{K\}\}$, thus $G_{M+1,k} = \{k\}$. The group structure for a toy example with $K=8$ responses and $M=1$ groupings is displayed in Figure \ref{fig:regularizer_structure}. 
	We note that in this toy example, the groups are overlapping (e.g. $G_{1,1} \subset G_{0,1}$) because the groups are hierarchically structured. In our motivating application, the group structure we use involves such overlapping groups. 
	
	In our setting the outcome groupings are pre-defined, but in general when such a pre-defined set of groups are not available \textit{a priori},
	the $M$ groupings can be formed by iteratively refined, hierarchical groupings of the responses, as illustrated in Figure \ref{fig:weefim_corrs}; for example, the $M$ groupings could be formed by clustering the responses at $M$ different levels in a hierarchical clustering of the responses.

For a particular group $G\in\calK$ with size $|G|$, define the length $|G|$ subvector of $\bbeta_{j,\cdot}$ limited to the outcomes in group $G$ as $\bbeta_{j,G}$.
We propose to add to the above objective function two regularizers that induce sparsity in a structured manner. Let $\betahat$ be the solution of the following problem
\begin{align}
	\argmin_{\bbeta} (2N)^{-1}\tr\left[(\bfY - \bfX\bbeta)(\bfY - \bfX\bbeta)^\top\right] + \lambda_1 P_1(\bbeta) + \lambda_2 P_2(\bbeta) \label{overlapping_linear_model_objective}
\end{align}
where $$P_1(\bbeta) = \sum_{j=1}^p\sum_{m=0}^{M+1}\sum_{G \in \mathcal{G}_m} \lambda_{1,j,G} ||\bbeta_{j,G}||_2$$ and 
$$P_2(\bbeta) = \sum_{j=1}^p\sum_{G\in \calG_M}\sum_{l, o \in G: l \neq o } \lambda_{2,j,l,o}|{\beta}_{j,l} - {\beta}_{j,o}|,$$
where the penalty $\lambda_{1,j,G} ||\bbeta_{j,G}||_2$ is an overlapping group lasso penalty encourages joint selection or removal of the effects of variable $j$ across the responses in response group $G$. All possible {zero patterns} of coefficients can be represented as unions of groups due to the results of Jenatton, et al \cite{jenatton2011}; hence {nonzero patterns} of coefficients can be thought of as complements of unions of groups. Thus, the grouping $\calG_0$ encourages joint selection and removal of effects for a variable across all outcomes simultaneously and grouping $\calG_{M+1}$ allows for individual selection or removal for each outcome separately, while the groupings $\calG_{1}, \dots, \calG_M$ allow for selection or removal of effects across related groups.
The second penalty $\lambda_{2,j,l,o}|{\beta}_{j,l} - {\beta}_{j,o}|$ is a fused lasso penalty that encourages the effect of variable $j$ on response $l$ and on response $o$ to be more similar. Thus, $P_1(\bbeta)$ and $P_2(\bbeta)$, respectively, help borrow strength across the $K$ responses by 1) leveraging the natural groupings of the responses and taking advantage of joint significance of predictors across related responses and 2) encouraging the effect estimates for a particular variable to be similar across the most related responses according to the response groupings.  These two regularizers incorporate structural knowledge about outcomes in two manners by utilizing two different types of structured sparsity inducing penalties. The overall structure of the regularizers are depicted in Figure \ref{fig:regularizer_structure} in a toy example with $K=8$ responses. In our formulation, we only add a fused lasso penalty within the final grouping $\calG_M$, reflective of the notion that with an iteratively refined grouping, the final groups are likely to have more similar outcomes within each group. However, the fused lasso penalty can be added within any grouping and can even be set so that fused lasso terms are included for every possible pair of responses; our asymptotic results cover any arbitrary set of fused pairs. In our application, the multivariate outcomes are measured on the same scale, but our approach can be adapted to outcomes that have different scales by standardizing the outcomes prior to analysis. Further, as we list below, our adaptive penalization allows for the fused lasso penalty to adapt to observed differences in effects across outcomes.

To perform adaptive penalization, we take $\lambda_{1,j,G} = ||\betahat^{OLS}_{j,G}||_2^{-\gamma_1}$ and $\lambda_{2,j,l,m} = |\hat{\beta}_{j,l}^{OLS} - \hat{\beta}_{j,m}^{OLS}|^{-\gamma_2}$, where $\gamma_1, \gamma_2 > 0$ and $\betahat^{OLS} = (\bfX^\top\bfX)^{-1}\bfX^\top\bfY$ is the ordinary least squares estimate of $\bbeta$. We study this adaptive version of our estimator in Section \ref{sec:asymptotics}. For non-adaptive penalization, we take $\lambda_{1,j,G} = |G|^{1/2}$, as is used in Jenatton, et al. \cite{jenatton2011} and, Huling et al. \cite{huling2018risk} and $\lambda_{2,j,l,m}=1$, however for this choice selection consistency is not guaranteed without an irrepresentable condition \citep{jenatton2011}.
In order to get reliable estimates of out-of-sample performance using cross validation when the group structure is derived in a data-driven manner, the procedure used to estimate the group structure should be applied in each cross validation fold instead of being fixed.

\subsection{Implementation Details}\label{sec:impl_details}

In our implementation, we use a generic ADMM algorithm described in the next section for computation and feed into it the response vector $\wtbfY$ and design matrix $\wtbfX$, where the latter is treated as a sparse matrix object that only stores the values and locations of the non-zero entries of $\wtbfX$, which both dramatically saves space, memory used, and reduces computation time. The sparse matrix object is of the type provided in the \texttt{Eigen C++} linear algebra library \citep{eigenweb} with interface to an \texttt{R} package through \texttt{RcppEigen} \citep{rcppeigen}, allowing for highly efficient sparse-matrix manipulations. Our \texttt{R} package implementation of our method is \texttt{groupFusedMulti} available in the open source repository \url{https://github.com/jaredhuling/groupFusedMulti}, which uses an interface similar to the interface used in \texttt{glmnet}.

For ease of tuning the tuning parameters, we utlize the following re-parameterization of $\lambda_1$ and $\lambda_2$. Instead of using the penalty $\lambda_1 P_1(\bbeta) + \lambda_2 P_2(\bbeta)$, we use $\lambda (1-\alpha) P_1(\bbeta) + \lambda\alpha P_2(\bbeta)$, where $\lambda \geq 0$ and $\alpha\in [0,1]$ so that $\alpha$ controls the proportion of the total penalty that the fused lasso comprises.

\subsection{Computation via a Multi-block ADMM Algorithm}\label{sec:admm}

We utilize an alternating direction method of multipliers (ADMM) \citep{glowinski1975, gabay1976, boyd2011} algorithm for optimization. The ADMM algorithm works by decomposing an objective function and solving the decomposed subproblems iteratively, where each subproblem has a simple and computationally tractable form. ADMM solves problems of the form
\begin{align*}
	& \mbox{minimize } f(\bbeta) + P(\bgamma) \mbox{ subject to } \bfA\bbeta + \bfB\bgamma = \bfc
\end{align*}
where $\bfA$ and $\bfB$ are constraint matrices, both $f$ and $P$ are convex functions, $\bbeta$ is the parameter vector of interest, and typically $f$ represents some loss function and $P$ represents a penalty. In the simplest case the constraint is of the form $\bbeta = \bgamma$ and the purpose of the constraint is to find a new variable for which the penalty is equivalent, but separable across loss and penalty in the terms of the new variable. To optimize our objective, we require the following multi-block version of ADMM 
\begin{align*}
	& \mbox{minimize } f(\bbeta) + P_1(\bgamma)  + P_2({\boldsymbol \eta}) \mbox{ subject to } \bfA\bbeta + \bfB\bgamma + \bfC{\boldsymbol \eta} = \bfc,
\end{align*}
where $\bfC$ is an additional constraint matrix that allows the second penalty to have a decomposed form with loss and the two penalties all separable from each other.
To solve the above problem, the augmented Lagrangian is formed as:
\begin{align*}
	L_\rho(\bbeta, \bgamma, {\boldsymbol \eta}, \bnu) = {} &  f(\bbeta) + P_1(\bgamma)  + P_2({\boldsymbol \eta}) + \bnu^\top (\bfA\bbeta + \bfB\bgamma +  \bfC{\boldsymbol \eta} - \bfc) \\
	& + (\rho/2)||\bfA\bbeta + \bfB\bgamma +  \bfC{\boldsymbol \eta}- \bfc||^2_2,
\end{align*}
where $\rho$ is any strictly positive number.
The multi-block ADMM algorithm iterates by alternatingly minimizing with respect to $\bbeta$, $\bgamma$, and ${\boldsymbol \eta}$ and following these minimizations, updating the Lagrangian parameter $\bnu$:
\begin{align}
	\bbeta^{(t + 1)} = {} &  \argmin_{\bbeta}L_\rho(\bbeta, \bgamma^{(t)}, {\boldsymbol \eta}^{(t)}, \bnu^{(t)}) \label{eqn:betamin}\\
	\bgamma^{(t + 1)} = {} & \argmin_{\bgamma}L_\rho(\bbeta^{(t + 1)}, \bgamma, {\boldsymbol \eta}^{(t)}, \bnu^{(t)})\label{eqn:gammamin} \\
	{\boldsymbol \eta}^{(t+1)} = {} & \argmin_{\boldsymbol \eta}L_\rho(\bbeta^{(t + 1)}, \bgamma^{(t + 1)}, {\boldsymbol \eta}, \bnu^{(t)})\label{eqn:etamin} \\
	\bnu^{(t + 1)} = {} &  \bnu^{(t)} + \rho (\bfA\bbeta^{(t + 1)}  + \bfB\bgamma^{(t + 1)}  + \bfC {\boldsymbol \eta}^{(t+1)} - \bfc) \nonumber
\end{align}
where $t$ indexes the iteration number. The standard ADMM has been shown to converge for any $\rho > 0$ and the multi-block ADMM algorithm above has been shown to converge under certain conditions on $\bfA$, $\bfB$, and $\bfC$. These conditions, from Chen, et al \cite{chen2016direct}, are met if either $\bfA^\top \bfB = \bfzero$,  $\bfB^\top \bfC = \bfzero$, or $\bfA^\top \bfC = \bfzero$.

The following describes the multi-block ADMM algorithm applied to the overlapping group lasso with fused lasso problem. Let $m = \sum_{G \in \mathcal{G}}|G|$, $g = |\mathcal{G}|$, and suppose $\mathcal{G}=\{G_1,\cdots,G_g\}$, let $\bfF = (\bfF_1, \dots, \bfF_g)$ be a matrix of dimension $m \times Kp$ where $\bfF_l$ is a $|G_l|\times Kp$ matrix with $(i,j)$th entry equal to 1 if $j$ is the $i^{th}$ element of group $G_l$, and 0 otherwise, $\forall j=1,\dots,g$. Then $\bfF\bbeta$ is a vector of length $m$ comprised components of $\bbeta$ where each element of $\bbeta$ appears in $\bfF\bbeta$ the total number of times it appears in any group. For example, if p=1, K=3, $ \bbeta = (\beta_1, \beta_2, \beta_3)^\top $ and $\mathcal{G} = \{ \{ 1, 2\}, \{2, 3 \} \}$, then
$$
\bfF =
\begin{pmatrix}
	1 & 0 & 0 \\
	0 & 1 & 0 \\
	0 & 1 & 0 \\
	0 & 0 & 1
\end{pmatrix}
\mbox{ and }
\bfF\bbeta =
\begin{pmatrix}
	\beta_1 \\
	\beta_2 \\
	\beta_2 \\
	\beta_3
\end{pmatrix}.
$$
In this example, the penalty $P_1(\bgamma)$ is
$
=\lambda_1(\lambda_{1,G_1}||(\gamma_1,\gamma_2)||_2+\lambda_{1,G_2}||(\gamma_3,\gamma_4)||_2),
$ which is a standard group lasso penalty on the variable $\bgamma$.
In general, the overlapping group lasso penalty $P_1(\bgamma)$ can be written as
$
P_1(\bgamma)=\lambda_1\sum_{l=1}^g\lambda_{1,G_l}||\bgamma_{l}||_2,
$
where $\bgamma=(\bgamma_{1},\dots,\bgamma_{g})$ and $\bgamma_{l}$ is a $|G_l|$-dimensional vector.
Note that for the nonoverlapping group lasso, $\bfA = \bfI_{Kp}$, where $\bfI_{Kp}$ is the identity matrix of dimension $Kp\times Kp$.
To accommodate the fused lasso penalty, we allow for a more general penalty, the generalized fused lasso penalty. Denote the matrix $\bfD$ with $e$ rows and $Kp$ columns where each row represents a pair of effects which have a fused lasso penalty applied. Thus $e$ is the total number of fused lasso terms/pairs. For example, if the pair of effects $\beta_{j,\ell}$ and $\beta_{j,m}$ are have a fused lasso applied, then the row of $\bfD$ corresponding to this pair will have $\ell$th position taking the value 1 and the $m$th position taking the value $-1$. In the toy example in Figure \ref{fig:regularizer_structure}, if $\beta_1$ has a fused lasso penalty with $\beta_2$ and $\beta_2$ has a fused lasso penalty with $\beta_3$, then 
$$
\bfD =
\begin{pmatrix}
	1 & -1 & \hphantom{-}0 \\
	0 & \hphantom{-}1 & -1
\end{pmatrix},
\mbox{ and }
\bfD\bbeta =
\begin{pmatrix}
	\beta_1 - \beta_2\\
	\beta_2 - \beta_3 
\end{pmatrix}.
$$
In this example, the $P_2(\boldsymbol\eta)=\lambda_2(\lambda_{2,1}|\eta_1| + \lambda_{2,2}|\eta_2|)$, which is a standard lasso penalty on the variable $\boldsymbol \eta$.
In general, the fused lasso penalty $P_2(\boldsymbol\eta)$ can be written as
$
P_2(\beta)=\lambda_2\sum_{l=1}^e\lambda_{2,l}|\eta_l|,
$
where $\boldsymbol\eta$ is an $e$-dimensional vector.
The ADMM algorithm for the overlapping group lasso plus the fused lasso is constructed by taking $$\bfA = \begin{pmatrix}
	\bfF \\
	\bfD
\end{pmatrix}, \bfB = \begin{pmatrix}
	-\bfI_m \\
	\boldsymbol 0
\end{pmatrix}, \bfC = \begin{pmatrix}
	\boldsymbol 0 \\
	-\bfI_e
\end{pmatrix}, 
\text{ and } \bfc = \bfzero,$$ which meets the condition that $\bfB^\top\bfC=\bfzero$, meaning a multi-block ADMM algorithm using this setup is valid and convergent. When $f(\bbeta) = \frac{1}{2}|| \wtbfY - \wtbfX\bbeta||^2_2$, step (\ref{eqn:betamin}) for the overlapping group lasso is simply the solution of $(\wtbfX^\top\wtbfX + \rho (\bfF^\top \bfF+ \bfD^\top \bfD) )\bbeta = \wtbfX^\top\wtbfY + (\bfF^\top, \bfD^\top) \bnu^{(t)} + \rho (\bfF^\top \bgamma^{(t)} + \bfD^\top{\boldsymbol \eta}^{(t)})$.  When $f(\bbeta)$ is the negative log-likelihood, step (\ref{eqn:betamin}) can be carried out by Newton-Raphson or other standard optimization techniques. As step (\ref{eqn:gammamin}) is group-separable, it can be minimized by minimizing with respect to each group $\bgamma_{l}$ independently. This is achieved by the block soft-thresholding operator $S_{\lambda_1 \lambda_{G_l} / \rho}((\bfF\bbeta^{(t+1)})_{l} - \bnu_{l}^{(t)}/ \rho)$, where $S_\lambda(\bfu) = \bfu\left(  1 - \lambda / ||\bfu||_2  \right)_+$ and $(\bfF\bbeta^{(t+1)})_{l}$ and $\bnu_{l}^{(t)}$ are defined in the same way as $\bgamma_l$. Since \eqref{eqn:etamin} is separable and is equivalent to a type of lasso penalization problem, it can be similarly minimized simply via soft thresholding: $S_{\lambda_2 \lambda_{j} / \rho}((\bfD\bbeta^{(t+1)})_{j} - \bnu_{j}^{(t)}/ \rho)$, where with some abuse of notation $(\bfD\bbeta^{(t+1)})_{j}$ is the $j$th element of $(\bfD\bbeta^{(t+1)})$,  $\bnu_{j}^{(t)}$ is the $j$th element of $\bnu^{(t)}$, and $S_\lambda(u) = \text{sign}(u)\left(  |u| - \lambda  \right)_+$, which is the equivalent of the block soft thresholding operator applied to a scalar.
Our convergence criterion is the same as suggested in Section 3.3.1 of Boyd, et al \cite{boyd2011} with $\epsilon^{\mbox{abs}} = \epsilon^{\mbox{rel}} = 10^{-5}$. The parameter $\rho$ for the ADMM algorithm used is the adaptive scheme described in Section 3.4.1 of Boyd, et al \cite{boyd2011}.

\subsection{Connections with existing literature}\label{sec:lit_connections}

The methods of Rothman, et al \cite{rothman2010sparse} and Li, et al \cite{li2015multivariate} both also address scenarios where high dimensional covariates are used to predict a set of multivariate outcomes in a penalized regression framework. These two methods in addition to ours borrow strength in information across outcomes through penalization techniques, albeit in different ways.

If the multivariate outcomes are only correlated due to correlations between the outcomes themselves, then the multivariate regression with covariance estimation (MRCE) approach  \cite{rothman2010sparse} may be most appropriate as it directly incorporates the covariance of the responses via a penalized log-likelihood; here, the lasso penalty induces the estimated coefficients to be a function of the covariance of the responses, unlike the unpenalized maximum likelihood estimator. However, when many covariates  are expected to have effects of similar magnitude across the multivariate outcomes, our additional fused lasso penalty is designed to make use of this by borrowing information across outcomes for such covariates at a more granular level. Further, when there is strong group structure of whether or not a given covariate has an effect across groups of outcomes, but the magnitudes of the effects are unrelated, then both our approach and the multinomial logistic regression with sparse group lasso (MSGL) approach  \cite{li2015multivariate} are appropriate options and their performances should be expected to be similar, despite our additional fused lasso penalty, whereas MRCE is not designed explicitly to make use of such information. However, if indeed there are many covariates with similar effects across outcomes and a group structure, our approach may be expected to perform well. If there is no relationship whatsoever across the multivariate outcomes, an approach of simply fitting a lasso-penalized model separately for each outcome may have reasonable performance.


\subsection{Asymptotic Properties}\label{sec:asymptotics}

We now present new asymptotic results for our penalization methods pertaining to semiparametric linear models. We show results for a more general, encompassing form for the overlapping group lasso penalty $P_1$ and the fused lasso penalty $P_2$. We allow $P_1(\bbeta) = \sum_{j=1}^p\sum_{g\in\calG}\lambda_{1,j,G}||\bbeta_{j,G}||_2$ for an arbitrary, potentially overlapping group structure $\calG$ and we allow $P_2(\bbeta) = \sum_{j=1}^p\sum_{(l,o) \in \calF} \lambda_{2,j,l,o}|{\beta}_{j,l} - {\beta}_{j,o}|$ for an arbitrary fusing set $\calF$ that contains pairs of indices of coefficients to be fused, where the elements of $\calF$ are of the form $(l,o)$ with $l,o\in\calK$. The penalty form introduced in Section \ref{sec:methods_def} is a special case of the form we study in this section, as our Theorem \ref{thm:linear_model_oracle_property_mis} applies to any group structure, overlapping or not. 
We also denote the set of all pairs of coefficients which are equal and non-zero for the $j$th variable by $\calE_{j,\cdot} = \{ (l,m) \in \mathcal{F}: \beta_{j,l}^0 = \beta_{j,m}^0 \neq 0  \}$. Similarly, define $\hat{\calE}_{j,\cdot} = \{ (l,m) : l,m \in \mathcal{F} \mbox{ and } \hat{\beta}_{j,l} = \hat{\beta}_{j,m} \neq 0  \}$. Denote the union of these sets over all variables as $\calE = \bigcup_{j=1}^p\calE_{j,\cdot}$ and $\hat{\calE} = \bigcup_{j=1}^p\hat{\calE}_{j,\cdot}$.

To facilitate our explanation of the asymptotic results, for any vector ${\bf a}_{\cdot,k}\in \mathbbm{R}^p$ associated with response $k$ and a index set ${\calI}_{\cdot,k}\subset\{1,\dots,p\}$ of size $|{\calI}_{\cdot,k}|$ associated with response $k$, ${\bf a}_{{\calI}_{\cdot,k}}$ represents a $|{\calI}_{\cdot,k}|$ dimensional sub-vector with elements in ${\bf a}_{\cdot,k}$ indexed by ${\calI}_{\cdot,k}$. 
Now let $\wtbfX = \bfI_K\ \otimes \bfX  = \mbox{diag}(\bfX, \dots, \bfX)$ be the block diagonal design matrix with $K$ blocks, one for each outcome, with the $k$th block as the design matrix $\bfX$, where $\otimes$ is the Kronecker product; similarly let $\widetilde{\bbeta}^0 = \vec(\bbeta^0) = ({\betatrue_{\cdot,1}}^\top, \dots, {\betatrue_{\cdot,K}}^\top)^\top$ be the vectorization of the true coefficients for all $K$ responses, let $\wtbfY^\top = (\bfY_{\cdot,1}^\top, \dots, \bfY_{\cdot,K}^\top)$. Using this notation, we can re-express the model \eqref{lin_model} as $\wtbfY = \wtbfX \widetilde{\bbeta}^0 + \wtbepsilon$, where $\wtbepsilon$ is the stacked error vector defined similarly as $\wtbfY$.

Now let $\calJ = (\calJ_{\cdot,1}, \dots, \calJ_{\cdot,K}) \subseteq \{1, \dots, Kp\}$, where $\calJ_{\cdot,k} = \{j+(k-1)p: j\in\{1,\dots,p\} \text{ and } \beta^0_{j,k} \neq 0 \}$, to be the set of indices of all non-zero effects in $\widetilde{\bbeta}^0$ and let $\calH = \text{Hull}(\calJ) = \left\{  \cup_{G\in \mathcal{G}, G\cap \calJ = \varnothing}G  \right\}^c$ be the hull of the nonzero pattern $\calJ$, where for a set $\calS$, $\calS^c$ denotes its complement. The hull of the non-zero pattern is essentially the smallest set of groups in $\calG$ that contains all elements in $\calJ$.  Similarly, denote $\hat{\calJ} = (\hat{\calJ}_{\cdot,1}, \dots, \hat{\calJ}_{\cdot,K})$ to be the set of indices of all nonzero variables in $\betahat$. Note that by construction of $P_1(\cdot)$ due to each effect of each variable across the outcomes having its own group in $\calG_{M+1}$,  $\text{Hull}(\hat{\calJ}) = \hat{\calJ}$, however we present our results in terms of the hull so as to be fully general with respect to the group structure. We denote $\wtbfX_H$ as the columns in $\wtbfX$ corresponding to variables in $\calH$ and similarly denote $\wtbbeta^0_H$ as the values in $\wtbbeta^0$ corresponding to elements in $\calH$. 
Then denote $\wtbfX_\calH^*$  constructed by dropping and collapsing columns of $\wtbfX_\calH$ corresponding to the \textit{distinct}, non-zero values of $\wtbbeta^0_\calH$. Specifically, for each $j$, let $\{c_{j,1}, \dots, c_{j,L}\}$ for $L\leq K$ denote the unique \textit{non-zero} values of $\{\widehat{\beta}_{j,1}, \dots, \widehat{\beta}_{j,k}\}$ and denote $\calC_{j,\ell} = \{k\in\calK:  \widehat{\beta}_{j,k} = c_{j,\ell}  \}$. Then for each $c_{j,\ell}\in\{c_{j,1}, \dots, c_{j,L}\}$
all columns $j+(k-1)p$ of $\wtbfX$ such that $k\in \calC_{j,\ell}$, $j\in\{1,\dots,p\}$ are collapsed and added together into a single column, i.e. $\sum_{k\in \calC_{j,\ell}} \wtbfX_{\cdot,j+(k-1)p}$ . As an illustration, denote $\wtbfX_{\cdot,j+(k-1)p}$ as the $j$th column of the design matrix for response $k$. Then if the coefficient for variable $j$ is in $\calH$ for responses $k, \ell$, and $K$ and $\beta^0_{j,k} = \beta^0_{j,\ell} = \beta^0_{j,K} \neq 0$, this results in the following to the corresponding columns in $\wtbfX$ in the formulation of $\bfX^*_\calH$:
$$
\stackrel{\mbox{$\vphantom{\wtbfX_{\cdot,j+(k-1)p}}$}}{%
\begin{matrix}
	\vphantom{\boldsymbol 0} \\[15pt]
	 \vphantom{\vdots} \\
	\mbox{Block } k\rightarrow \\ 
	\vphantom{\vdots} \\
	\mbox{Block } \ell\rightarrow \\ 
	 \vphantom{\vdots} \\
	\mbox{Block } K\rightarrow \\  \\
\end{matrix}
}
\stackrel{\mbox{$\wtbfX_{\cdot,j+(k-1)p}$}}{%
\begin{pmatrix}
\boldsymbol 0 \\
\vdots \\
\bfx_{\cdot,j} \\ 
\vdots \\
\boldsymbol 0 \\
\vdots \\
\boldsymbol 0 \\
\end{pmatrix}
} 
+ 
\stackrel{\mbox{$\wtbfX_{\cdot,j+(\ell-1)p}$}}{%
\begin{pmatrix}
\boldsymbol 0 \\
\vdots \\
\boldsymbol 0 \\
\vdots \\
\bfx_{\cdot, j} \\ 
\vdots \\
\boldsymbol 0 \\
\end{pmatrix}
}
 + 
 \stackrel{\mbox{$\wtbfX_{\cdot,j+(K-1)p}$}}{%
\begin{pmatrix}
\boldsymbol 0 \\
\vdots \\
\boldsymbol 0 \\
\vdots \\
\boldsymbol 0 \\
\vdots \\
\bfx_{\cdot, j} \\
\end{pmatrix} 
}
\rightarrow
\begin{pmatrix}
\boldsymbol 0 \\
\vdots \\
\bfx_{\cdot, j} \\
\vdots \\
\bfx_{\cdot, j} \\
\vdots \\
\bfx_{\cdot, j} \\
\end{pmatrix}
$$
In other words, the columns in $\wtbfX$ corresponding to these three variables are collapsed and added together.  Note that there exist matrices $\bfH$ and $\bfE$ such that $\wtbfX_\calH^* = \wtbfX\bfH\bfE$, where $\bfH$ is formed by removing all columns of positions in the indices of $\calH$ from the identity matrix of dimension $pK\times pK$ and $\bfE$ is formed by taking the identity matrix of dimension $|\calH|\times |\calH|$ and collapsing and summing columns corresponding to non-zero coefficients that are equal to each other and are in $\calF$, the fusing set used in $P_2$.

We assume the following standard regularity conditions

\begin{enumerate}

	\item[] (D.1) $\lim_{N \to \infty} \frac{1}{N}\bfX^\top \bfX \to \bQ$ where $\bQ$ is positive definite

	\item[] (D.2) The errors $\boldsymbol\epsilon_i$ for all samples $i=1\dots,n$ are $i.i.d.$ random vectors with mean zero and finite, positive definite variance-covariance matrix $\bSigma$.  

\end{enumerate}

The following result pertains to cases where the group structure has been misspecified. As such, we define ${\bbeta^*}^0$ to be the \textit{distinct} values of ${\bbeta}^0_\calH$ (i.e. the union of the distinct values in $\{\beta^0_{j,k},\beta^0_{j,\ell}: (k,\ell) \in \calE \text{ or } (\ell,k) \in \calE \text{ and }  (k,\ell) \in \calF \text{ or } (\ell,k) \in \calF\}$ and all remaining values $\beta^0_{j,k} \neq 0$)  appended with zeros corresponding to the elements of ${\bbeta}^0_{\calH^c}$. Similarly, we define $\betahat^*$ to be the distinct values of ${\betahat}_{\hat{\calJ}}$ appended with zeros corresponding to the elements of ${\betahat}_{\hat{\calJ}^c}$.

\begin{theorem} \label{thm:linear_model_oracle_property_mis}

$\mbox{}$

Assume the data are generated under the model described in (\ref{lin_model}) and that our estimator is given by (\ref{overlapping_linear_model_objective}). Furthermore,
assume conditions (D.1) and (D.2) and let $\lambda_{1,j,G} = ||\betahat^{OLS}_{j,G}||_2^{-\gamma_1}$ and $\lambda_{2,j,l,m} = |\hat{\beta}_{j,l}^{OLS} - \hat{\beta}_{j,m}^{OLS}|^{-\gamma_2}$, where $\gamma_1, \gamma_2 > 0$ such that $N^{(\gamma_\ell + 1)/2}\lambda \to \infty$ for $\ell=1,2$. If $\sqrt{N}\lambda \to 0$, then we have the following:
\begin{align}
	& P(\hat{\calJ}_{\cdot,j} = {\calH}_{\cdot,j}) \to 1 \mbox{ as }  N \to \infty,  \label{linear_model_selection_consistency_mis} \\
	 & P(\hat{\calE}_{\cdot,j} = {\calE}_{\cdot,j}) \to 1  \mbox{ as }  N \to \infty, \label{linear_model_fused_selection_consistency_mis}
\end{align}
for each $j=1,\dots,p$	and
\begin{align}
	 \sqrt{N}(\betahat^* - {\bbeta^*}^0) \xrightarrow{d} \bfZ^*, \label{linear_model_asympt_distr_mis}
\end{align}	
	  where $\bfZ^* = ({{}\bfZ_{\calH}^*}^\top, {\boldsymbol 0}^\top)^\top$  and $\bfZ_{\calH}^* \sim N_{|\calH^*|}(0,  {\bfQ^*}_{\calH}^{-1}{\bfV^*}_{\calH}{\bfQ^*}_{\calH}^{-1} )$, where  $\bfV^*_\calH = \bfE^\top\bfH^\top \left(\bSigma \otimes \bfQ\right)\bfH\bfE$, $\bfQ^*_\calH = \bfE^\top\bfH^\top \left(\bfI_K \otimes \bfQ\right)\bfH\bfE$, and $|\calH^*|$ denotes the number of distinct non-zero elements in the true coefficients ${\bbeta^*}^0$.

\end{theorem}

The proof of Theorem \ref{thm:linear_model_oracle_property_mis} is provided in the Supplementary Material Appendix A. We reiterate that Theorem \ref{thm:linear_model_oracle_property_mis} applies to \textit{any} group structure, including those with overlapping groups.
We note that the variance-covariance matrix in the distribution of $\bfZ^*$ does not simplify due to the allowance of selection of variables individually by outcome in combination with the Kronecker-product variance structure of the response vector $\wtbfY$; without either it would simplify considerably. The terms $\bfH$ and $\bfE$ in the asymptotic variance of the estimates are due to selection of variables and collapsing of effects of a single variable across outcomes, respectively; without any selection these matrices are removed and the variance simplifies to the usual asymptotic variance of multivariate response regression via least squares. However, the asymptotic variance matches that of the estimator where all and only all truly non-zero variable effects are included in a model and all variables with equal effects across a subset of outcomes are accordingly collapsed. The result of Theorem \ref{thm:linear_model_oracle_property_mis} does not require that $\calE \subseteq \calF$, however, if indeed $\calE \subseteq \calF$, i.e.
if $\calF$ contains indices corresponding to all pairs of coefficients in $\bbeta^0_{\calH}$ that are equal to each other, then due to \eqref{linear_model_asympt_distr_mis}, when using adaptive terms for the tuning parameters, our approach yields selection consistency for the hull of the non-zero terms; when the group structure has individual groups for each of the individual coefficients in the model, the convex hull of the non-zero coefficients is simply the set of non-zero coefficients. Thus, when such individual groups are included, we have selection consistency for the non-zero coefficients. Our theory also shows that with probability tending to one, our approach will fuse together the coefficients that are truly equal to each other. Finally, our theory shows that our estimate will converge to the same asymptotic distribution as if we had known which coefficients were non-zero and which coefficients were equal to each other.


\section{Numerical Experiments}\label{sec:simulation}

\subsection{Estimators used}\label{sec:comparator_methods}

In this section, we conduct simulation experiments to assess the small sample operating characteristics of our proposed doubly-structured sparsity inducing estimation approach in comparison with several other state-of-the-art approaches for multivariate regression in high-dimensions. In our simulations, we vary both the dimensionality of the problem and the sample size. We also consider data-generating mechanisms with varying degrees of sparsity of variable effects that align with grouping of the outcomes and further consider scenarios where the effects of each variable have either minimal similarity across outcomes or a varying degree of similarity across outcomes. This allows our studies to explore under what data-generating mechanisms our two penalties indeed help in estimation. We use adaptive and non adaptive versions of our method, denoted {OGFM(adapt)} and {OGFM}, respectively, where OGFM indicates Overlapping Group $+$ Fused Multivariate regression. For the adaptive version, we set $\gamma_1=\gamma_2=0.5$ and for high dimensional settings with $p\geq n$, we used marginal regression estimates for the adaptive weights as in Huang, et al. \cite{huang2008}.
We compare our approach with the approach of Li, et al. \cite{li2015multivariate} (denoted as {MSGL}), which allows for overlapping group lasso penalties. For {MSGL}, we use the \texttt{MSGLasso} \texttt{R} package version 2.1 We further compare with a simple approach of fitting a separate lasso-penalized linear regression model for each outcome, with the tuning parameter chosen separately for each outcome (denoted as {Sep-Lasso}) and do so using the \texttt{glmnet} \texttt{R} package version 4.1-3. We also compare with the approach of Rothman, et al. \cite{rothman2010sparse} (denoted {MRCE}) implemented in the  \texttt{MRCE} \texttt{R} package version 2.1. For all methods, the tuning parameters are chosen by 10-fold cross validation; {MSGL} has two tuning parameters (one for a lasso penalty, one for a group lasso penalty), {MRCE} has two tuning parameters (one for a lasso penalty for the coefficients for the variable effects on responses, another for a lasso penalty on the elements of the inverse covariance matrix of the residuals), {Sep-Lasso, for separate lasso} has a tuning parameter for each outcome for a lasso penalty on coefficients, and the {OGFM} approaches have two tuning parameters (one for the overlapping group lasso penalty and another for the fused lasso penalty).  For both the {OGFM} approaches and {MSGL}, the group lasso penalty applied corresponds to the true underlying group structure of the data-generating mechanism.

\subsection{Data generation}

For each replication of the simulation, we generate data under model \eqref{lin_model}, where $\bfx_i$ are generated as i.i.d. multivariate normal random variables with mean vector $\bzero$ and covariance matrix $\bSigma_{\bfX} = [\sigma_{Xjk}]_{j,k=1}^p$, where $\sigma_{Xjk} = 0.5^{|j-k|}$, as was used in the simulations of Yuan and Lin \cite{yuan2007model} and Rothman, et al. \cite{rothman2010sparse}. The responses are generated according to model \eqref{lin_model}, where the error vectors $\bepsilon_i$ are generated as i.i.d. multivariate normal random variables with mean vector $\bzero$ and covariance matrix $\bSigma_{\bepsilon} = [\sigma_{\epsilon jk}]_{j,k=1}^K$, where $\sigma_{\epsilon jk} = 4\left(0.5^{|j-k|}\right)$. In the Supplementary Material Appendix B.2, we explore a simulation setting with binary covariates and a Likert scale outcome similar to our motivating data.
The dimensionality of the outcome/error vector is $K=8$ and the outcomes form 3 groups, with the first three outcomes forming group 1, the fourth and fifth forming group 2, and the last three outcomes forming group 3. The variable effects are generated as
\vspace{10pt}

\begin{normalsize}
\begin{equation*}
	\bbeta^0 = 
\begin{pmatrix}
	\bovermat{$G_{1,1}$}{ \xi_{1,1} \xi^G_{1,1} \eta_{1,1} & \xi_{1,2} \xi^G_{1,1} \eta_{1,2} & \xi_{1,3} \xi^G_{1,1} \eta_{1,3} } & \bovermat{$G_{1,2}$}{\xi_{1,4} \xi^G_{1,2} \eta_{1,4} & \xi_{1,5} \xi^G_{1,2} \eta_{1,5}} & \bovermat{$G_{1,3}$}{\xi_{1,6} \xi^G_{1,3} \eta_{1,6} & \xi_{1,7} \xi^G_{1,3} \eta_{1,7} & \xi_{1,8} \xi^G_{1,3} \eta_{1,8}} \\
	\vdots & & & & & & & \vdots \\
	\xi_{z,1} \xi^G_{z,1} \eta_{z,1} & \xi_{z,2} \xi^G_{z,1} \eta_{z,2} & \xi_{z,3} \xi^G_{z,1} \eta_{z,3} & \xi_{z,4} \xi^G_{z,2} \eta_{z,4} & \xi_{z,5} \xi^G_{z,2} \eta_{z,5} & \xi_{z,6} \xi^G_{z,3} \eta_{z,6} & \xi_{z,7} \xi^G_{z,3} \eta_{z,7} & \xi_{z,8} \xi^G_{z,3} \eta_{z,8} \\
	0 & \cdots & & & & &  \cdots & 0 \\
	\vdots & \ddots & & & & & & \vdots \\
	0 & \cdots & & & & &  \cdots & 0
\end{pmatrix},
\end{equation*}
\end{normalsize}
where the last $p-z$ rows of $\bbeta^0$ have all elements as 0, the terms $\xi_{j,k}\sim$ Bernoulli$(0.9)$ induce sparsity at the individual effect level, the terms $\xi^G_{j,1}, \xi^G_{j,2}, \xi^G_{j,3} \sim$ Bernoulli$(1-p_{\text{HS}})$ induce sparsity at the group level, the variable effect size terms $\eta_{j,k}$ are distributed i.i.d. uniformly from $\{-1, -0.5, -0.25, -0.125, 0.125, 0.25, 0.5, 1\}$ to create both small and large effects, and for each variable $j$ separately the terms $\eta_{j,k}$ for $k$ in the same group are set to be all equal to each other with probability $p_{\text{GE}}/2$ and independently, the terms $\eta_{j,k}$ for all outcomes $k=1,\dots,8$ are set to be all equal to $\eta_{j,1}$ with probability $p_{\text{GE}}/2$. The latter process induces effects for some variables within a group to be equal to each other and induces effects for some variables to be equal for all outcomes. 

We explore dimensions of $p=50, 100,$ and $200$ and set $z=25, 50, 50$ for each dimension setting, respectively. We explore hierarchical sparsity parameters from $p_{\text{HS}} = 0, 0.25$ and $0.5$ and fusing probabilities $p_{\text{GE}} = 0, 0.5, 0.75$ and $0.95$; when $p_{\text{HS}} = 0$, there is no group-wise sparsity and thus any group lasso penalty applied is superfluous. For each replication of the simulation, we additionally generate an independent test set of size 10000 for use in evaluation of predictive performance.

\subsection{Performance evaluation}

We evaluate methods in terms of three metrics: root mean squared error of predictions (RMSE) on a large, independent test set of size 10000 generated anew for each simulation replication, model error defined below, and balanced accuracy which is the average of the true positive rate (TPR) and true negative rate (TNR), described in detail in the Supplementary Material {Appendix B.1 and not to be confused with standard classification accuracy}. As our primary interest is performance of predictions on validation data,  it is our primary focus of evaluation. For the RMSE, we compute the RMSE for each outcome separately and then average the RMSEs across the outcomes. Model error for a given estimate is defined as ME$(\widehat{\bbeta}, \bbeta^0) = \text{tr}\left[  (\widehat{\bbeta} - \bbeta^0)^\top \bSigma_{\bfX} (\widehat{\bbeta} - \bbeta^0) \right]$. 

\subsection{Results}

\begin{figure}[!htpb]
	\centering
	\includegraphics[width=1\textwidth]{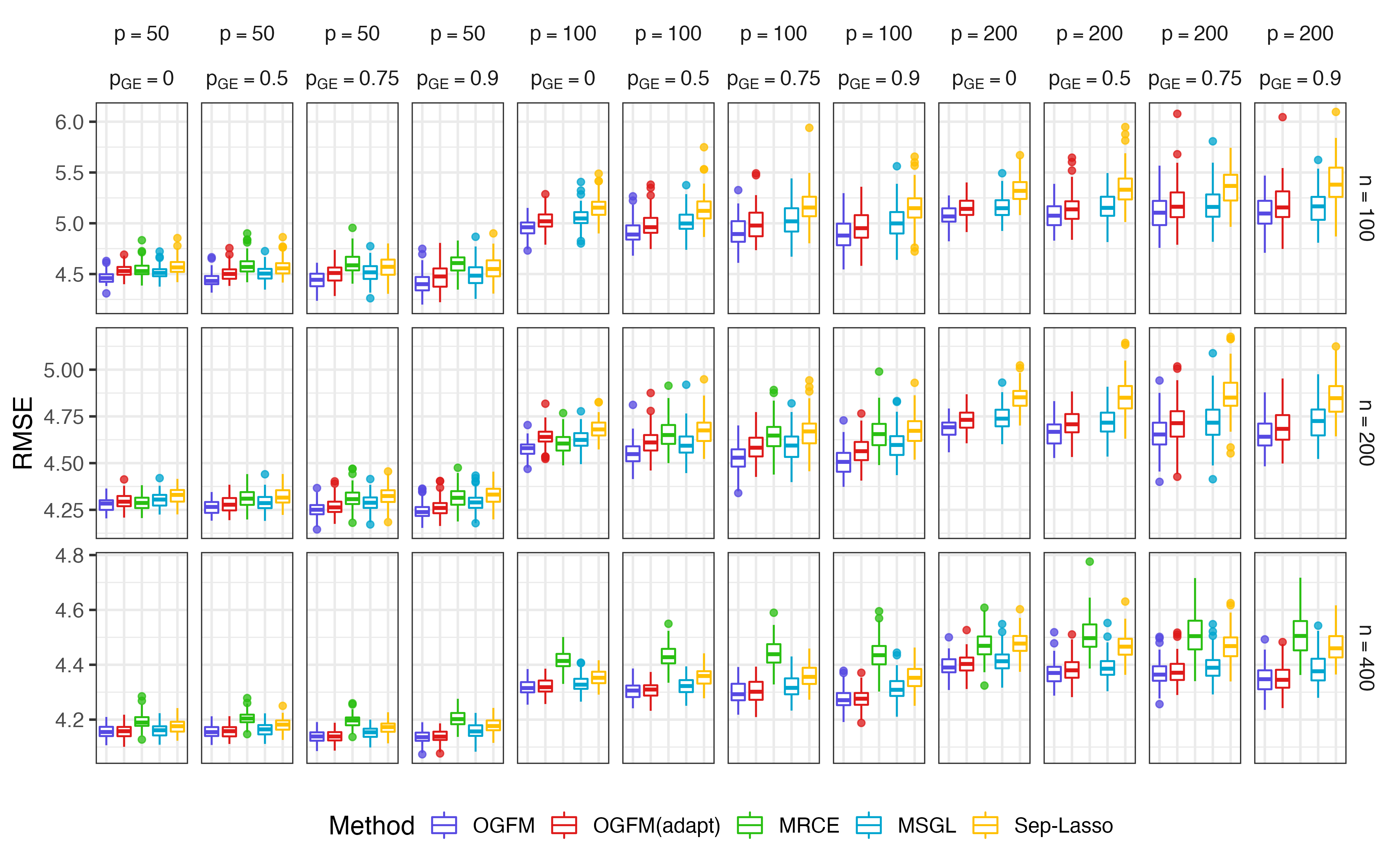}
	\caption{Validation RMSEs for all methods across 100 replications of the simulation experiment holding the parameter $p_{\text{HS}}=0$ so that there is no group-level sparsity.}	
	\label{fig:sim_res_rmse_0hsp}
\end{figure}

\begin{figure}[!htpb]
	\centering
	\includegraphics[width=1\textwidth]{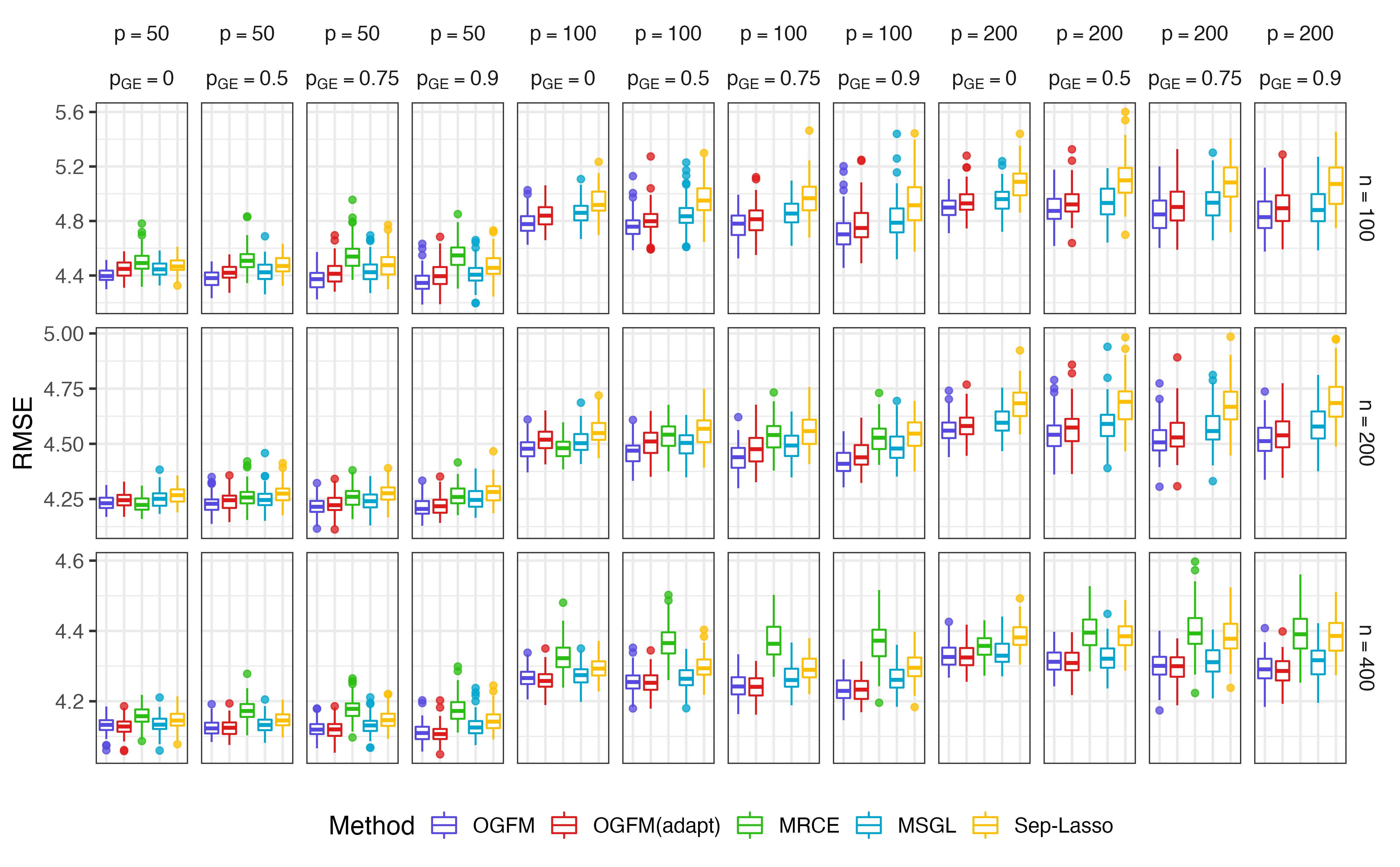}
	\caption{Validation RMSEs for all methods across 100 replications of the simulation experiment holding the parameter $p_{\text{HS}}=0.25$ so that there is moderate group-level sparsity.}	
	\label{fig:sim_res_rmse_025hsp}
\end{figure}

The RMSE results fixing the parameter at $p_{\text{HS}}=0$ are displayed in Figure \ref{fig:sim_res_rmse_0hsp} and the RMSE results for $p_{\text{HS}}=0.25$ are displayed in Figure \ref{fig:sim_res_rmse_025hsp}. When the sample size is small ($n=100$) and the hierarchical sparsity probability is 0 (Figure \ref{fig:sim_res_rmse_0hsp}), our proposed approach OGFM results in the smallest validation RMSE on average across the replications across all settings including moderate dimensions ($p=50,100$) and high dimensions ($p=200$), with MSGL as second best across the majority of settings and OGFM(adapt) a close third. The separate lasso tends to perform worse than all competing approaches, except for in some settings (e.g. when $n=400$) where it performs marginally better than MRCE. We note that the code for MRCE throws an error when $p\geq n$, although in principle MRCE can work for high dimensional settings. In the same settings but with $n=200$, OGFM still tends to work the best across all settings, however when the probability of coefficients being equal to each other is zero ($p_{\text{GE}}=0$), and the dimensionality is moderate, the MRCE approach performs nearly as well. In the high dimensional setting ($p=200$), MSGL performs on par with OGFM and sometimes better, however this benefit slightly attenuates for the larger sample size setting ($n=400$). In general the adaptive version of our approach, OGFM(adapt) performs worse than OGFM in small sample size settings and performs better with larger sample sizes. The results with the hierarchical sparsity probability set to 0.25 (Figure \ref{fig:sim_res_rmse_025hsp}) roughly mirror the results with no hierarchical sparsity, however the performance for all methods is slightly better in terms of RMSE as there are fewer overall non-zero coefficients in the data-generating process.
The model error results largely track with the validation RMSE results, albeit on a different scale, and are thus shown in the Supplementary Material Appendix B.1. 

\begin{figure}[!htpb]
	\centering
	\includegraphics[width=0.85\textwidth]{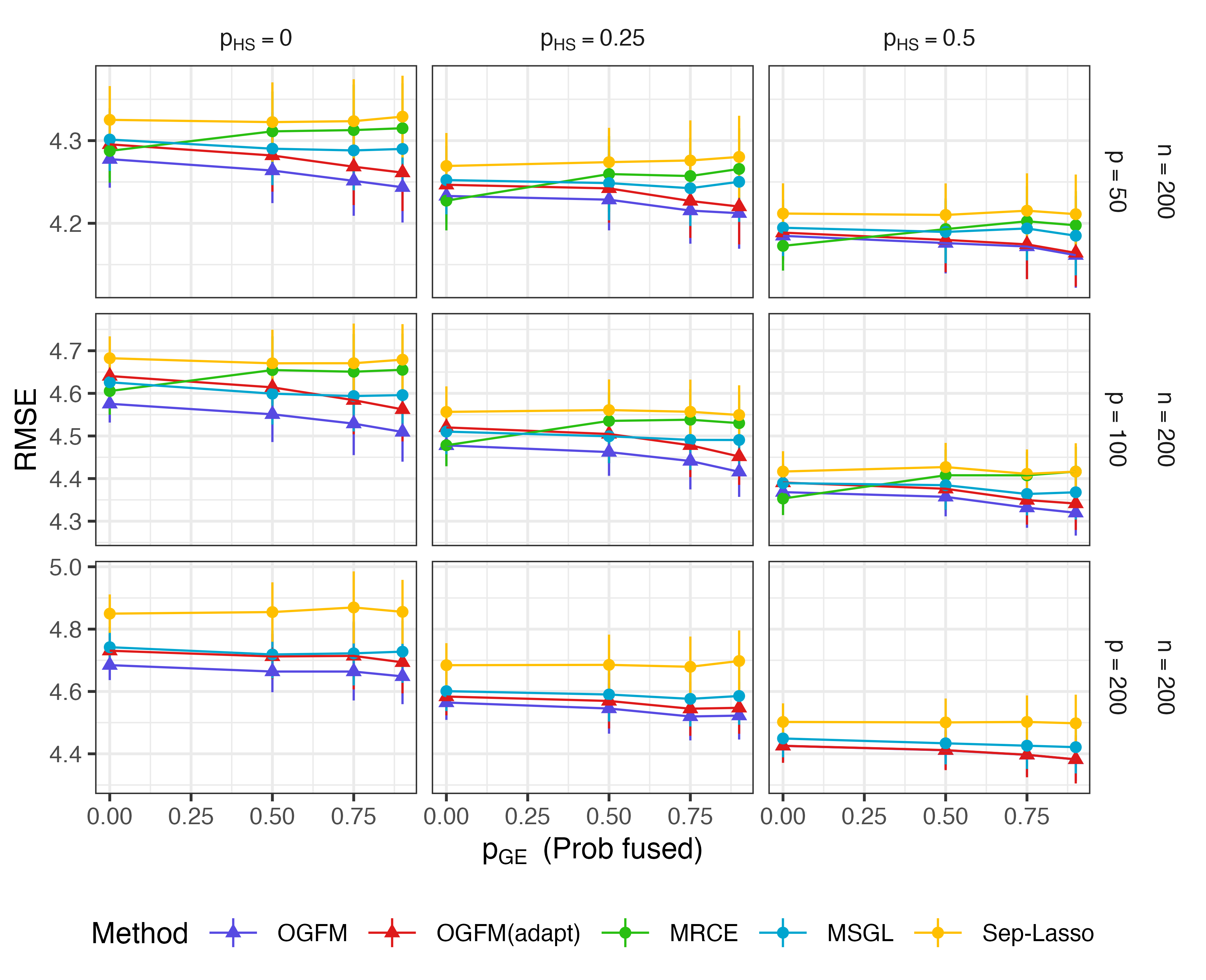}
	\caption{Validation RMSEs for all methods across 100 replications of the simulation experiment holding fixing the sample size $n=200$ and varying all other simulation parameters. The points are the average RMSEs across the 100 replications and error bars are plus and minus 1 standard deviation of this average.}	
	\label{fig:sim_res_rmse_sparsity_view}
\end{figure}

Figure \ref{fig:sim_res_rmse_sparsity_view} shows the same results as Figures \ref{fig:sim_res_rmse_0hsp} and \ref{fig:sim_res_rmse_025hsp}, but is re-organized to focus on the effect of the fused coefficients probability ($p_{\text{GE}}$) and the hierarchical sparsity probability ($p_{\text{HS}}$). In this figure, we fix the sample size to be 200. In general, we can see that as the probability of coefficients being equal/fused within groups increases, the performance of OGFM and OGFM(adapt) relative to the Sep-Lasso, MSGL, and MRCE tends to improve, with the trend being pronounced in moderate dimensional settings. Only the OGFM methods improve as the probability of coefficients being equal increases, as they are the only approaches which explictly allow for shrinkage of effect sizes to each other across outcomes. The relative performance of all methods tends to stay relatively consistent as $p_{\text{HS}}$ is varied. However, we note that MRCE tends to perform best in the setting with the fewest true non-zero coefficients ($p_{\text{HS}}=0.25$) and smallest amount of truly equal coefficients ($p_{\text{GE}}=0$). 

In the Supplementary Material Appendix B.1, we show simulation results in terms of the average of the TPR and TNR. From these results, it can be seen that while OGFM performs best in terms of prediction performance, it tends to over-select terms, resulting in a high TPR but low TNR and thus lower average TPR and TNR value. On the other hand, for larger sample sizes OGFM(adapt) performs very well in terms of the average TPR and TNR as it selects fewer variables with a higher proportion of selected variables being ones with truly non-zero coefficients. Overall, MSGL performs best in terms of the average TPR and TNR and MRCE worst, the latter trend having been concurrently observed in the original work of Rothman, et al. \cite{rothman2010sparse}, who noted that MRCE tends to result in better model error but with less benefit in terms of TPR and TNR. In the Supplementary Material Appendix B.1 we also present computation times for all methods. The OGFM approaches are highly competitive computationally as the sample size increases, however their performance deteriorates as the dimensionality increases.

\section{Analysis of Pediatric Functional Status Data}\label{sec:application}

Our motivating study consists of 1897 children, adolescents, and young adults who were admitted to an inpatient rehabilitation unit with a diagnosis of neurologic injury or illness at a major Midwestern children's hospital for neurological injury or illness between the years 2000 and August of 2020. The sample consisted of individuals between the ages of six months to 32 years with a median age of 12 years and 1st and 3rd quartiles of 6 and 16 years, respectively. The covariates used as predictors are (numbering 608 in total) included basic sociodemographic information including age and sex, and billing codes including CPT codes, ICD-9 and -10 codes, pharmaceutical codes, and data indicating durable medical equipment (DME) use. The billing codes represent information from a single episode of care involving an admission to an inpatient rehabilitation unit. The billing code information pertain only to information collected prior to the WeeFIM assessment. CPT and all ICD-9 and -10 codes were linked together to unique concept unique identifiers (CUIs) using the UMLS metathesaurus mapping system \citep{bodenreider2004unified}. Concept Unique Identifiers (CUIs) are distinct medical concepts (codes, diseases, etc) identified by the UMLS metathesaurus. By linking all billing codes to CUIs, we are able to in many cases handle the switch-over from ICD-9 to ICD-10 codes. An added benefit of CUIs is that they can be linked to other coding systems.
We utilize all 18 WeeFIM\textregistered{} components as responses; these 18 components are divided into three domains. The self-care domain describes how well a child is able to feed themselves, groom, bath, dress, and complete toileting tasks including the management of their bowel and bladder. The mobility domain describes how well a child is able to transfer on and off a toilet, in and out of a bathtub or shower, and in and out of a chair or wheelchair. The mobility domain also describes a child's ability to walk, crawl, or use a wheelchair, and to move up or down stairs. The cognition domain describes how well a child can express themselves, understand information, interact with peers, solve daily problems, and recall information. Together, they describe the ability of children to function in routine and important aspects of daily life. 

As our motivating use case of developing a model for the WeeFIM\textregistered{} components is to apply it to future data to assess functional ability across a health system population, we validate all developed models by splitting data into training and validation sub-datasets, where the training dataset is from years prior to data from the validation dataset. We use data prior to 2018 for training and data from 2018 to 2020 as validation data, leaving 1592 observations for training and 305 observations for validation.  For validation, we compare methods in terms of the response-specific mean squared error for each response and the average validated mean squared error across all 18 responses. We also compute the corresponding validation R-squared values, i.e. the proportion of the validation responses explained by the out-of-sample predictions. 
We exclude any covariates that have no variation in either the training or validation datasets. After this screening, the combined dimensionality of all the predictors was 608, a high-dimensional scenario given the sample size.
We apply all methods described in Section \ref{sec:comparator_methods} and use the approaches described therein for selection of tuning parameters. For OGFM(adapt), we use $\gamma_1=\gamma_2=0.5$, as use of $\gamma_1=\gamma_2=1$ resulted in some extreme weights given the high dimensionality of the data. For both the OGFM approaches and MSGL, the group lasso penalty applied corresponds to the three pre-defined domains of the WeeFIM\textregistered{} components (self-care, mobility, and cognition). 

\begin{figure}[!htpb]
	\includegraphics[width=1\textwidth]{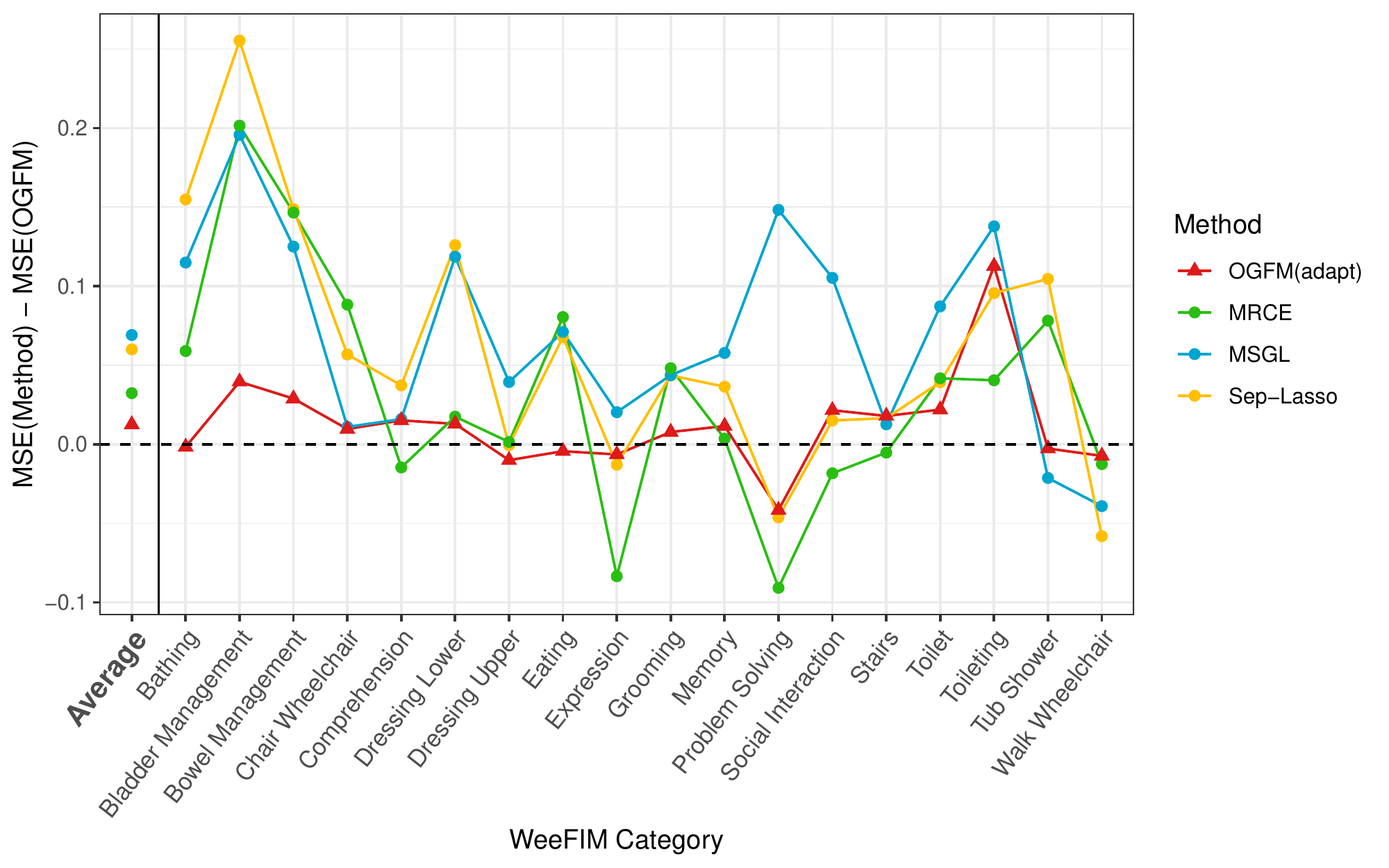}
	\caption{Comparisons of the validation MSE for each of the 18 WeeFIM components and on average (left) across the 18 components}	
	\label{fig:mse_rsq_diffs}
\end{figure}

The validation MSEs and R-squared values are displayed in Tables \ref{tab:validation_mse} and \ref{tab:validation_rsq}, respectively. Our OGFM approach has the best performance on the validation data on average across the 18 responses and also most often performs best for the individual responses, with the adaptive version of OGFM with the second lowest validation MSE on average. For responses where OGFM does not perform best, most often its performance does not differ much from the best MSE across the remaining methods, excluding OGFM(adapt). When OGFM performs best for particular responses, its improvement in terms of RMSE and R-squared is often large, as is depicted in Figure \ref{fig:mse_rsq_diffs}, which shows the difference between OGFM and competing approaches in terms of MSE on the validation data.

\begin{figure}[!htpb]
	\includegraphics[width=1\textwidth]{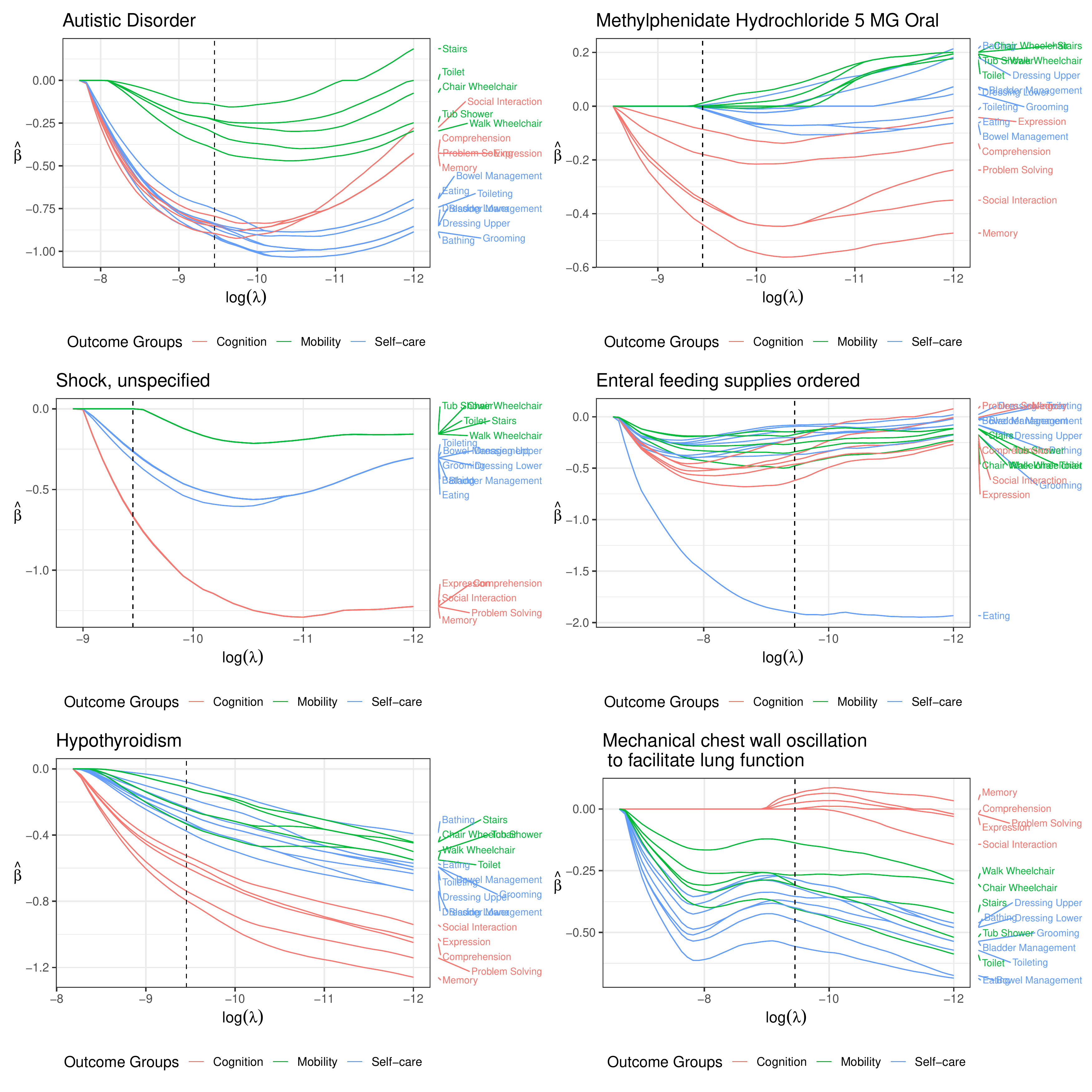}
	\caption{Coefficient path plots across all 18 WeeFIM components for six variables from the pediatric functional status data. The dashed vertical line indicates the value of $\lambda$ that minimizes the cross-validation error. }	
	\label{fig:weefim_coefs_example}
\end{figure}

For both OGFM approaches, the fused lasso mixing parameter $\alpha$ was chosen by cross-validation to be $1\times10^{-5}$, indicating only a small amount of fused lasso was required by the data. However, this small amount of fused lasso penalty had the effect of shrinking a large number of coefficients either close to each other or exactly to each other. In Figure \ref{fig:weefim_coefs_example}, we display the coefficient paths versus $\lambda$ (described in Section \ref{sec:impl_details}) for 6 variables, where each plot is the coefficient path for a particular variable across the 18 outcomes. For each of the coefficient path plots, we fix the mixing tuning parameter $\alpha$ at the value that minimizes the cross validation error. The patterns of these coefficient paths demonstrate that our approach fuses coefficients for a variable to be the same across multiple responses when appropriate, allows effects to differentiate when warranted by the data, an allows for joint selection of effects across an entire domain of related responses. For the ICD code ``shock, unspecified'', the fused lasso grouped the effects for all cognition outcomes together along the entire path, all mobility coefficients together, and grouped the self-care coefficients into two groups. As can be seen for other variables, different amounts fusing occurred for different groups of outcomes, and for some outcomes effects were shrunk towards each other but without exact fusing.

\begin{table}[ht]
\centering
\resizebox{0.8\textwidth}{!}{%
\begin{tabular}{r|r|rrrrrrrrrr}
  \toprule
  & Domain &  \multicolumn{8}{c}{Self-Care WeeFIM Responses} & \\
  \cmidrule{2-10}
 Method & Average & Ea & G & B & D-U & D-L & T-g & Bl-M & Bo-M & \\
  \midrule
OGFM & \textbf{2.153} & 2.074 & \textbf{1.816} & 2.034 & 1.673 & \textbf{2.099} & \textbf{2.052} & \textbf{3.533} & \textbf{3.492} & \\ 
 OGFM(adapt) & 2.165 & \textbf{2.070} & 1.824 & \textbf{2.032} & \textbf{1.663} & 2.112 & 2.165 & 3.573 & 3.521 &  \\ 
 MRCE & 2.185 & 2.155 & 1.864 & 2.093 & 1.674 & 2.117 & 2.093 & 3.735 & 3.638 &  \\ 
 MSGL & 2.222 & 2.145 & 1.860 & 2.149 & 1.712 & 2.218 & 2.190 & 3.729 & 3.617 &  \\ 
 Sep-Lasso & 2.213 & 2.142 & 1.860 & 2.189 & 1.672 & 2.225 & 2.148 & 3.788 & 3.641 &  \\ 
  \cmidrule{2-12}
  & Domain & \multicolumn{5}{|c|}{Mobility WeeFIM Responses} & \multicolumn{5}{c}{Cognition WeeFIM Responses}  \\
  & &C-W  & T & T-S & W-W & \multicolumn{1}{r|}{S} & C & Ex & S-I & P-S & M \\
   \cmidrule{3-12}
OGFM &&  \textbf{1.483}  & \textbf{1.770} & 1.811 & 1.344 & \multicolumn{1}{r|}{2.169} & 2.140 & 2.165 & 2.159 & 2.537 & \textbf{2.395} \\ 
 OGFM(adapt) && 1.492 & 1.792 & 1.808 & 1.336 & \multicolumn{1}{r|}{2.187} & 2.155 & 2.159 & 2.180 & 2.495 & 2.407 \\ 
 MRCE && 1.571  & 1.811 & 1.889 & 1.331 & \multicolumn{1}{r|}{\textbf{2.164}} & \textbf{2.126} & \textbf{2.081} & \textbf{2.140} & \textbf{2.446} & 2.399 \\ 
 MSGL && 1.494 & 1.857 & \textbf{1.790} & 1.305 & \multicolumn{1}{r|}{2.182} & 2.156 & 2.185 & 2.264 & 2.685 & 2.453 \\ 
 Sep-Lasso && 1.539 & 1.809 & 1.915 & \textbf{1.286} & \multicolumn{1}{r|}{2.186} & 2.178 & 2.152 & 2.174 & 2.491 & 2.432 \\ 
   \bottomrule
\end{tabular}%
}
\caption{Mean squared errors (MSEs) on the 18 WeeFIM components in the validation data and average MSE across the 18 components. Bold indicates the smallest MSE across the different methods for a particular component. Ea: Eating, G: Grooming, B: Bathing, D-U: Dressing Upper, D-L: Dressing Lower, T-g: Toileting, Bl-M: Bladder Management, Bo-M: Bowel Management, C-W: Chair Wheelchair, T: Toilet, T-S: Tub Shower, W-W: Walk Wheelchair, S: Stairs, C: Comprehension, Ex: Expression, S-I: Social Interaction, P-S: Problem Solving, M: Memory.}
\label{tab:validation_mse}
\end{table}

\begin{table}[ht]
	\centering
	\resizebox{0.8\textwidth}{!}{%
		\begin{tabular}{r|r|rrrrrrrrrr}
			\toprule
  			& Domain &  \multicolumn{8}{c}{Self-Care WeeFIM Responses} & \\
			\cmidrule{2-10}
			Method & Average & Ea & G & B & D-U & D-L & T-g & Bl-M & Bo-M &  \\
			\midrule
			OGFM & \textbf{0.432} & 0.526 & \textbf{0.536} & 0.476 & 0.446 & \textbf{0.447} & \textbf{0.498} & \textbf{0.449} & \textbf{0.429} & \\ 
 OGFM(adapt) & 0.429 & \textbf{0.527} & 0.534 & \textbf{0.477} & \textbf{0.449} & 0.444 & 0.470 & 0.443 & 0.425 &  \\ 
 MRCE & 0.425 & 0.507 & 0.524 & 0.461 & 0.446 & 0.442 & 0.488 & 0.418 & 0.405 &  \\ 
 MSGL & 0.416 & 0.509 & 0.525 & 0.447 & 0.433 & 0.416 & 0.464 & 0.419 & 0.409 & \\ 
 Sep-Lasso & 0.419 & 0.510 & 0.525 & 0.437 & 0.446 & 0.414 & 0.474 & 0.410 & 0.405 &  \\ 
			\cmidrule{2-12}
  & Domain & \multicolumn{5}{|c|}{Mobility WeeFIM Responses} & \multicolumn{5}{c}{Cognition WeeFIM Responses}  \\
			& & C-W & T & T-S & W-W & \multicolumn{1}{r|}{S} & C & Ex & S-I & P-S & M \\
			\cmidrule{3-12}
			OGFM &&  \textbf{0.433}  & \textbf{0.522} & 0.447 & 0.287 & \multicolumn{1}{r|}{0.298} & 0.410 & 0.405 & 0.400 & 0.375 & \textbf{0.384} \\ 
 OGFM(adapt) &&  0.430 & 0.516 & 0.448 & 0.290 & \multicolumn{1}{r|}{0.292} & 0.406 & 0.406 & 0.394 & 0.385 & 0.381 \\ 
 MRCE && 0.400 & 0.510 & 0.423 & 0.293 & \multicolumn{1}{r|}{\textbf{0.299}} & \textbf{0.414} & \textbf{0.428} & \textbf{0.405} & \textbf{0.397} & 0.383 \\ 
 MSGL && 0.429 & 0.498 & \textbf{0.454} & 0.307 & \multicolumn{1}{r|}{0.294} & 0.406 & 0.399 & 0.371 & 0.338 & 0.369 \\ 
 Sep-Lasso && 0.412 & 0.511 & 0.415 & \textbf{0.317} & \multicolumn{1}{r|}{0.292} & 0.400 & 0.408 & 0.396 & 0.386 & 0.374 \\ 
			\bottomrule
		\end{tabular}%
	}
\caption{R-squared values on the 18 WeeFIM components in the validation data and average R-squared value across the 18 components. Bold indicates the larger R-squared across the different methods for a particular components.}
\label{tab:validation_rsq}
\end{table}

From the coefficient paths in Figure \ref{fig:weefim_coefs_example}, our approach yields benefit in terms of predictive performance and also results in clinically sensible estimated coefficients. We chose these 6 variables to demonstrate effects of several classes of billing codes in the data: ICD codes, DME codes, pharmaceutical codes, and CPT codes. The three figures on the left of Figure \ref{fig:weefim_coefs_example} represent examples of diagnostic codes from billing data. The top panel represents the functional domains’ coefficients path plot for children coded with autistic disorder.  Children with autism spectrum disorders are diagnosed based on deficits in areas of social communication and behavior. The degree of impairment in social skills as well as comorbid neurodevelopmental disorders often leads to functional deficits in self-care.  Mobility is typically unaffected in children with autism. Another example is the diagnosis code for shock. Shock is a life-threatening condition in which the body is not getting enough blood flow, resulting in end-organ damage, including damage to the brain.   Consistent with this, in the figure we see from the coefficient path plots that the diagnosis code of shock is most related  with the functional domain of  cognition/communication as compared to mobility and self-care skills. Hypothyroidism is a disorder caused when the thyroid gland does not make enough hormone. In children, it has numerous etiologies including autoimmune reactions, brain injury, or radiation treatment. Hypothyroidism causes slow growth, decreased strength, and impaired cognition. The associated plot shows some relationship to the self-care and mobility domains, but a stronger relationship to the cognition/communication domain. The three figures in the right of the panel represent billing codes for a medication, a durable good, and a procedure. Methylphenidate Hydrochloride, a stimulant medication used to treat attention deficit hyperactivity disorder, showed minimal relationships with mobility and self-care.  However a pronounced relationship with cognition/communication is observed, as would be expected in a child with being treated with this medication. An example of durable medical equipment are enteral feeding supplies. These are ordered for patients who require feeding through a tube inserted into the stomach. This plot shows a clear relationship between feeding supplies and eating outcomes, but markedly less strong relationship with other functional outcome domains. Mechanical chest wall oscillation is used in patients who have an impaired lung function such as in children with cystic fibrosis or other diseases which lead to general physicial debilitation or loss of strength. The plot demonstrates that this procedure code holds associations with  mobility and self-care limitations as would be expected, but does not relate to cognition/communication. 

\section{Discussion}\label{sec:discussion}

In this paper we introduced a doubly structured sparsity regression approach for modeling multivariate outcomes with known groupings to borrow information across related outcomes. Our approach allows for overlapping groupings of the responses and borrows strength by joint selection of the effects of variables across groups of related outcomes and encourages shrinkage of the effects of variables to be more similar across related outcomes. We prove an oracle property for an adaptive version of our penalty, showing that our approach results in estimates with the same asymptotic distribution as if the true non-zero coefficients and \textit{which} variable effects are truly equal across related responses were both known in advance. The results pertaining to our motivating study indicate our approach yields benefit in terms of predictive performance and yields estimated coefficients that are clinically sensible, as was demonstrated in our analysis of the pediatric functional status data.
Our work is motivated by modeling functional status scores in a pediatric population however the methods are applicable more generally. Since our approach does not rely on normality of the multivariate outcomes, it can be straightforwardly extended to more general models, such as generalized linear models both in theory and computationally. Incorporation of covariance information in addition to borrowing strength through joint selection and effect shrinkage may be a promising direction for additionally borrowing strength of information across outcomes. However, in this work we have not focused on this as it would introduce another tuning parameter and another layer of complexity to computation, which may limit the applicability of such an approach to large scale and high dimensional settings.



\section*{Acknowledgments}

Research reported in this publication was supported by the Eunice Kennedy Shriver National Institute of Child Health \& Human Development of the National Institutes of Health under Award Number R03HD101083. The content is solely the responsibility of the authors and does not necessarily represent the official views of the National Institutes of Health. \vspace*{-8pt}

{
 \bibliography{paper_draft_11-22-2022}

\begin{thebibliography}{10}
\providecommand \doibase [0]{http://dx.doi.org/}%

\bibitem{lo2009pediatric}
Lo W, Stephens J, Fernandez S. Pediatric stroke in the {United States} and the
  impact of risk factors. {\it Journal of Child Neurology} 2009\string;
  24(2)\string: 194--203.

\bibitem{patel2014pediatric}
Patel S, Bhatnagar A, Wear C, et al. Are pediatric brain tumors on the rise in
  the {USA}? {S}ignificant incidence and survival findings from the {SEER}
  database analysis. {\it Child's Nervous System} 2014\string; 30(1)\string:
  147--154.

\bibitem{taylor2017traumatic}
Taylor CA, Bell JM, Breiding MJ, Xu L. Traumatic brain injury--related
  emergency department visits, hospitalizations, and deaths—{United States},
  2007 and 2013. {\it MMWR Surveillance Summaries} 2017\string; 66(9)\string:
  1.

\bibitem{dhillon2017us}
Dhillon JK, Shi J, Janezic A, Wheeler KK, Xiang H, Leonard JC. {US} estimates
  of pediatric spinal cord injury: implications for clinical care and research
  planning. {\it Journal of Neurotrauma} 2017\string; 34(12)\string:
  2019--2026.

\bibitem{schneier2006incidence}
Schneier AJ, Shields BJ, Hostetler SG, Xiang H, Smith GA. Incidence of
  pediatric traumatic brain injury and associated hospital resource utilization
  in the United States. {\it Pediatrics} 2006\string; 118(2)\string: 483--492.

\bibitem{szekely2007measuring}
Sz{\'e}kely GJ, Rizzo ML, Bakirov NK. Measuring and testing dependence by
  correlation of distances. {\it The Annals of Statistics} 2007\string;
  35(6)\string: 2769--2794.

\bibitem{tibshirani96}
Tibshirani R. Regression shrinkage and selection via the lasso. {\it Journal of
  the Royal Statistical Society. Series B (Methodological)} 1996\string;
  58\string: 267--288.

\bibitem{rothman2010sparse}
Rothman AJ, Levina E, Zhu J. Sparse multivariate regression with covariance
  estimation. {\it Journal of Computational and Graphical Statistics}
  2010\string; 19(4)\string: 947--962.

\bibitem{yuan2007model}
Yuan M, Lin Y. Model selection and estimation in the Gaussian graphical model.
  {\it Biometrika} 2007\string; 94(1)\string: 19--35.

\bibitem{sofer2014variable}
Sofer T, Dicker L, Lin X. Variable selection for high dimensional multivariate
  outcomes. {\it Statistica Sinica} 2014\string; 24(4)\string: 1633.

\bibitem{fan2001variable}
Fan J, Li R. Variable selection via nonconcave penalized likelihood and its
  oracle properties. {\it Journal of the American Statistical Association}
  2001\string; 96(456)\string: 1348--1360.

\bibitem{zhang2010nearly}
Zhang CH. Nearly unbiased variable selection under minimax concave penalty.
  {\it The Annals of Statistics} 2010\string; 38(2)\string: 894--942.

\bibitem{li2015multivariate}
Li Y, Nan B, Zhu J. Multivariate sparse group lasso for the multivariate
  multiple linear regression with an arbitrary group structure. {\it
  Biometrics} 2015\string; 71(2)\string: 354--363.

\bibitem{jenatton2011}
Jenatton R, Audibert JY, Bach F. Structured variable selection with
  sparsity-inducing norms. {\it The Journal of Machine Learning Research}
  2011\string; 12\string: 2777--2824.

\bibitem{yuan2006}
Yuan M, Lin Y. Model selection and estimation in regression with grouped
  variables. {\it Journal of the Royal Statistical Society: Series B
  (Statistical Methodology)} 2006\string; 68(1)\string: 49--67.

\bibitem{tibshirani2005sparsity}
Tibshirani R, Saunders M, Rosset S, Zhu J, Knight K. Sparsity and smoothness
  via the fused lasso. {\it Journal of the Royal Statistical Society: Series B
  (Statistical Methodology)} 2005\string; 67(1)\string: 91--108.

\bibitem{zou2006}
Zou H. {The Adaptive Lasso and Its Oracle Properties}. {\it Journal of the
  American Statistical Association} 2006\string; 101(476)\string: 1418--1429.

\bibitem{wang2008note}
Wang H, Leng C. A note on adaptive group lasso. {\it Computational Statistics
  \& Data Analysis} 2008\string; 52(12)\string: 5277--5286.

\bibitem{huling2018risk}
Huling JD, Yu M, Liang M, Smith M. Risk prediction for heterogeneous
  populations with application to hospital admission prediction. {\it
  Biometrics} 2018\string; 74(2)\string: 557--565.

\bibitem{viallon2013adaptive}
Viallon V, Lambert-Lacroix S, H{\"o}fling H, Picard F. Adaptive generalized
  fused-lasso: Asymptotic properties and applications. {\it Technical Report}
  2013.

\bibitem{eigenweb}
Guennebaud G, Jacob B, others . Eigen v3. http://eigen.tuxfamily.org;  2010.

\bibitem{rcppeigen}
Bates D, Eddelbuettel D. Fast and Elegant Numerical Linear Algebra Using the
  {RcppEigen} Package. {\it Journal of Statistical Software} 2013\string;
  52(5)\string: 1--24.

\bibitem{glowinski1975}
Glowinski R, Marroco A. Sur l'approximation, par {\'e}l{\'e}ments finis d'ordre
  un, et la r{\'e}solution, par p{\'e}nalisation-dualit{\'e} d'une classe de
  probl{\`e}mes de Dirichlet non lin{\'e}aires. {\it Revue fran{\c{c}}aise
  d'automatique, informatique, recherche op{\'e}rationnelle. Analyse
  num{\'e}rique} 1975\string; 9(2)\string: 41--76.

\bibitem{gabay1976}
Gabay D, Mercier B. A dual algorithm for the solution of nonlinear variational
  problems via finite element approximation. {\it Computers \& Mathematics with
  Applications} 1976\string; 2(1)\string: 17--40.

\bibitem{boyd2011}
Boyd S, Parikh N, Chu E, Peleato B, Eckstein J. Distributed optimization and
  statistical learning via the alternating direction method of multipliers.
  {\it Foundations and Trends{\textregistered} in Machine Learning}
  2011\string; 3(1)\string: 1--122.

\bibitem{chen2016direct}
Chen C, He B, Ye Y, Yuan X. The direct extension of {ADMM} for multi-block
  convex minimization problems is not necessarily convergent. {\it Mathematical
  Programming} 2016\string; 155(1-2)\string: 57--79.

\bibitem{huang2008}
Huang J, Ma S, Zhang CH. Adaptive Lasso for sparse high-dimensional regression
  models. {\it Statistica Sinica} 2008\string: 1603--1618.

\bibitem{bodenreider2004unified}
Bodenreider O. The unified medical language system ({UMLS}): integrating
  biomedical terminology. {\it Nucleic Acids Research} 2004\string;
  32(suppl\_1)\string: D267--D270.

\end{thebibliography}


\begin{thebibliography}{6}
\providecommand{\natexlab}[1]{#1}
\providecommand{\url}[1]{\texttt{#1}}
\expandafter\ifx\csname urlstyle\endcsname\relax
  \providecommand{\doi}[1]{doi: #1}\else
  \providecommand{\doi}{doi: \begingroup \urlstyle{rm}\Url}\fi

\bibitem[Geyer(1994)]{geyer1994}
C.~Geyer.
\newblock On the asymptotics of constrained $m$-estimation.
\newblock \emph{The Annals of Statistics}, 22\penalty0 (4):\penalty0
  1993--2010, 12 1994.
\newblock \doi{10.1214/aos/1176325768}.
\newblock URL \url{http://dx.doi.org/10.1214/aos/1176325768}.

\bibitem[Jenatton et~al.(2011)Jenatton, Audibert, and Bach]{jenatton2011}
R.~Jenatton, J.-Y. Audibert, and F.~Bach.
\newblock Structured variable selection with sparsity-inducing norms.
\newblock \emph{The Journal of Machine Learning Research}, 12:\penalty0
  2777--2824, 2011.

\bibitem[Knight and Fu(2000)]{knight2000}
K.~Knight and W.~Fu.
\newblock Asymptotics for lasso-type estimators.
\newblock \emph{The Annals of Statistics}, 28\penalty0 (5):\penalty0
  1356--1378, 10 2000.
\newblock \doi{10.1214/aos/1015957397}.
\newblock URL \url{http://dx.doi.org/10.1214/aos/1015957397}.

\bibitem[Lee and Xing(2014)]{lee2014}
S.~Lee and E.~P. Xing.
\newblock Screening rules for overlapping group lasso.
\newblock Technical report, 2014.

\bibitem[Rothman et~al.(2010)Rothman, Levina, and Zhu]{rothman2010sparse}
A.~J. Rothman, E.~Levina, and J.~Zhu.
\newblock Sparse multivariate regression with covariance estimation.
\newblock \emph{Journal of Computational and Graphical Statistics}, 19\penalty0
  (4):\penalty0 947--962, 2010.

\bibitem[Viallon et~al.(2013)Viallon, Lambert-Lacroix, H{\"o}fling, and
  Picard]{viallon2013adaptive}
V.~Viallon, S.~Lambert-Lacroix, H.~H{\"o}fling, and F.~Picard.
\newblock Adaptive generalized fused-lasso: Asymptotic properties and
  applications.
\newblock \emph{Technical Report}, 2013.

\end{thebibliography}
 }

\section*{Supporting Information}

Web Appendix A, referenced in Section~\ref{sec:methods} and Web Appendix B, referenced in Section~\ref{sec:simulation}, and \cb{code for our simulation studies} are available online.\vspace*{-8pt}

\label{lastpage}

\end{document}


\date{}

\maketitle

\section{Proofs}\label{sec:proofs}

In this section, we prove Theorem 1 of our paper.

\begin{proof}[Proof of Theorem \ref{thm:linear_model_oracle_property_mis}]
We first prove asymptotic results for the linear model and then extend them to generalized linear models later.
We begin by showing result (\ref{linear_model_asympt_distr_mis}) of the main text. Let $\bbeta_{\cdot,k} = \betatrue_{\cdot,k} + \frac{1}{\sqrt{N}}\bfu_{\cdot,k}$, where $\bfu_{\cdot,k} \in \mathbbm{R}^p$. We can write the objective function (\ref{overlapping_linear_model_objective}) of the main text multiplied by $N$ as a function of $\bfu=(\bfu_{\cdot,1},\dots,\bfu_{\cdot,K})$ as follows:
\begin{align*}
F_N(\bfu) = {} &  \sum_{k=1}^K\frac{1}{2}\blnorm-\frac{1}{\sqrt{N}}\bfX\bfu_{\cdot,k} + \boldsymbol\epsilon_k\brnorm^2_2 \\
& + \lambda_1 N \sum_{j=1}^p \sum_{G\in \mathcal{G}}\lambda_{1,j,G}||\bbeta_{j,G}^{0}+\frac{1}{\sqrt{N}}\circ\bfu_{j,G}||_2 \\
& + \lambda_2N\sum_{j=1}^p\sum_{(l, m) \in \mathcal{F} } \lambda_{2,j,l,m}\blvert{\beta}^0_{j,l} - {\beta}^0_{j,m} + \vphantom{\frac{1}{\sqrt{n_l + n_m}}}\vphantom{\brvert} \\
& \qquad\qquad\qquad + \vphantom{\blvert}\frac{1}{\sqrt{N}}\left(  u_{j,G} -  u_{j,m} \right) \brvert
\end{align*}
where $u_{j,k} = \sqrt{N}(\beta_{j,k} - \beta^0_{j,k})$.
Let $\hat{\bfu}^{(N)} = \argmin_{\bfu}F_N(\bfu)$ and note that $\hat{\bfu}^{(N)}_{k,\cdot} = \sqrt{N}(\betahat_{\cdot,k} - \betatrue_{\cdot,k})${, where $\betahat$ is the minimizer of the objective function (\ref{overlapping_linear_model_objective})  of the main text. Thus, to investigate the asymptotic distribution of $\betahat$ is equivalent to investigating the asymptotic distribution of $\hat{\bfu}^{(N)}$.} 

Now, we let
\begin{align}
D_N(\bfu) = {} & F_N(\bfu) - F_N(\boldsymbol 0) \label{eqn:gram_matrix}\\
 = {} & \sum_{k=1}^K \left(   \frac{1}{2} \bfu^\top_{\cdot,k}\left( \frac{1}{N}{\bfX}^\top \bfX \right)\bfu_{\cdot,k} - \frac{1}{\sqrt{N}}\bfu_{\cdot,k}^\top\bfX^\top\boldsymbol\epsilon_k  \right) \nonumber\\
 {} &  + \sqrt{N}\lambda_1 \sqrt{N} \left(  \sum_{j=1}^p \sum_{G\in \mathcal{G}}\lambda_{1,j,G}||\bbeta_{j,G}^{0}+\frac{1}{\sqrt{N}}\bfu_{j,G}||_2  \right.  \nonumber \\
 & \qquad\qquad\qquad   - \left. \sum_{j=1}^p \sum_{G\in \mathcal{G}}\lambda_{1,j,G}||\bbeta_{j,G}^{0}||_2 \right) \label{eqn:diff_f_group}  \\
 {} & + \sqrt{N}\lambda_1 \sqrt{N} \left(  \sum_{j=1}^p\sum_{(l, m) \in \mathcal{F} } \lambda_{2,j,l,m}\blvert{\beta}^0_{j,l} - {\beta}^0_{j,m} + \vphantom{\blvert}\frac{1}{\sqrt{N}}\left[  u_{j,l} - u_{j,m} \right] \brvert  \right. \nonumber \\
 {} & \qquad\qquad\qquad \left. -\sum_{j=1}^p\sum_{(l, m) \in \mathcal{F}} \lambda_{2,j,l,m}\blvert{\beta}^0_{j,l} - {\beta}^0_{j,m} \brvert \right) \label{eqn:diff_f_fused}  \\
 = {} & \sum_{k=1}^K \left(   \frac{1}{2} \bfu_{\cdot,k}^\top\left( \frac{1}{N}{\bfX}^\top \bfX \right)\bfu_{\cdot,k} - \frac{1}{\sqrt{N}}\bfu_{\cdot,k}^\top\bfX^\top\boldsymbol\epsilon_k  \right) \nonumber\\
  {} &  + \sqrt{N}\lambda_1 \sum_{j=1}^p\sum_{G \in \mathcal{G}_{H_{\cdot,j}}} \lambda_{1,j,G}\sqrt{N} \left(  \blnorm\bbeta_{j,G}^{0} + \frac{1}{\sqrt{N}}\circ\bfu_{j,G}\brnorm_2 - \blnorm\bbeta_{j,G}^{0}\brnorm_2  \right)    \\
   {} &  + \sqrt{N}\lambda_1 \sum_{j=1}^p\sum_{G \in \mathcal{G}_{H_{\cdot,j}^c}} \lambda_{1,j,G}\sqrt{N} \left(  \blnorm\bbeta_{j,G}^{0} + \frac{1}{\sqrt{N}}\circ\bfu_{j,G}\brnorm_2 - \blnorm\bbeta_{j,G}^{0}\brnorm_2  \right)  \\
    {} & + \sqrt{N}\lambda_2 \sqrt{N} \left(  \sum_{j=1}^p\sum_{(l, m) \in \mathcal{F}, {\beta}^0_{j,l} \neq {\beta}^0_{j,m} } \lambda_{2,j,l,m}\blvert{\beta}^0_{j,l} - {\beta}^0_{j,m} + \vphantom{\frac{1}{\sqrt{n_l + n_m}}}\vphantom{\brvert} \right. \nonumber \\
   {} &  \qquad\qquad\qquad + \vphantom{\blvert}\frac{1}{\sqrt{N}}\left[ u_{j,l} - u_{j,m} \right] \brvert \nonumber \\
   {} & \qquad\qquad\qquad \left. \sum_{j=1}^p\sum_{(l, m) \in \mathcal{F},{\beta}^0_{j,l} \neq {\beta}^0_{j,m} } \lambda_{2,j,l,m}\blvert{\beta}^0_{j,l} - {\beta}^0_{j,m} \brvert \right) + \\
    {} & + \sqrt{N}\lambda_2 \sqrt{N} \left(  \sum_{j=1}^p\sum_{(l, m) \in \mathcal{F}, {\beta}^0_{j,l} = {\beta}^0_{j,m} } \lambda_{2,j,l,m} \right. \nonumber \\
{} &  \qquad\qquad\qquad \times \left. \blvert\frac{1}{\sqrt{N}}\left[  u_{j,l} - u_{j,m} \right] \brvert   \vphantom{\sum_{l, m \in \mathcal{K}: l \setminus m = 1, {\beta}^0_{j,l} = {\beta}^0_{j,m} }} \right)  
\end{align}
where $\mathcal{G}_{\calH_{j,\cdot}}$ is the set of all groups $G$ with any $ k \in G$ such that $\beta^0_{j,k}\not=0$ and $\mathcal{G}_{\calH_{j,\cdot}^c}$ is the set of all groups  $G$ such that $\beta^0_{j,k}=0$ for all $k \in G$. {We obtain the asymptotic distribution of $\hat{\bfu}^{(N)}$, by first investigating the asymptotic properties of $D_N(\bfu)$ for every fixed $\bfu\in \mathbbm{R}^{Kp}$.}

For all $G \in \mathcal{G}_{\calH_{j,\cdot}}$, we have $\lambda_{1,j,G} \xrightarrow{p} || \bbeta_{j,G}^{0} ||_2^{-\gamma_1} $ and by taking the directional derivative in the direction of $\bfu_{j,G}$, we have
\begin{align*}
\sqrt{N} \left(  \blnorm\bbeta_{j,G}^{0} + \frac{1}{\sqrt{N}}\bfu_{j,G}\brnorm_2 - \blnorm\bbeta_{j,G}^{0}\brnorm_2  \right)  = \sqrt{N}\frac{1}{\sqrt{N}}\frac{{\bfu_{j,G}}^\top\bbeta_{j,G}^{0}}{||\bbeta_{j,G}^{0}||_2} + o_p(1).
\end{align*}
Then because $\sqrt{N}\lambda = o(1)$, we have by Slutsky's theorem that
\begin{align*}
\sqrt{N}\lambda_1\lambda_{1,j,G}\sqrt{N} \left(  \blnorm\bbeta_{j,G}^{0} + \frac{1}{\sqrt{N}}\bfu_{j,G}\brnorm_2 - \blnorm\bbeta_{j,G}^{0}\brnorm_2  \right) = o_p(1).
\end{align*}
For all $G \in \mathcal{G}_{H_{\cdot,j}^c}$, because $N^{\gamma_1/2}||\betahat^{MLE}_{j,G}||^{\gamma_1}_2 = O_p(1)$  and \newline $\sqrt{N} \left(  \blnorm\bbeta_{j,G}^{0} + \frac{1}{\sqrt{N}}\bfu_{j,G}\brnorm_2 - \blnorm\bbeta_{j,G}^{0}\brnorm_2  \right) = ||\sqrt{N}\frac{1}{\sqrt{N}}\bfu_{j,G}||_2$, we have
\begin{align}
\lambda_1\lambda_{1,j,G}\sqrt{N} ||\sqrt{N}\frac{1}{\sqrt{N}}\bfu_{j,G}||_2 = ||\sqrt{N}\frac{1}{\sqrt{N}}\bfu_{j,G}||_2\lambda_1 \frac{N^{(\gamma_1 + 1)/2}}{(\sqrt{N}|| \betahat^{MLE}_{j,G} ||_2)^{\gamma_1}} \xrightarrow{p} \infty, \label{lambda_g_infty1}
\end{align}
{
if $\bfu_{j,G}\not =\boldsymbol 0$, and,
\begin{align}
\lambda_1\lambda_{1,j,G}\sqrt{N} ||\sqrt{N}\frac{1}{\sqrt{N}}\bfu_{j,G}||_2 = ||\sqrt{N}\frac{1}{\sqrt{N}}\bfu_{j,G}||_2\lambda_1 \frac{N^{(\gamma_1 + 1)/2}}{(\sqrt{N}|| \betahat^{MLE}_{j,G} ||_2)^{\gamma_1}} = o_p(1), \label{lambda_g_infty2}
\end{align}
if $\bfu_{j,G}=\boldsymbol 0$.}

Similarly, for all $(l,m) \in \mathcal{F}$ with ${\beta}^0_{j,l} \neq {\beta}^0_{j,m}$, we have $\lambda_{2,j,l,m} \xrightarrow{p} | {\beta}^0_{j,l} - {\beta}^0_{j,m} |^{-\gamma_2}$ and

\begin{align*}
& \sqrt{N}\left(  \blvert \beta_{j,l}^0 - \beta_{j,m}^0 +  \frac{1}{N}\left(  u_{j,l} - u_{j,m} \right)   \brvert  - \blvert  \beta_{j,l}^0 - \beta_{j,m}^0   \brvert  \right)  \\
{} & = \left( u_{j,l} - u_{j,m} \right)  \mbox{sign}\left({\beta}^0_{j,l} - {\beta}^0_{j,m}\right) + o_p(1)
\end{align*}

Then since $\sqrt{N}\lambda = o(1)$, we have by Slutsky's theorem that

\begin{align*}
& \sqrt{N}\lambda_2\lambda_{2,j,l,m}\sqrt{N}\left(  \blvert \beta_{j,l}^0 - \beta_{j,m}^0 +  \frac{1}{N}\left( u_{j,l} -  u_{j,m} \right)   \brvert  -  \blvert  \beta_{j,l}^0 - \beta_{j,m}^0   \brvert  \right) = o_p(1).
\end{align*}

Now for all $(l,m) \in \mathcal{F}$ with ${\beta}^0_{j,l} = {\beta}^0_{j,m}$, we have

\begin{align*}
& \sqrt{N}\lambda_2\lambda_{2,j,l,m} \sqrt{N}   \blvert\frac{1}{\sqrt{N}}\left(  u_{j,l} - u_{j,m} \right) \brvert \\
{} & = \sqrt{N}   \blvert\frac{1}{\sqrt{N}}\left(  u_{j,l} - u_{j,m} \right) \brvert \lambda_2 \frac{N^{(\gamma_2 + 1)/2 }}{\left(  \sqrt{N}\lvert \hat{\beta}^{MLE}_{j,l} - \hat{\beta}^{MLE}_{j,m} \rvert  \right)^{\gamma_2}} \xrightarrow{p} \infty
\end{align*}
if $ u_{j,l} \neq u_{j,m}$ and

\begin{align*}
& \sqrt{N}\lambda_2\lambda_{2,j,l,m} \sqrt{N}   \blvert\frac{1}{\sqrt{N}}\left(  u_{j,l} - u_{j,m} \right) \brvert \\
{} & = \sqrt{N}   \blvert\frac{1}{\sqrt{N}}\left(  u_{j,l} - u_{j,m} \right) \brvert \lambda_2 \frac{N^{(\gamma_2 + 1)/2 }}{\left(  \sqrt{N}\lvert \hat{\beta}^{MLE}_{j,l} - \hat{\beta}^{MLE}_{j,m} \rvert  \right)^{\gamma_2}} = o_p(1)
\end{align*}
if $  u_{j,l} =  u_{j,m}$.

By our condition on the rates of $n_j$ for $j \in \mathcal{K}$ and $N$, we have that $N^{-1}{\bfX}^\top\bfX \to \bfQ$ and thus $N^{-1}{\wtbfX}^\top\wtbfX \to \wtbfQ \equiv \bfI_K\otimes \bfQ$, where $\boldsymbol Q$ and $\wtbfQ$ are positive definite. Note that there exist matrices $\bfH$ and $\bfE$ such that $\wtbfX_\calH^* = \wtbfX\bfH\bfE$, so that $N^{-1}{{\wtbfX_\calH{}}^*}^\top{\wtbfX_\calH{}}^* = N^{-1}\bfE^\top\bfH^\top\wtbfX^\top\wtbfX\bfH\bfE \to \boldsymbol Q_\calH^* \equiv \bfE^\top\bfH^\top\wtbfQ\bfH\bfE$, where $\boldsymbol Q_\calH^*$ is positive definite and is constructed from $\wtbfQ$ in a manner corresponding to the pattern of collapsed and dropped columns of $\wtbfX$ but also dropping and collapsing rows in the same pattern. Further, $N^{-1/2}\boldsymbol\epsilon^\top\wtbfX \xrightarrow{d} \boldsymbol W$ with $\boldsymbol W \sim N(\bfZero, \bSigma \otimes \bfQ)$,  $N^{-1/2}\boldsymbol\epsilon^\top\wtbfX_\calH^* = N^{-1/2}\boldsymbol\epsilon^\top\wtbfX\bfH\bfE \xrightarrow{d} \boldsymbol W_\calH^*$ with $\bfW_\calH^* \sim N(\bfZero, \bfV^*_\calH)$, where $\bfV^*_\calH = \bfE^\top\bfH^\top \left(\bSigma \otimes \bfQ\right)\bfH\bfE$.   Hence $\bfu_\calH$ is comprised of the elements in $\bfu$ which correspond to the nonzero elements in $\bbeta^0$. Further, denote $\bfu_\calH^*$ to be the unique elements in $\bfu_\calH$ collapsed in the same manner as $\bfX_\calH^*$.

Since the term (\ref{eqn:gram_matrix}) converges in distribution to 
$\frac{1}{2}\bfu^\top\wtbfQ\bfu + \bfu^\top\boldsymbol W$,
using Slutsky's theorem we have that $D_N(\bfu) \xrightarrow{d} D(\bfu)$ {for each $\bfu$}, where

\[ D(\bfu) =
 \begin{cases}
      \frac{1}{2}{\bfu^*}_\calH^\top\boldsymbol Q_\calH^*\bfu^*_{\calH} + {\bfu^*}_\calH^\top\boldsymbol W_\calH^* & \mbox{if } {\bfu_{\calH_{k,\cdot}^c}  = \boldsymbol 0}, \forall k=1,\dots,K,  \\
      & \mbox{and } u_{j,l} = u_{j,m} \mbox{ for all } (l,m) \in {E_{\cdot,j}} \\
      \infty & \mbox{otherwise.}
   \end{cases}
\]

It is clear that $D_N(\bfu)$ is convex and the unique minimum of $D(\bfu)$ is $((\bQ_\calH^*)^{-1} \bfW_\calH^*, \boldsymbol 0)$. By the epiconvergence results of \cite{geyer1994} and \cite{knight2000}, we have the following:
\begin{align}  
{{}\hat{\bfu}^*_{\calH}}^{(N)} \xrightarrow{d} (\bQ_\calH^*)^{-1} \bfW^*_{\calH}   \mbox{ and } \hat{\bfu}_{\calH^c}^{(N)} \xrightarrow{d} \boldsymbol 0
\end{align}
where ${\bQ_\calH^*}^{-1} \bfW_{\calH}^* \sim N_{|\calH^*|}(0,{\bQ_\calH^*}^{-1}\bfV^*_\calH {\bQ_\calH^*}^{-1} )$, where $|\calH^*|$ represents the number of columns in $\bQ_\calH^*$. Hence result (\ref{linear_model_asympt_distr_mis}) of the main text is verified.

We now show selection consistency. For any {$j \in \calH_{\cdot,k}$}, we have by the asymptotic normality (\ref{linear_model_asympt_distr_mis}) of the main text and thus it follows that $P(k \in \hat{\calJ}_{\cdot,j}) \to 1$. To verify result (\ref{linear_model_selection_consistency_mis}) of the main text it is equivalent to show that for any $k'$ such that $j \in \calH_{\cdot,k'}^c$ which implies that $k'\not\in \mbox{Hull}(\calJ_{j,\cdot})$, $P(k' \in \hat{\calJ}_{j,\cdot}) \to 0$. Suppose $\hat{\beta}_{j,k'} \neq  0$ and $j \in \calH_{\cdot,k'}^c$, then there exists at least one $G\subset \{1,\dots,K\}$ such that $k'\in G\subset \calH_{j,\cdot}^c$. For any $k \in \mathcal{K}$ define $\mathcal{G}_{j,k} = \{ G \subset \mathcal{G} : k \in G \subset \calH_{j,\cdot}^c  \}$. 
Let $G_{j,k} = \argmax_{G \in \mathcal{G}_{j,k}} |G| $, i.e. $G_{j,k}$ is the biggest group in the complement of the true hull for variable $j$ which contains $k$. 
Assume without loss of generality that $G_{j,k}=\{1,\dots,k_0\}$ for some $k_0\in\calK$ subject to re-ordering of the response labels. For any $k \in \mathcal{K}$ define $C_{j,k} = \{l \in G_{j,k} : l = k \mbox{ or } (l, k) \in \calF \mbox{ or } (k,l) \in \calF \}$, in other words $C_{j,k}$ is indices of all coefficients fused to the coefficient for the $j$th variable in the $k$th subpopulation. Further, let $C_{j,k}^0 = \{l \in C_{j,k} : \beta^0_{j,l} = 0 \}$. Suppose there exists some $l \in C_{j,k}^0$ such that $\hat{\beta}_{j,l} \neq 0$. Then at least one of $S_{j,k,neg} = \{m \in C_{j,k}^0 : \hat{\beta}_{j,m} < 0 \}$ and $S_{j,k,pos} = \{m \in C_{j,k}^0 : \hat{\beta}_{j,m} > 0 \}$ is nonempty. If $S_{j,k,neg}  \neq \varnothing$, then let $\hat{\beta}_{j,k,min} = \min_{m \in S_{j,k,neg}}\hat{\beta}_{j,m}$ be the largest magnitude nonzero coefficient in $S_{j,k,neg}$ and $S_{j,k,neg} \supset L_{j,k,min} = \{m \in S_{j,k,neg} : \hat{\beta}_{j,m} = \hat{\beta}_{j,k,min} \}$. Clearly  $L_{j,k,min} \neq \varnothing$. Then by combining ideas from \cite{lee2014, jenatton2011, viallon2013adaptive}, based on the KKT optimality conditions summed up over the indices in $L_{j,k,min}$ we have 
\begin{align*}
\boldsymbol 0 = {} & 
\begin{pmatrix}
-\sum_{m \in L_{j,1,min}}{\bfX_{j}}^\top(\bfY_{\cdot,m} - \bfX_m\betahat_{\cdot,m})\\
\vdots\\
-\sum_{m \in L_{j,k_0,min}}{\bfX_{j}}^\top(\bfY_{\cdot,m} - \bfX_{m}\betahat_{\cdot,m})\\
\end{pmatrix}+\lambda_1
\begin{pmatrix}
\mathbf{v}_{j,1}\\
\vdots\\
\mathbf{v}_{j,k_0}\\
\end{pmatrix}+\lambda_2
\begin{pmatrix}
\mathbf{b}_{j,1}\\
\vdots\\
\mathbf{b}_{j,k_0}\\
\end{pmatrix} \\
= {} & \Phi + \lambda_1 \mathbf{v} + \lambda_2 \mathbf{b},
\end{align*}

where $\bfX_{j}$ is the vector of length $N$ corresponding to the $j$th covariate, 
\begin{align*}
\mathbf{v}_{j,k'} = {} & N \cdot \sum_{m \in L_{j,k',min}}\sum_{G\in \mathcal{G} \mbox{ s.t. } m\in G} \lambda_{j,G}t_{j,m,G},
\end{align*}
where $t_{j,m,G} = \frac{\hat{\beta}_{j,m}}{||\betahat_{j,G}||_2}$ for any $j\in \{1,\dots,p\},m\in\calK, G\in\calG$ with $\betahat_{j,G} \neq \bfzero$ and $t_{j,m,G} \in [-1,1]$ for any  $j\in \{1,\dots,p\},m\in\calK, G\in\calG$ with $\betahat_{j,G} = \bfzero$,
 and
\begin{align}
 \mathbf{b}_{j,k'} = {} & N\cdot\sum_{m \in L_{j,k',min}}\sum\limits_{\substack{l \in \mathcal{K} \mbox{ s.t. } (l,m)\in\calF \mbox{ or } (m,l)\in\calF \\ \mbox{ and } \beta_{j,l}^0 \neq 0}} \lambda_{2,j,m,l} \cdot r_{j,m,l}  \label{eqn:_b1} \\
 & + N\cdot\sum_{m \in L_{j,k',min}}\sum\limits_{\substack{l \in \mathcal{K} \mbox{ s.t. } (l,m)\in\calF \mbox{ or } (m,l)\in\calF \\  \mbox{ and } \beta_{j,l}^0 = 0, \hat{\beta}_{j,k',min} < \hat{\beta}_{j,l}    }} \lambda_{2,j,m,l} \cdot r_{j,m,l} ,\label{eqn:_b2}
 \end{align}
where $r_{j,m,l} = \mbox{sign}(\hat{\beta}_{j,m} - \hat{\beta}_{j,l})$ for any $j\in \{1,\dots,p\}, (m,l)\in \calF$ with $\hat{\beta}_{j,m} \neq \hat{\beta}_{j,l}$ and $r_{j,m,l} \in [-1,1]$ for any $j\in \{1,\dots,p\}, (m,l)\in \calF$ with $\hat{\beta}_{j,m} = \hat{\beta}_{j,l}$.
  Due to the asymptotic normality of $\sqrt{n_k}(\betatrue_{k,\cdot} - \betahat_{k,\cdot})$ and conditions (D.1) - (D.3), we have
\begin{align*}
\blnorm \frac{1}{\sqrt{N}}\Phi\brnorm_2 = O_p(1).
\end{align*}
In addition, by the same arguments that show (\ref{lambda_g_infty1}) and definition of $G_0$, we have
\[
{||\sqrt{N}\lambda_1\lambda_{j,G}\frac{\betahat_{j,G_0}}{||\betahat_{j,G_0}||_2}||_2} \xrightarrow{p} \infty
\]
and for $m \in L_{k',j,min}$ and $l \in \mathcal{K} \mbox{ s.t. } (l,m)\in\calF \mbox{ or } (m,l)\in\calF$  when $\beta_{j,l}^0 \neq 0$ we have  $\beta_{j,l}^0 \neq \beta_{j,m}^0$ because $\beta_{j,m}^0 \neq 0$ and 
$$\frac{\lambda_2}{\sqrt{N}} \frac{\mbox{sign}(\hat{\beta}_{m,j} - \hat{\beta}_{l,j})}{| \hat{\beta}^{MLE}_{j,m} - \hat{\beta}^{MLE}_{j,l} |^{\gamma_2}} \xrightarrow{p} 0 $$ as $N \to \infty$. Since $L_{k',j,min} \subset S_{k',j,neg}$, we have that $\frac{\hat{\beta}_{m,j}}{||\betahat_{G,j}||_2}< 0$  and by the construction of $L_{k',j,neg}$ we have $\mbox{sign}(\hat{\beta}_{m,j} - \hat{\beta}_{l,j}) = -1$. Then since $\lambda_1N^{\gamma_1/2} / \sqrt{N}$ and $\lambda_2N^{\gamma_2/2} / \sqrt{N}$ both $\to \infty$ as $N \to \infty$, we have that  $||\frac{1}{\sqrt{N}}\Phi||_2 \to -\infty$ as $N \to \infty$ and hence we have arrived at a contradiction. Thus $Pr(S_{k',j,neg} = \varnothing) \to 1$. However if $S_{k',j,neg} = \varnothing$ we have $S_{k',j,pos}$ must be nonempty. Yet if we carry out the above arguments for 
 $S_{k',j,pos}$ we similarly arrive at a contradiction. Thus we can see that the probability of the KKT conditions holding vanishes,
\begin{align*}
P(k'\in \hat{J}_{\cdot,j}) \to 0.
\end{align*}
Hence, we have established selection consistency.

We now seek to show result (\ref{linear_model_fused_selection_consistency_mis}) of the main text, i.e. that we consistently estimate all coefficients for each variable $j$ which are equal to each other and are in adjacent subpopulations to be equal to each other. Specifically, we need to show that for all $(l,m) \not\in \calE_{j,\cdot}$, $P((l,m) \not\in \hat{\calE}_{j,\cdot}) \to 1$ and for all $(l,m) \in \calE_{j,\cdot}$ that $P((l,m) \not\in \hat{\calE}_{j,\cdot}) \to 0$. If $(l,m) \not\in \calE_{j,\cdot}$ then $\beta_{j,l}^0 \neq 0$ and $\beta_{j,m}^0 \neq 0$ and $\beta_{j,l}^0 \neq \beta_{j,m}^0$. Then by our result of asymptotic normality, we have that $\hat{\beta}_{j,l} - \hat{\beta}_{j,m} \xrightarrow{p} \beta_{j,l}^0 - \beta_{j,m}^0 \neq 0$ and hence $P((l,m) \not\in \hat{\calE}_{\cdot, j}) \to 1$. 

We will use an approach similar to our proof for selection consistency to show for all $(l,m) \in \calE_{j,\cdot}$ that $P((l,m) \not\in \hat{\calE}_{j,\cdot}) \to 0$. Let $k' \in \mathcal{K}$ be such that there exists an $m\in \mathcal{K}$ such that $(k',m) \in \calE_{j,\cdot}$.  Suppose there is some $m \in \mathcal{K}$ with $(k',m) \in \calE_{j,\cdot}$ such that $\hat{\beta}_{j,k'} \neq \hat{\beta}_{j,m}$. Similar as before, define $\hat{c}_{j,k,min} = \min_{m \in \mathcal{K} \mbox{ with } (k,m) \in \calE_{j,\cdot}}\hat{\beta}_{j,k} - \hat{\beta}_{j,m}$ and $$m_{j,k,min} = \argmin_{m \in \mathcal{K} \mbox{ with } (k,m) \in \calE_{j,\cdot}}\hat{\beta}_{j,k} - \hat{\beta}_{j,m}.$$ Let $L_{j,k,min} = \{  m \in \mathcal{K}: \hat{\beta}_{j,m} = \hat{\beta}_{j,m_{j,k,min}} \mbox{ and there is a connected path } (l,l+1,\dots, m_{min}, \dots, l+q) \mbox{ s.t.  either }  (l_{i}, l_{i+1}) \in \calF \mbox{ or  }  (l_{i+1}, l_{i}) \in \calF  \mbox{ and } \hat{\beta}_{j,l_i} = \hat{\beta}_{j,l_{i+1}} \mbox{ for } i = 0,\dots,q  \}$, in other words, $L_{k,j,min} $ is the indices of all coefficients which have been fused to each other along a path of pairwise fusings and have their coefficients equal to $\hat{\beta}_{j,m_{j,k,min}}$. Then similar to the previous KKT conditions and with $G_{j,k}$ defined as in the proof of selection consistency, we have

\begin{align*}
\boldsymbol 0 = {} & 
\begin{pmatrix}
-\sum_{m \in L_{j,1,min}}{\bfX_{j}^\top}(\bfY_{\cdot,m} - \bfX\betahat_{\cdot,m})\\
\vdots\\
-\sum_{m \in L_{j,k_0,min}}{\bfX_{j}^\top}(\bfY_{\cdot,m} - \bfX\betahat_{\cdot,m})\\
\end{pmatrix}+\lambda_1
\begin{pmatrix}
\mathbf{v}_{j,1}\\
\vdots\\
\mathbf{v}_{j,k_0}\\
\end{pmatrix}+\lambda_2
\begin{pmatrix}
\mathbf{b}_{j,1}\\
\vdots\\
\mathbf{b}_{j,k_0}\\
\end{pmatrix} \\
\equiv {} & \Phi + \lambda_1 \mathbf{v} + \lambda_2 \mathbf{b},
\end{align*}

where 
$$\mathbf{v}_{j,k'} = N \cdot \sum_{m \in L_{j,k',min}}\sum_{G\in \mathcal{G} \mbox{ s.t. } m\in G} \lambda_{G,j}t_{j,m,G},$$ and
\begin{align}
\mathbf{b}_{j,k'} = {} & N\cdot\sum_{m \in L_{j,k',min}}\sum\limits_{\substack{l \in \mathcal{K} \mbox{ s.t. } (l,m)\in\calF \mbox{ or } (m,l)\in\calF \\ \mbox{ and } \beta_{j,l}^0 \neq \beta_{j,m}^0}} \lambda_{2,j,l,m} \cdot r_{j,m,l}  \label{eqn:_bb1} \\
& + N\cdot\sum_{m \in L_{j,k',min}}\sum\limits_{\substack{l \in \mathcal{K} \mbox{ s.t. } (l,m)\in\calF \mbox{ or } (m,l)\in\calF \\  \mbox{ and } \beta_{j,l}^0 = \beta_{j,m}^0, \hat{\beta}_{j,m_{j,k',min}} < \hat{\beta}_{j,l}    }} \lambda_{2,j,l,m} \cdot r_{j,m,l}.\label{eqn:_bb2}
\end{align}

As before, $|| N^{-1/2}\Phi ||_2 = O_p(1)$. By similar  arguments as before, each term in ${N}^{-1/2}\mathbf{v}_{j,k'}$ converges to zero in probability and each term in (\ref{eqn:_bb1}) divided by $\sqrt{N}$ converges to 0 in probability, whereas the terms in (\ref{eqn:_bb2}) divided by $\sqrt{N}$ converge to $-\infty$ in probability and hence we have arrived at a similar contradiction as in the proof of selection consistency. Repeating these arguments for an appropriately defined $L_{k,j,max}$ results in the conclusion that for $(l,m) \in \calE_{\cdot, j}$ we have  $P((l,m) \in \hat{\calE}_{\cdot, j}) \to 0$ and hence we have completed the proof.

\end{proof}


\section{Additional simulation results}
\label{sec:additional_sims}

\subsection{Additional results under settings of main text}
\label{sec:additional_sims_sub1}

In this section we present additional simulation results under the settings described in Section \ref{sec:simulation} of the main text. In particular, we show results in terms of the model error metric, balanced accuracy ($0.5($TPR $+$ TNR$)$, where TPR is the true positive rate and TNR is the true negative rate, both defined below), and computation times. TPR and TNR are defined as in \citet{rothman2010sparse}:
\begin{align*}
	\text{TPR}(\widehat{\bbeta},\bbeta^0) = {} & \frac{\#\{(j,k): \widehat{\beta}_{j,k} \neq 0 \text{ and } \beta^0_{j,k} \neq 0\}}{\max(1, \#\{(j,k): \beta^0_{j,k} \neq 0\})} \text{ and} \\
	\text{TNR}(\widehat{\bbeta},\bbeta^0) = {} & \frac{\#\{(j,k): \widehat{\beta}_{j,k} = 0 \text{ and } \beta^0_{j,k} = 0\}}{\max(1, \#\{(j,k): \beta^0_{j,k} = 0\})}.
\end{align*}

\begin{figure}[H]
	\includegraphics[width=1\textwidth]{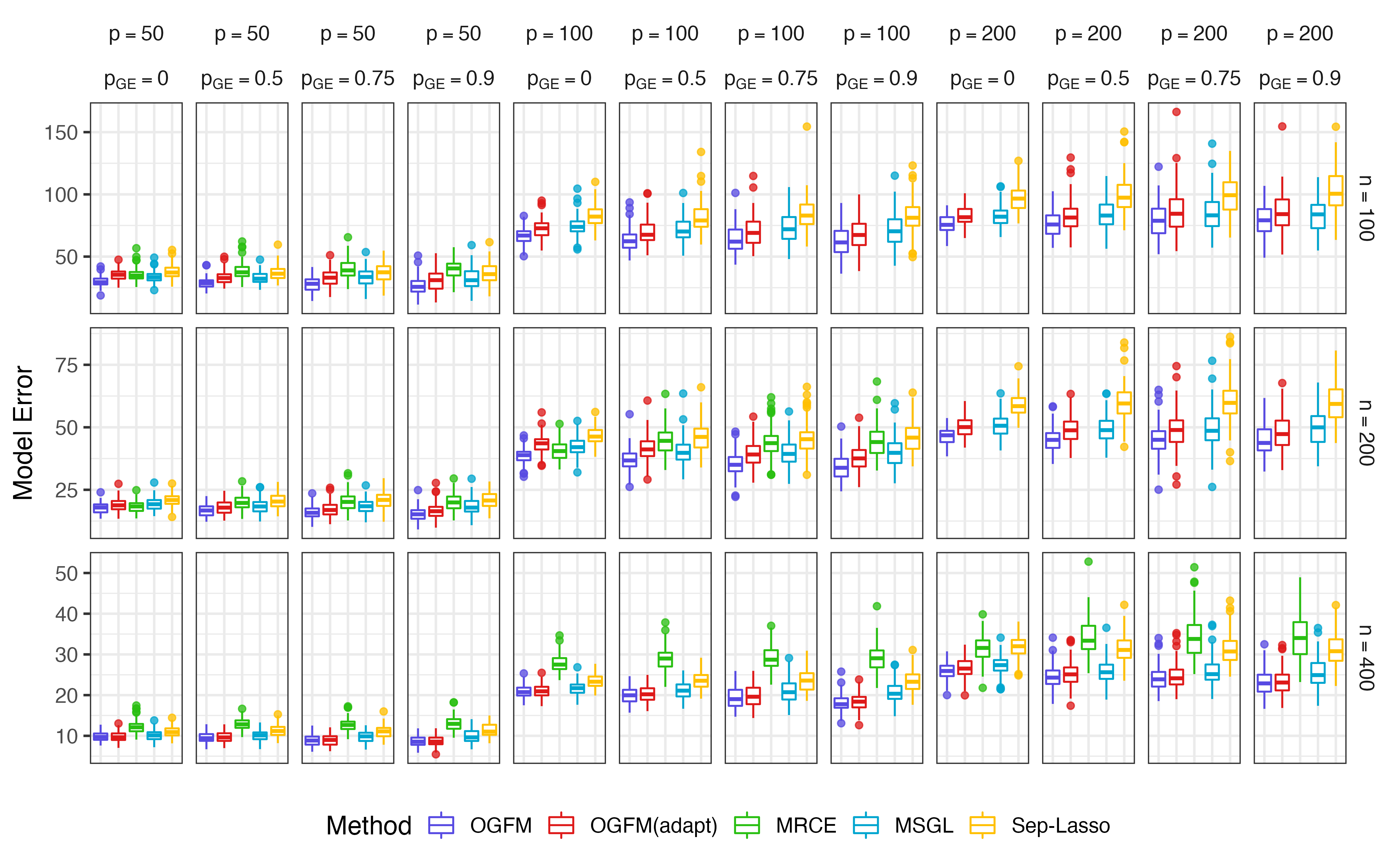}
	\caption{Model errors for all methods across 100 replications of the simulation experiment holding the parameter $p_{\text{HS}}=0$ so that there is no group-level sparsity.}	
	\label{fig:sim_res_model_error_0hsp}
\end{figure}

\begin{figure}[H]
	\includegraphics[width=1\textwidth]{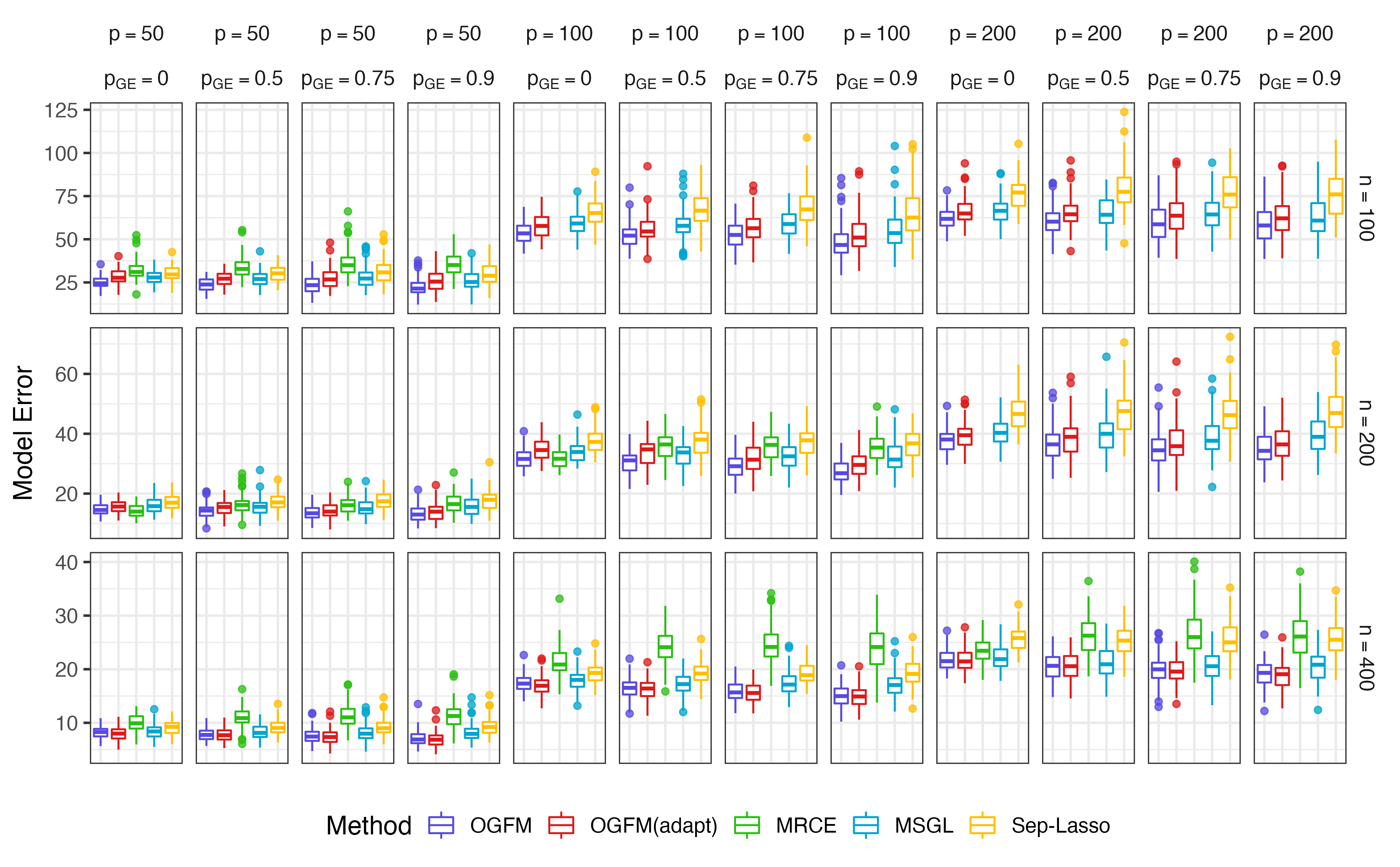}
	\caption{Model errors for all methods across 100 replications of the simulation experiment holding the parameter $p_{\text{HS}}=0.25$ so that there is no group-level sparsity.}	
	\label{fig:sim_res_model_error_025hsp}
\end{figure}

\begin{figure}[H]
	\includegraphics[width=1\textwidth]{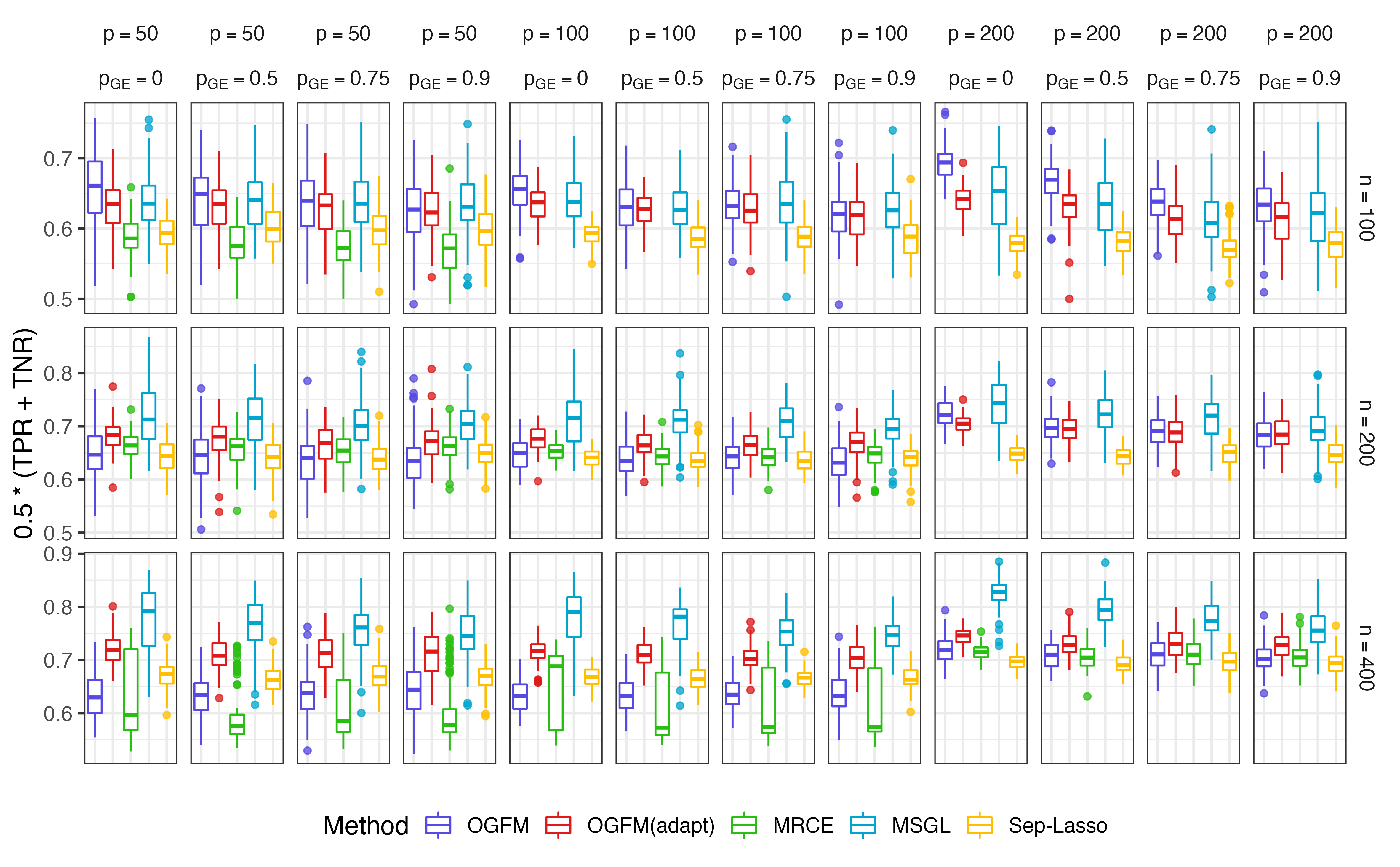}
	\caption{Balanced accuracy (average of TPR and TNR) for all methods across 100 replications of the simulation experiment holding the parameter $p_{\text{HS}}=0$ so that there is no group-level sparsity.}	
	\label{fig:sim_res_balacc_0hsp}
\end{figure}

\begin{figure}[H]
	\includegraphics[width=1\textwidth]{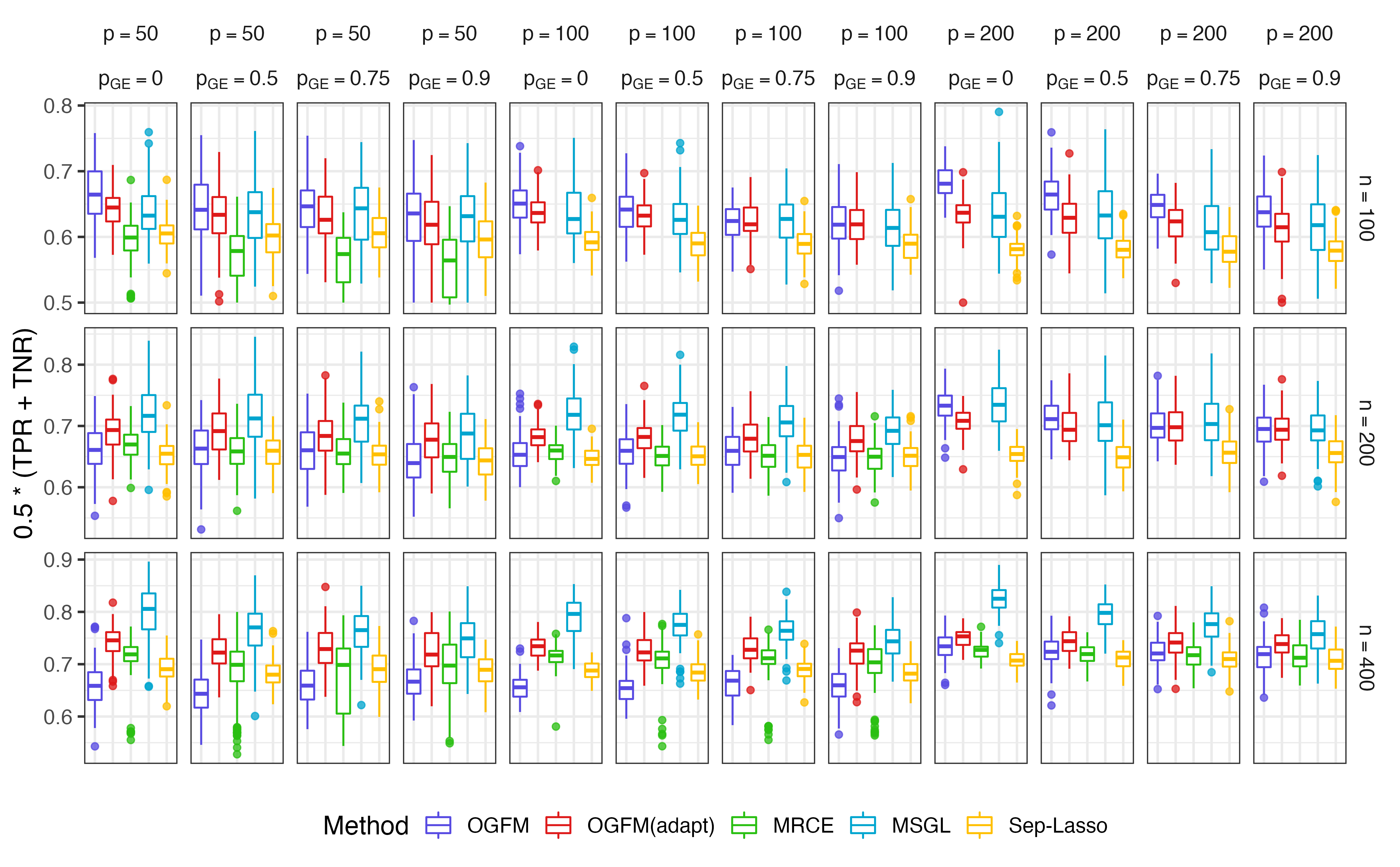}
	\caption{Balanced accuracy (average of TPR and TNR) for all methods across 100 replications of the simulation experiment holding the parameter $p_{\text{HS}}=0.25$ so that there is no group-level sparsity.}	
	\label{fig:sim_res_balacc_025hsp}
\end{figure}

The results in terms of the model errors fixing $p_{\text{HS}}=0$ are displayed in Figure \ref{fig:sim_res_model_error_0hsp} and fixing $p_{\text{HS}}=0.25$ are displayed in Figure \ref{fig:sim_res_model_error_025hsp}. The results in terms of balanced accuracy $p_{\text{HS}}=0$ are displayed in Figure \ref{fig:sim_res_balacc_0hsp} and \ref{fig:sim_res_balacc_025hsp}.

\begin{figure}[H]
	\includegraphics[width=1\textwidth]{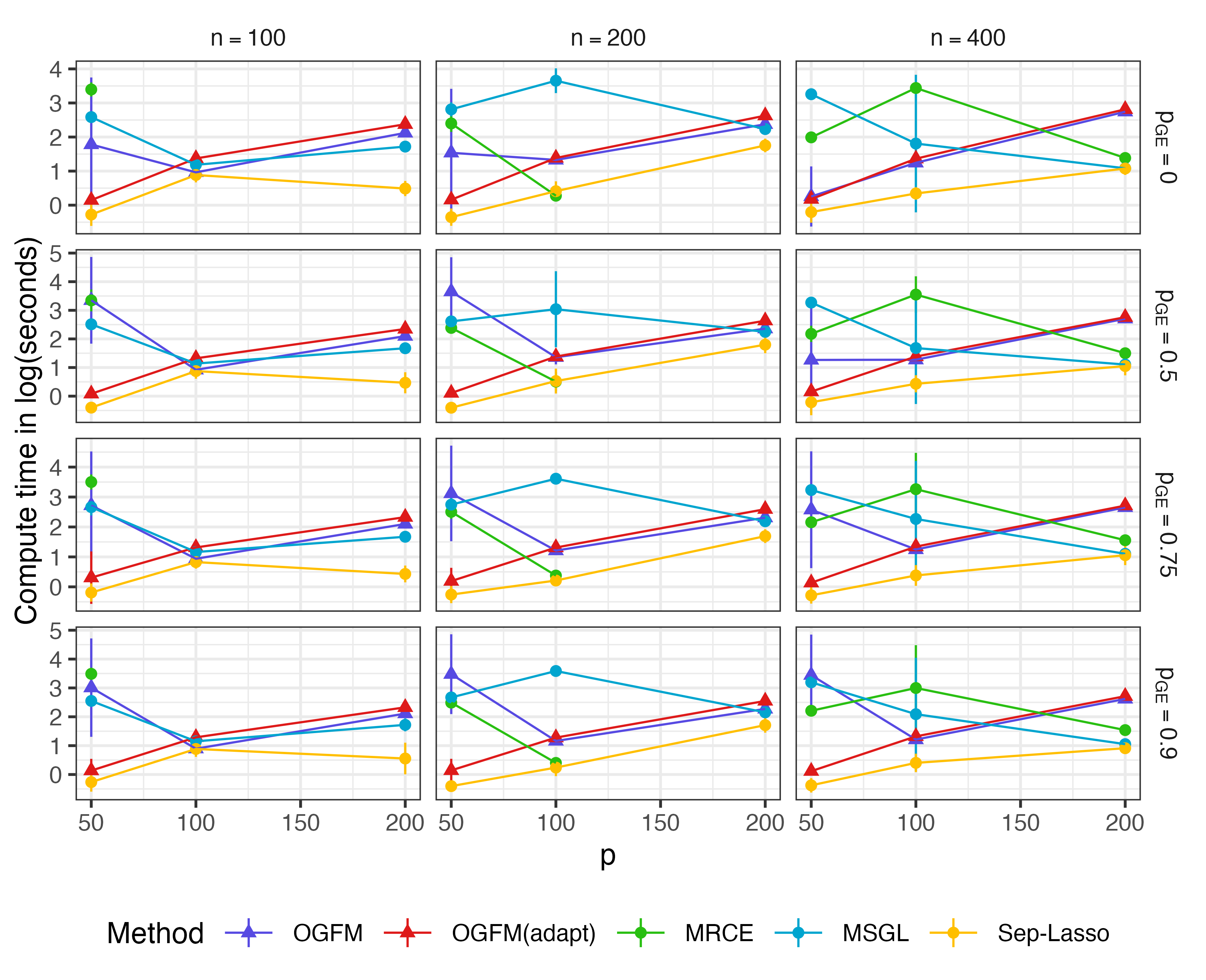}
	\caption{Computation times in terms of log seconds for all methods across 100 replications of the simulation experiment holding the parameter $p_{\text{HS}}=0$ so that there is no group-level sparsity. The points are the average log computation time in seconds and error bars are plus and minus 1 standard deviation.}	
	\label{fig:sim_res_comptime_0hsp}
\end{figure}

Since the setting with $p_{\text{HS}}=0$ has the most non-zero coefficients, it will in general require the greatest computational effort. As such we show computation times for this setting only. These results are displayed in Figure \ref{fig:sim_res_comptime_0hsp}.

\subsection{Simulations with discrete covariates and ordinal outcomes}
\label{sec:additional_sims_sub2}

\subsubsection{Data generation}

For each replication of the simulation, we generate data under model \eqref{lin_model}, where $\bfx_i$ are generated as i.i.d. Bernoulli$(0.2)$ random variables. The responses are generated as a multivariate ordered probit model where the observed responses $\bfy_i$ take 8 levels determined by an unobserved latent response $\bfy^*_i$, which is generated as $\bfy^*_i = {\betatrue}^\top\bfx_i + \boldsymbol\epsilon_i \text{ for } i=1,\dots,N$. The $j$th observed response for the $i$th individual is then determined by 
  \begin{equation*}
	\bfy_{ij} =
	\begin{cases}
		1, & \text{if}\  \bfy^*_{ij} \leq -2 \\
		2, & \text{if}\ -2 < \bfy^*_{ij} \leq -1 \\
		3, & \text{if}\ -1 < \bfy^*_{ij} \leq -1/2 \\
	    4, & \text{if}\ -1/2 < \bfy^*_{ij} \leq 0 \\
	    5, & \text{if}\ 0 < \bfy^*_{ij} \leq 1/2 \\
	    6, & \text{if}\ 1/2 < \bfy^*_{ij} \leq 1 \\
	    7, & \text{if}\ 1 < \bfy^*_{ij} \leq 2 \\
	    8, & \text{if}\ \bfy^*_{ij} > 2. \\
	\end{cases}
\end{equation*}
The error vectors $\bepsilon_i$ for the latent responses are generated as i.i.d. multivariate normal random variables with mean vector $\bzero$ and covariance matrix $\bSigma_{\bepsilon} = [\sigma_{\epsilon jk}]_{j,k=1}^K$, where $\sigma_{\epsilon jk} = 4\left(0.5^{|j-k|}\right)$. The dimensionality of the outcome/error vector is $K=8$ and the outcomes form 3 groups, with the first three outcomes forming group 1, the fourth and fifth forming group 2, and the last three outcomes forming group 3. The variable effects $\bbeta^0$ are generated in the exact same manner as in the simulation study in the main text, which for reference is
\vspace{10pt}

\begin{small}
	\begin{equation*}
		\bbeta^0 = 
		\begin{pmatrix}
			\bovermat{$G_{1,1}$}{ \xi_{1,1} \xi^G_{1,1} \eta_{1,1} & \xi_{1,2} \xi^G_{1,1} \eta_{1,2} & \xi_{1,3} \xi^G_{1,1} \eta_{1,3} } & \bovermat{$G_{1,2}$}{\xi_{1,4} \xi^G_{1,2} \eta_{1,4} & \xi_{1,5} \xi^G_{1,2} \eta_{1,5}} & \bovermat{$G_{1,3}$}{\xi_{1,6} \xi^G_{1,3} \eta_{1,6} & \xi_{1,7} \xi^G_{1,3} \eta_{1,7} & \xi_{1,8} \xi^G_{1,3} \eta_{1,8}} \\
			\vdots & & & & & & & \vdots \\
			\xi_{z,1} \xi^G_{z,1} \eta_{z,1} & \xi_{z,2} \xi^G_{z,1} \eta_{z,2} & \xi_{z,3} \xi^G_{z,1} \eta_{z,3} & \xi_{z,4} \xi^G_{z,2} \eta_{z,4} & \xi_{z,5} \xi^G_{z,2} \eta_{z,5} & \xi_{z,6} \xi^G_{z,3} \eta_{z,6} & \xi_{z,7} \xi^G_{z,3} \eta_{z,7} & \xi_{z,8} \xi^G_{z,3} \eta_{z,8} \\
			0 & \cdots & & & & &  \cdots & 0 \\
			\vdots & \ddots & & & & & & \vdots \\
			0 & \cdots & & & & &  \cdots & 0
		\end{pmatrix},
	\end{equation*}
\end{small}
where the last $p-z$ rows of $\bbeta^0$ have all elements as 0, the terms $\xi_{j,k}\sim$ Bernoulli$(0.9)$ induce sparsity at the individual effect level, the terms $\xi^G_{j,1}, \xi^G_{j,2}, \xi^G_{j,3} \sim$ Bernoulli$(1-p_{\text{HS}})$ induce sparsity at the group level, the variable effect size terms $\eta_{j,k}$ are distributed i.i.d. uniformly from $\{-1, -0.5, -0.25, -0.125, 0.125, 0.25, 0.5, 1\}$ to create both small and large effects, and for each variable $j$ separately the terms $\eta_{j,k}$ for $k$ in the same group are set to be all equal to each other with probability $p_{\text{GE}}/2$ and independently, the terms $\eta_{j,k}$ for all outcomes $k=1,\dots,8$ are set to be all equal to $\eta_{j,1}$ with probability $p_{\text{GE}}/2$. The latter process induces effects for some variables within a group to be equal to each other and induces effects for some variables to be equal for all outcomes. 

We explore dimensions of $p=50, 100,$ and $200$ and set $z=25, 50, 50$ for each dimension setting, respectively. We explore hierarchical sparsity parameters from $p_{\text{HS}} = 0, 0.25$ and $0.5$ and fusing probabilities $p_{\text{GE}} = 0, 0.5, 0.75$ and $0.95$; as mentioned in the main text, when $p_{\text{HS}} = 0$, there is no group-wise sparsity and thus any group lasso penalty applied is superfluous. For each replication of the simulation, we additionally generate an independent test set of size 10000 for use in evaluation of predictive performance.

\subsubsection{Simulation results}

\begin{figure}[!htpb]
	\centering
	\includegraphics[width=1\textwidth]{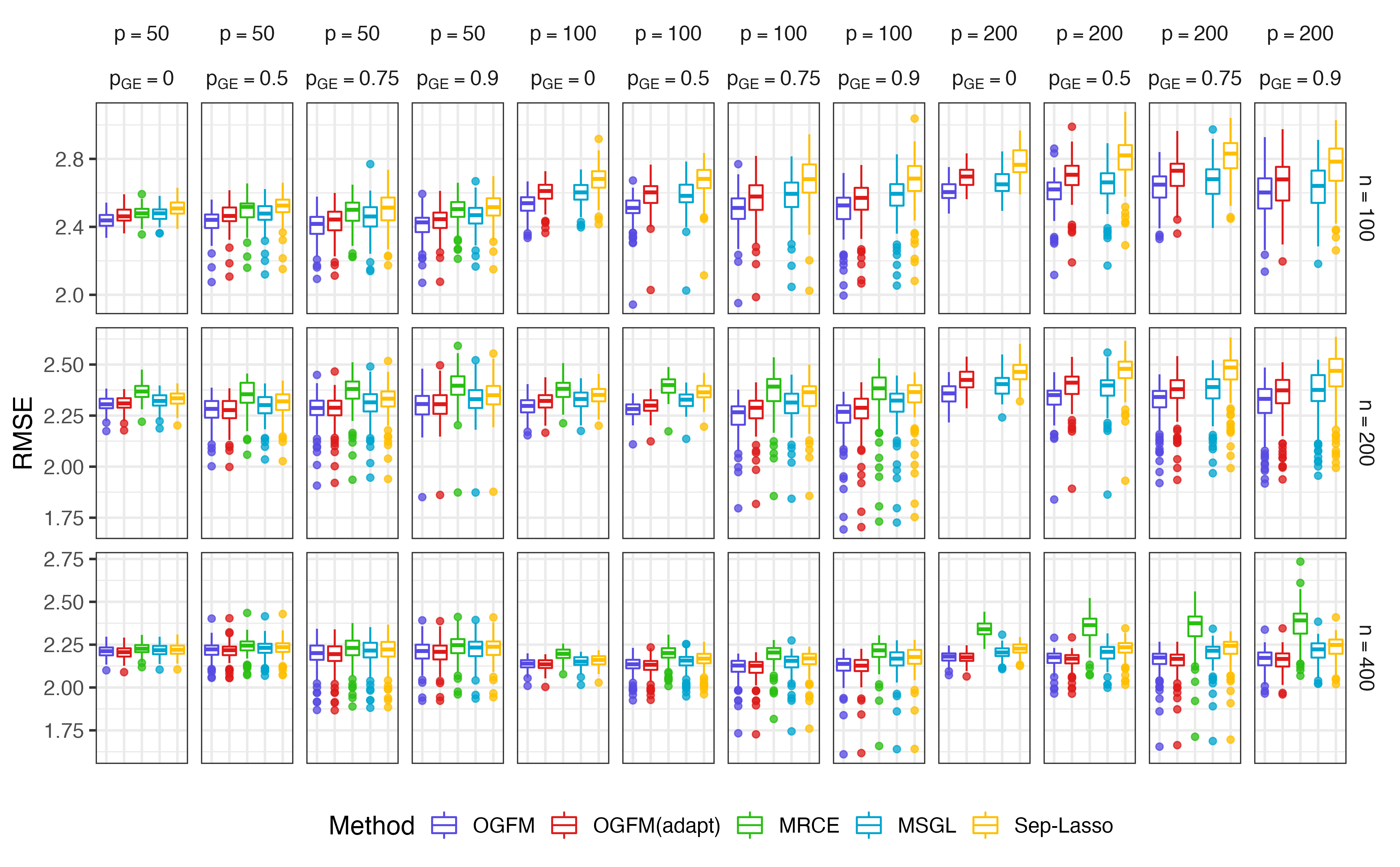}
	\caption{Validation RMSEs for all methods across 100 replications of the simulation experiment holding the parameter $p_{\text{HS}}=0$ so that there is no group-level sparsity. These results pertain to the simulation setting with binary covariates and an ordinal/Likert response.}	
	\label{fig:sim_res_rmse_0hsp_ordinal_y}
\end{figure}

\begin{figure}[!htpb]
	\centering
	\includegraphics[width=1\textwidth]{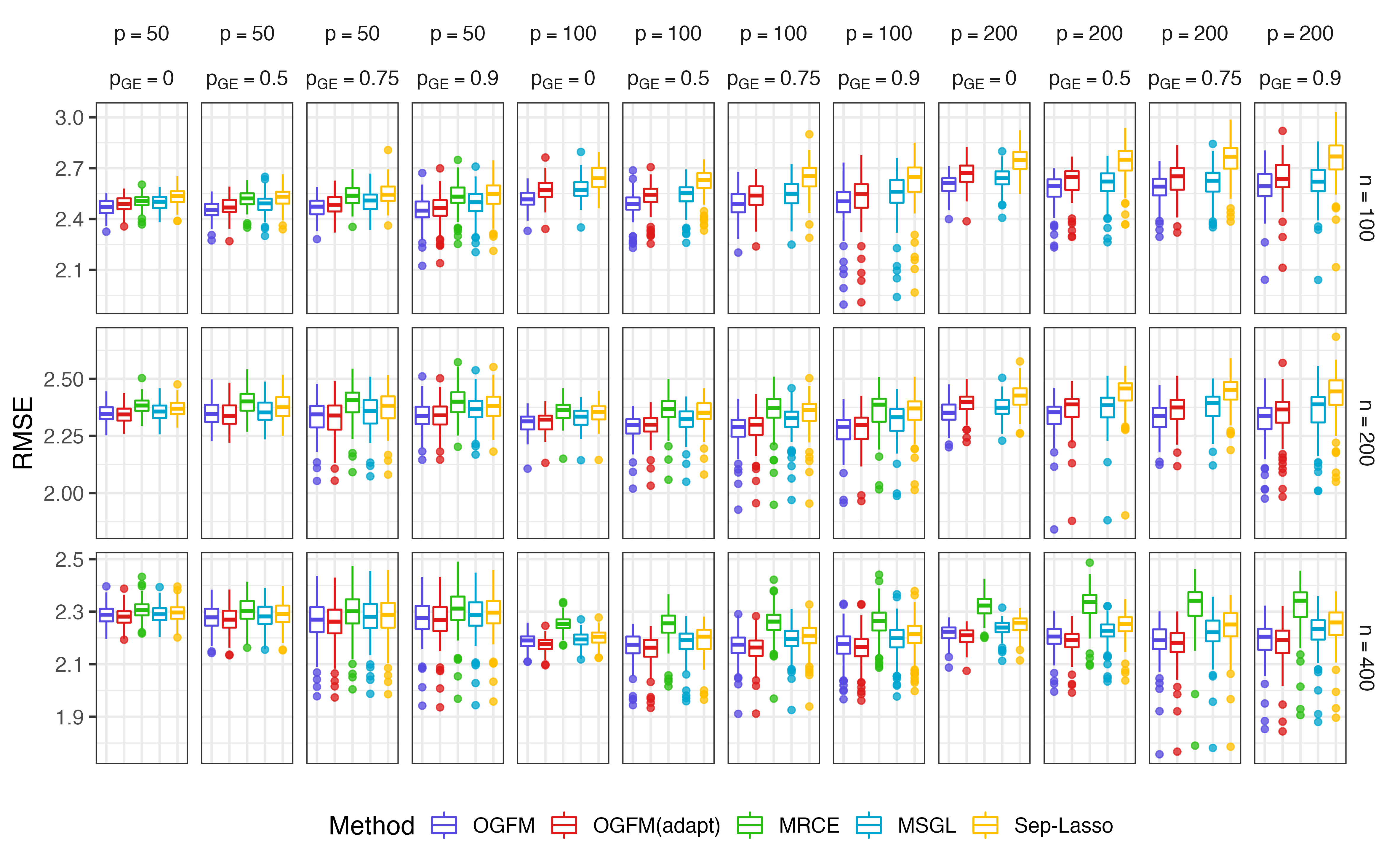}
	\caption{Validation RMSEs for all methods across 100 replications of the simulation experiment holding the parameter $p_{\text{HS}}=0.25$ so that there is moderate group-level sparsity. These results pertain to the simulation setting with binary covariates and an ordinal/Likert response.}	
	\label{fig:sim_res_rmse_025hsp_ordinal_y}
\end{figure}

The RMSE results fixing the parameter at $p_{\text{HS}}=0$ are displayed in Figure \ref{fig:sim_res_rmse_0hsp_ordinal_y} and the RMSE results for $p_{\text{HS}}=0.25$ are displayed in Figure \ref{fig:sim_res_rmse_025hsp_ordinal_y}. The larger trends in the results in terms of both RMSE and model error are quite similar to the results of the main simulation study and thus extended discussion is omitted. 

\begin{figure}[!htpb]
	\centering
	\includegraphics[width=0.85\textwidth]{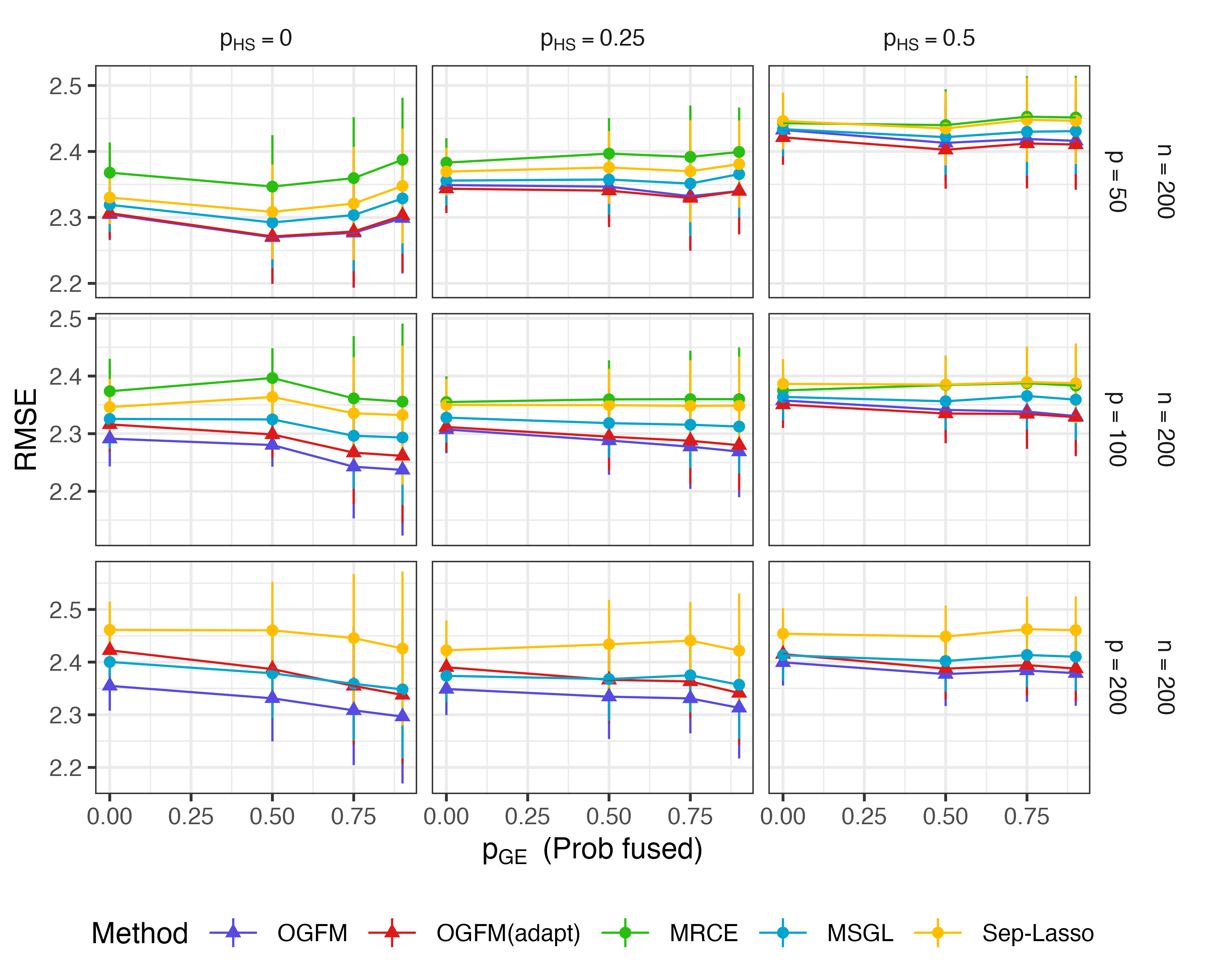}
	\caption{Validation RMSEs for all methods across 100 replications of the simulation experiment holding fixing the sample size $n=200$ and varying all other simulation parameters. The points are the average RMSEs across the 100 replications and error bars are plus and minus 1 standard deviation of this average. These results pertain to the simulation setting with binary covariates and an ordinal/Likert response.}	
	\label{fig:sim_res_rmse_sparsity_view_ordinal_y}
\end{figure}

Figure \ref{fig:sim_res_rmse_sparsity_view_ordinal_y} shows the same results as Figures \ref{fig:sim_res_rmse_0hsp_ordinal_y} and \ref{fig:sim_res_rmse_025hsp_ordinal_y}, but is re-organized to focus on the effect of the fused coefficients probability ($p_{\text{GE}}$) and the hierarchical sparsity probability ($p_{\text{HS}}$). In this figure, we fix the sample size to be 200.

\begin{figure}[H]
	\includegraphics[width=1\textwidth]{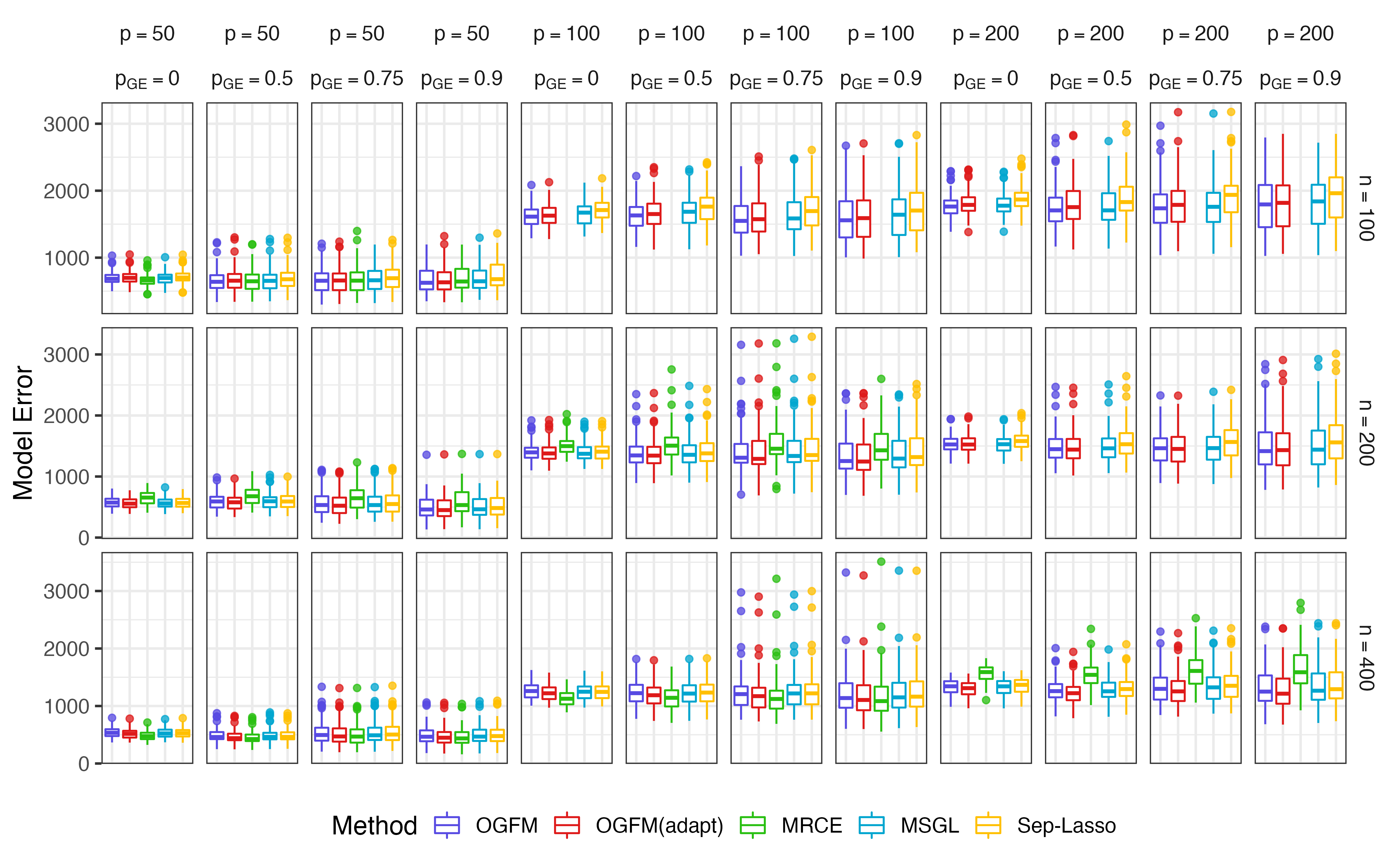}
	\caption{Model errors for all methods across 100 replications of the simulation experiment holding the parameter $p_{\text{HS}}=0$ so that there is no group-level sparsity. These results pertain to the simulation setting with binary covariates and an ordinal/Likert response.}	
	\label{fig:sim_res_model_error_0hsp_ordinal_y}
\end{figure}

\begin{figure}[H]
	\includegraphics[width=1\textwidth]{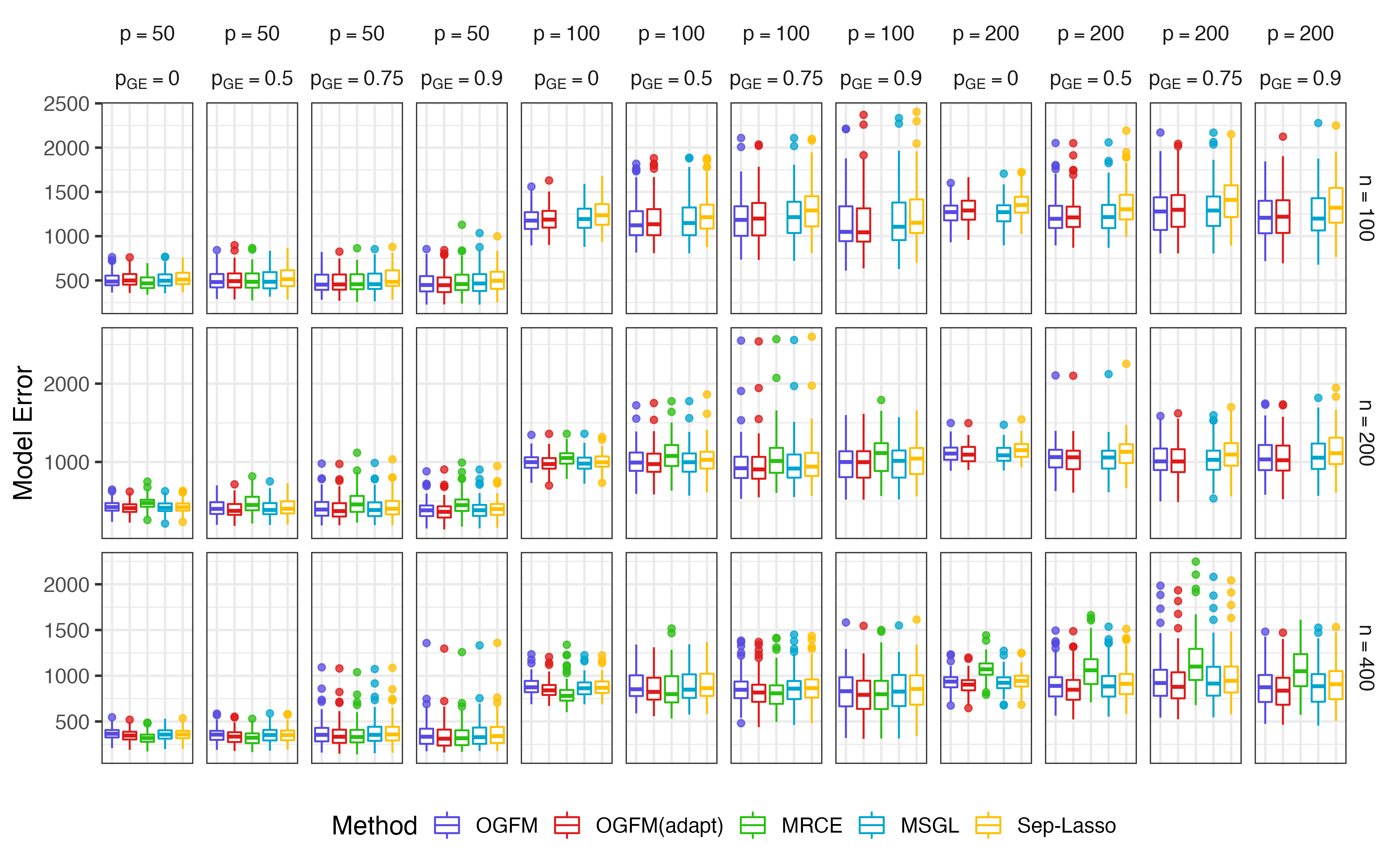}
	\caption{Model errors for all methods across 100 replications of the simulation experiment holding the parameter $p_{\text{HS}}=0.25$ so that there is no group-level sparsity. These results pertain to the simulation setting with binary covariates and an ordinal/Likert response.}	
	\label{fig:sim_res_model_error_025hsp_ordinal_y}
\end{figure}

\begin{figure}[H]
	\includegraphics[width=1\textwidth]{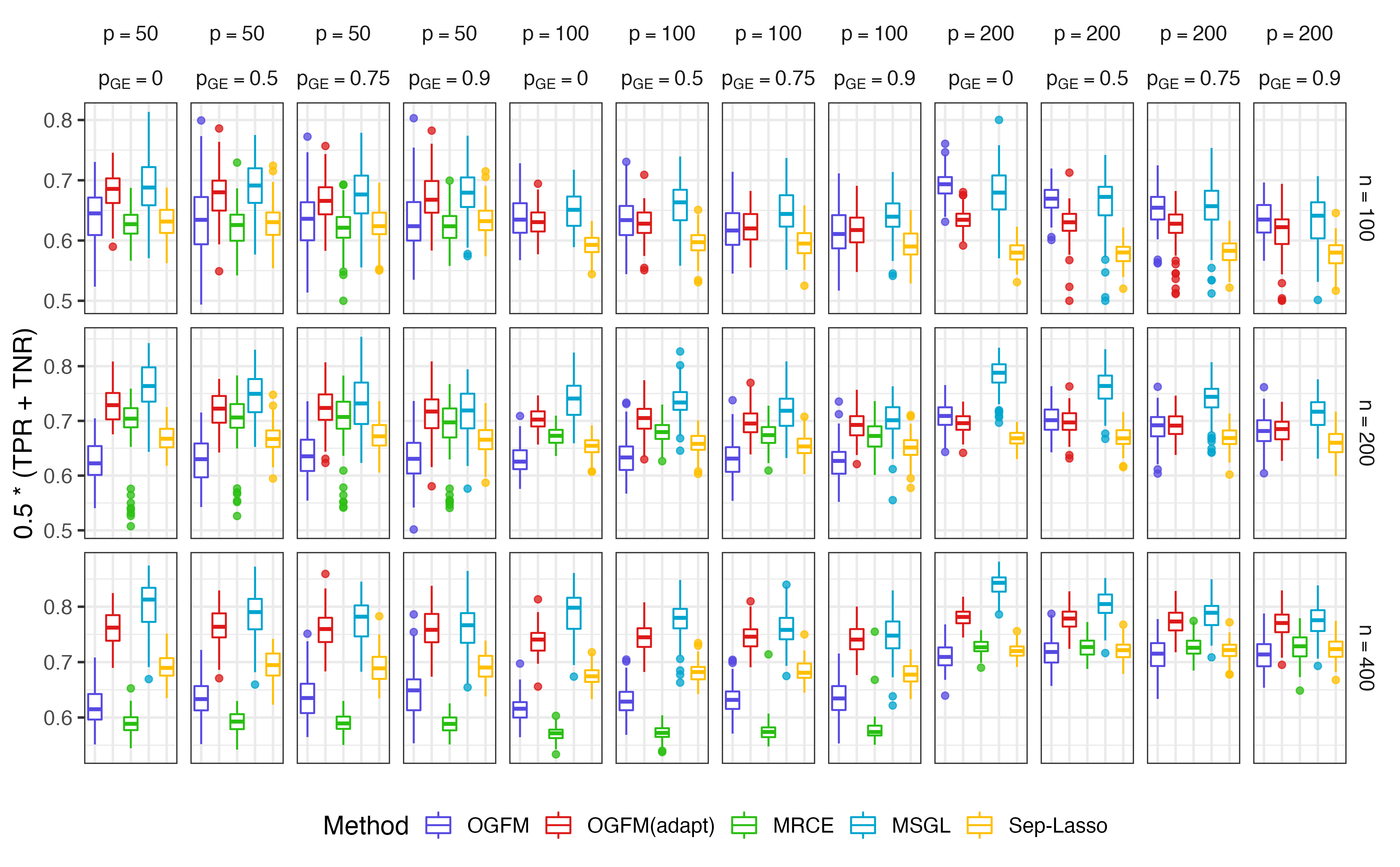}
	\caption{Balanced accuracy (average of TPR and TNR) for all methods across 100 replications of the simulation experiment holding the parameter $p_{\text{HS}}=0$ so that there is no group-level sparsity. These results pertain to the simulation setting with binary covariates and an ordinal/Likert response.}	
	\label{fig:sim_res_balacc_0hsp_ordinal_y}
\end{figure}

\begin{figure}[H]
	\includegraphics[width=1\textwidth]{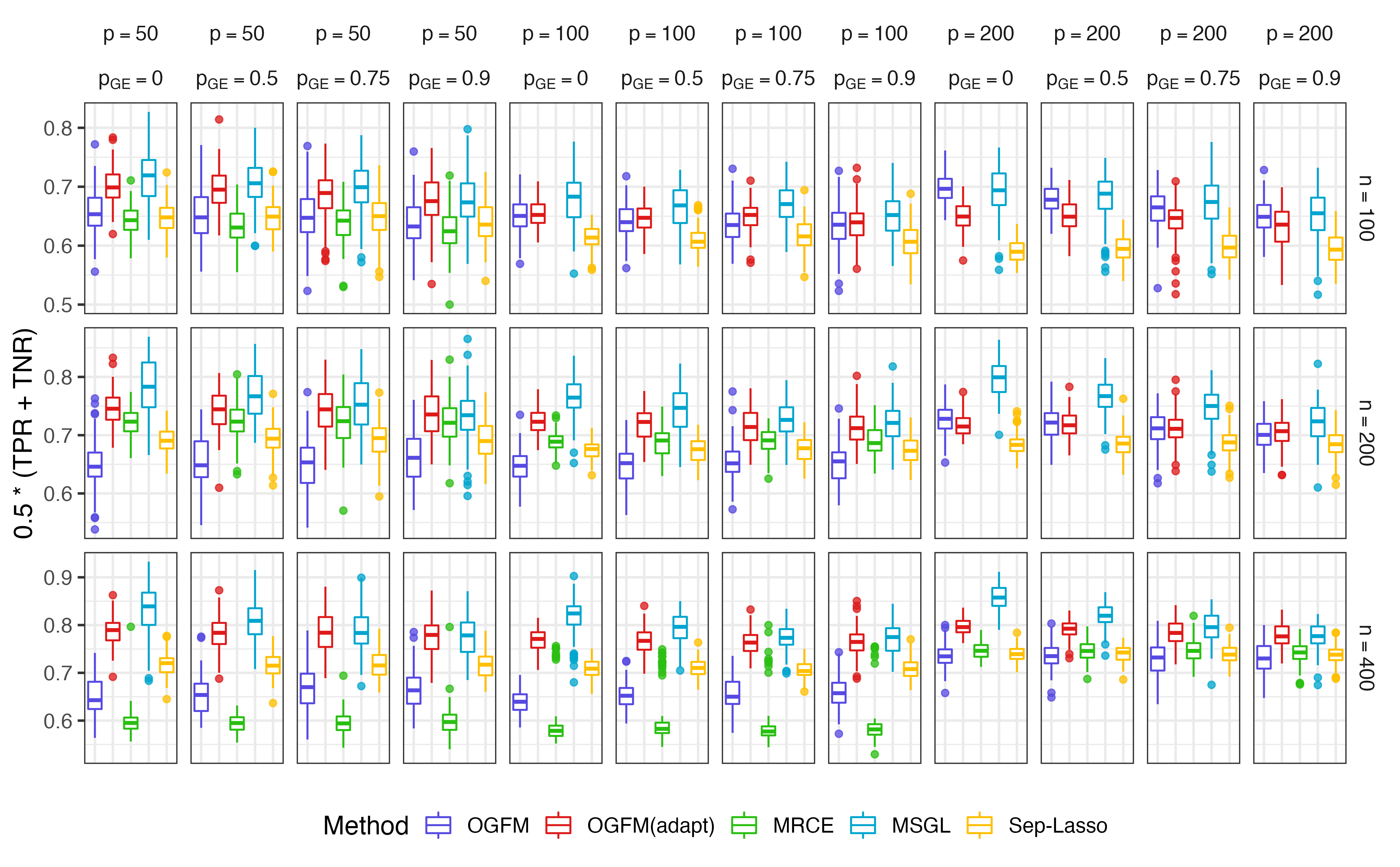}
	\caption{Balanced accuracy (average of TPR and TNR) for all methods across 100 replications of the simulation experiment holding the parameter $p_{\text{HS}}=0.25$ so that there is no group-level sparsity. These results pertain to the simulation setting with binary covariates and an ordinal/Likert response.}	
	\label{fig:sim_res_balacc_025hsp_ordinal_y}
\end{figure}

We also show below simulation results in terms of the average of the TPR and TNR. The results in terms of the model errors fixing $p_{\text{HS}}=0$ are displayed in Figure \ref{fig:sim_res_model_error_0hsp_ordinal_y} and fixing $p_{\text{HS}}=0.25$ are displayed in Figure \ref{fig:sim_res_model_error_025hsp_ordinal_y}. The results in terms of balanced accuracy $p_{\text{HS}}=0$ are displayed in Figure \ref{fig:sim_res_balacc_0hsp_ordinal_y} and \ref{fig:sim_res_balacc_025hsp_ordinal_y}.

\begin{figure}[H]
	\includegraphics[width=1\textwidth]{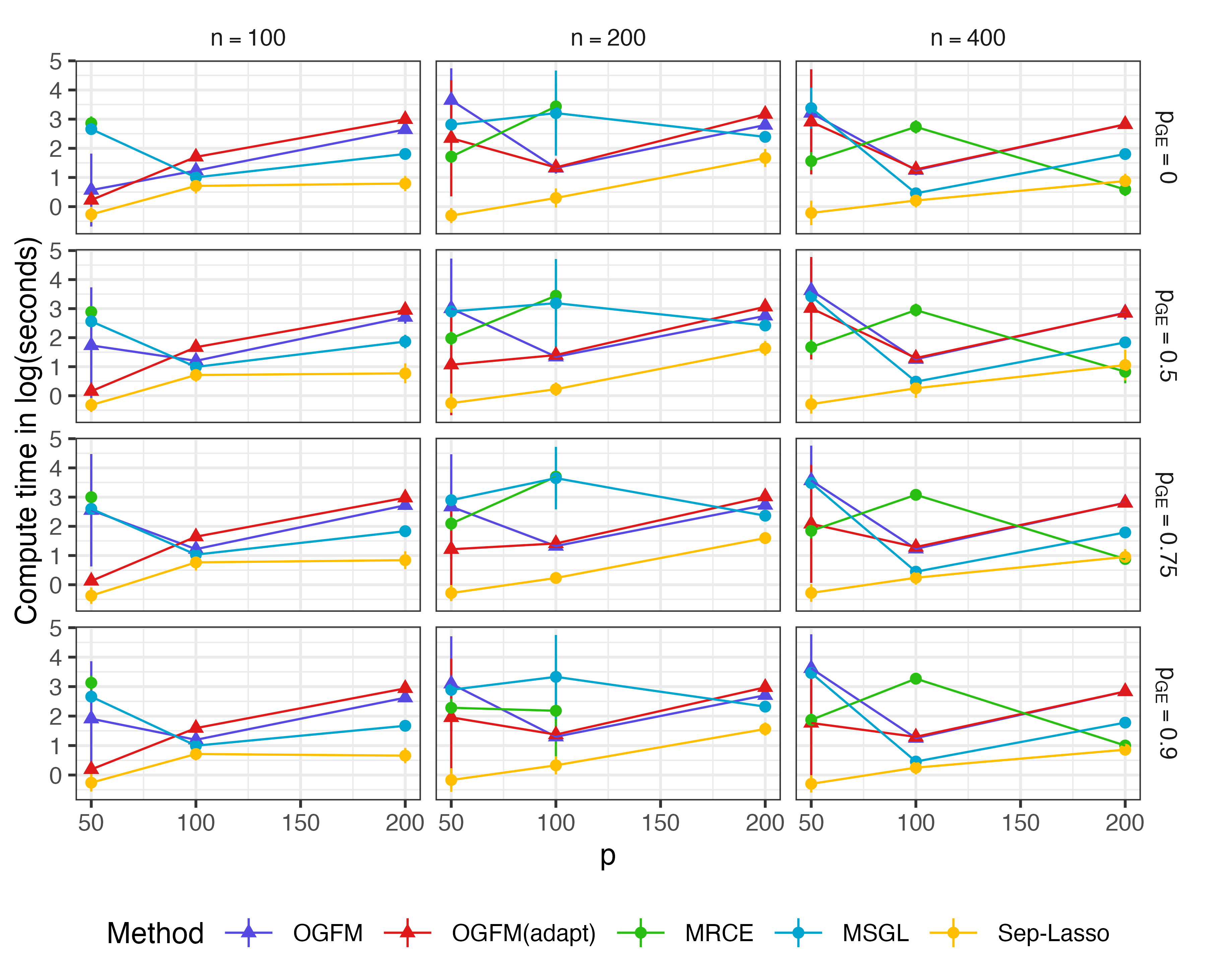}
	\caption{Computation times in terms of log seconds for all methods across 100 replications of the simulation experiment holding the parameter $p_{\text{HS}}=0$ so that there is no group-level sparsity. The points are the average log computation time in seconds and error bars are plus and minus 1 standard deviation. These results pertain to the simulation setting with binary covariates and an ordinal/Likert response.}	
	\label{fig:sim_res_comptime_0hsp_ordinal_y}
\end{figure}

Since the setting with $p_{\text{HS}}=0$ has the most non-zero coefficients, it will in general require the greatest computational effort. As such we show computation times for this setting only. These results are displayed in Figure \ref{fig:sim_res_comptime_0hsp_ordinal_y}.

{
 \bibliographystyle{abbrvnat}
 \bibliography{paper_draft_11-22-2022}
 }

\makeatletter\@input{xx.tex}\makeatother